\documentclass[useAMS,usenatbib]{mn2e}
\usepackage{amssymb}
\usepackage{deluxetable}
\usepackage{url}

\newcommand{\psim}{\lower.5ex\hbox{$\; \buildrel \propto \over\sim \;$}}
\newcommand{\lbar}{\lower.0ex\hbox{$\; \buildrel
{\lower0.0ex \hbox{-}} \over\lambda  \;$}}

\usepackage{booktabs}
\usepackage{epsfig}
\usepackage{subfloat}
\usepackage{subfig}
\usepackage{graphicx}

\title[The high-z $\gamma$-ray flaring FSRQ S5\,0836$+$710]{Radio VLBA polarization and multi-band monitoring of the high-redshift quasar S5\,0836$+$710 during a high activity period} 
\author[M. Orienti, et al.] {M. Orienti$^{1}$\thanks{E-mail: orienti@ira.inaf.it}, F. D'Ammando$^{1,2}$, M. Giroletti$^{1}$, D. Dallacasa$^{1,2}$, G. Giovannini$^{1,2}$,
  \newauthor S. Ciprini$^{3,4}$\\
$^1$INAF -- Istituto di Radioastronomia, Via Gobetti 101, I-40129, Bologna, Italy \\
$^{2}$Dipartimento di Fisica e Astronomia, Universi\`a degli Studi di Bologna, Via Gobetti 93/2, I-40129 Bologna, Italy\\
$^{3}$ Agenzia Spaziale Italiana (ASI), Space Science Data Center, I-00133 Roma, Italy\\
$^{4}$ Istituto Nazionale di Fisica Nucleare (INFN), Sezione di Perugia, I-06123 Perugia, Italy}
\date{Received \today; accepted ?}

\pagerange{\pageref{firstpage}--\pageref{lastpage}} \pubyear{2002}

\def\LaTeX{L\kern-.36em\raise.3ex\hbox{a}\kern-.15em
    T\kern-.1667em\lower.7ex\hbox{E}\kern-.125emX}

\begin{document}

\label{firstpage}

\maketitle


\begin{abstract}

We report on results of a multi-band monitoring campaign from radio to $\gamma$ rays of the high-redshift flat spectrum radio quasar S5\,0836$+$710 during a high activity period detected by the Large Area Telescope on board the {\it Fermi Gamma-ray Space Telescope}. Two major flares were detected, in 2015 August and November. In both episodes, the apparent isotropic $\gamma$-ray luminosity exceeds 10$^{50}$ erg s$^{-1}$, with a doubling time scale of about 3 hours. The high $\gamma$-ray activity may be related to a superluminal knot that emerged from the core in 2015 April at the peak of the radio activity and is moving downstream along the jet. The low variability observed in X-rays may indicate that X-ray emission is produced by the low-energy tail of the same electron population that produces the $\gamma$-ray emission. The analysis of full-polarization pc-scale radio observations suggests the presence of a limb-brightened polarization structure at about 1 mas from the core in which a rotation measure gradient with a sign change is observed transverse to the jet direction. These characteristics are consistent with a scenario in which Faraday rotation is produced by a sheath of thermal electrons with a toroidal magnetic field surrounding the emitting jet.

\end{abstract}


\section{Introduction}

High-redshift blazars ($z > 2$) are among the most powerful objects in the Universe. However, they are not commonly detected in $\gamma$ rays, and represent fewer than 10 per cent of the active galactic nuclei (AGN) detected by the Large Area Telescope (LAT) on board the {\it Fermi Gamma-ray Space Telescope} \citep{ackermann15}. The detection of high-redshift blazars during a
$\gamma$-ray flare is even more uncommon, and only 18 high-redshift
sources have been detected during a flare by {\it Fermi}-LAT so
far \footnote{https://fermi.gsfc.nasa.gov/ssc/data/access/lat/msl\_lc/}.
This may be related to the fact that it is hard for very distant objects to reach a high $\gamma$-ray flux that is needed to identify
  the source in a flaring state. In fact, high-$z$ blazars are
    difficult to detect by the {\em Fermi}-LAT because their inverse
    Compton (IC) peak is usually below the energy range covered by the LAT \citep{dammando16}. This implies that for high-$z$ objects we are observing the
    decreasing part of the IC bump in the GeV regime. However, a hardening of the high energy spectrum during a
flare may favour detection 
\citep[e.g.][]{mo14}. Furthermore,
    $\gamma$-ray emission from distant objects significantly interacts
    with the Extragalactic Background Light \citep[EBL; e.g.,][]{abdollahi18} via $\gamma$-$\gamma$ absorption and is difficult to detect above $\sim$20 GeV. Despite their small fraction in high-energy
catalogues, high-redshift blazars are important for the study of the
energetics and the emission mechanisms in such extreme objects and for
setting constraints on the EBL \citep{dominguez15}. \\ 
\indent Among the flaring high-redshift objects, the flat spectrum radio
quasar (FSRQ) S5\,0836+710 \citep[$z$ = 2.18][]{stickel93} has shown
variability in $\gamma$ rays since the 1990's during EGRET
observations \citep{thompson93} and has been detected during high
activity several times by {\it Fermi}-LAT
\citep[e.g.,][]{akyuz13,ciprini15}.\\
\indent High angular resolution radio observations indicate that
  the relativistic jet of S5\,0836$+$710 has a helical structure
  \citep{perucho12}, 
  and several knots with an apparent superluminal motion have been
  detected \citep{lister13}.
A possible spine-sheath structure
of the jet was suggested by \citet{asada10}. The spectral energy
distribution (SED) of the source is characterized by a strong big blue
bump due to the accretion disc emission peaking at $\sim$
8$\times$10$^{14}$ Hz \citep[e.g.][]{raiteri14}, a low-energy peak in
far infrared and a high-energy peak in the MeV regime
\citep[e.g.,][]{collmar06,sambruna07, tagliaferri15}. \\
\indent After a quiescent period with no significant activity at high energy,
the source entered in an active phase lasting from 2011 March to 2012
January reaching a daily apparent $\gamma$-ray luminosity of
8$\times$10$^{47}$ erg s$^{-1}$ \citep{akyuz13}. The
ejection of a jet component with an apparent superluminal motion of
$\sim$16$c$ was observed close in time with the $\gamma$-ray flare
\citep{jorstad17}. In 2015 August the source entered in a new high
activity phase in which two huge flares were detected by {\it
  Fermi}-LAT, peaking on August 2 \citep{ciprini15} and November
11, with the latter 
detected also by {\it AGILE} \citep{vercellone19}. After the first
flare we triggered a monitoring campaign with the Very Long Baseline
Array (VLBA) in full polarization at 15, 24 and 43 GHz spanning almost
one year. The study of the total intensity and polarization
  variability at different frequencies with high angular resolution
  is crucial for resolving the pc-scale structure of the radio source
  and for locating the variability region either in the core or along
  the jet. Polarimetric observations have proved to be effective also in the
study of magnetic fields associated with relativistic jets from AGN
\citep[e.g.][]{gomez11,hovatta12,gabuzda17}. If the emission is
optically thin, electric vector position angles (EVPA) are
perpendicular to the magnetic field 
\citep{pacho70}. Therefore,  
determining their distribution provides insights into the magnetic
field structure. However, if the radiation passes through a Faraday
screen of magnetized thermal (or mildly relativistic) plasma, its
polarization plane is rotated by:\\

\begin{equation}
\chi_{\rm obs} = \chi_{\rm int} + \frac{e^3 \lambda^2}{8 \pi^2
  \epsilon_0 m_{e}^2 c^3} \int n_{e} B_{||} dl = \chi_{\rm int} + RM
\lambda^2
\label{rm_equation}
\end{equation}

\noindent where $\chi_{obs}$ and $\chi_{int}$ are the observed and the intrinsic
polarization angle, respectively, RM is the rotation measure, $n_{e}$
and $B_{||}$ are the electron density and the magnetic field parallel to the
line of sight of the Faraday screen, $\lambda$ is the wavelength, $e$
is charge of the electron, 
$\epsilon_{0}$ is the vacuum permittivity, $m_{e}$ is the mass of the
electron, and $c$ is the speed of light. Therefore, to determine the
intrinsic orientation of the EVPA, and the structure of the magnetic
field along the jet, we must determine the RM in the various regions
of the radio source. The availability of multi-epoch observations
enables the study of possible variability of the RM and of the
location of the Faraday screen.\\
To complement the radio and high-energy data we have retrieved
{\it Swift} observations in X-rays, UV and optical bands,
in order to investigate the variability at different wavelengths. \\ 
Here we report on the main results achieved by our multi-band
observations of S5 0836+710 during its high activity period. \\

The paper is organized as follows. In Section 2 we present {\it
  Fermi}-LAT data, whereas in Sections 3 and 4 we report on the {\it Swift}
and VLBA observations, respectively. Results are presented
in Section 5 and discussed in Section 6, while a summary is given
in Section 7.\\  
Throughout this paper, we assume the following cosmology: $H_{0} =
71\; {\rm km \; s^{-1} \; Mpc^{-1}}$, $\Omega_{\rm M} = 0.27$ and 
$\Omega_{\rm \Lambda} = 0.73$, in a flat Universe.  At the redshift of
the target, $z$ = 2.218,  
the luminosity distance $D_{\rm L}$ is 17800 Mpc, and 1 milliarcsecond
= 8.37 pc. \\ 

\section{Observations and analysis}

\subsection{{\em Fermi}-LAT Data}
\label{FermiData}

\begin{figure}
\centering
\includegraphics[width=7.5cm]{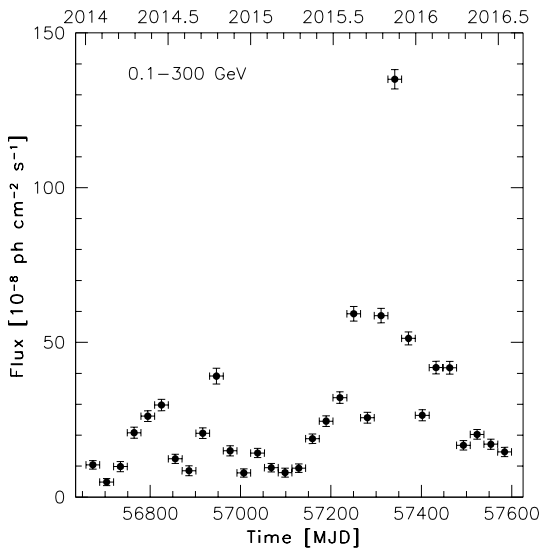}
\caption{Integrated flux LAT light curve of S5\,0836$+$710 obtained using an LP in the 0.1--300 GeV energy range during 2014 January--2016 July with 30-day time bins.} \label{LAT}
\end{figure}

{\em Fermi}-LAT  is a pair-conversion telescope operating from 20
MeV to $>$ 300 GeV. Details about {\em Fermi}-LAT are
given in \citet{atwood09}. The LAT data used in this paper were
collected from 2014 January 1 (MJD 56658) to 2016 July 31 (MJD 57600)
in the 0.1--300 GeV energy range. 
 Following the procedure reported in \citet{dammando16}\footnote{See also https://fermi.gsfc.nasa.gov/ssc/data/analysis/scitools/\\/binned\_likelihood\_tutorial.html for details.}, the analysis was performed with the \texttt{ScienceTools} software package version v10r0p5. We used Pass 8 data \citep{atwood13}, selecting events belonging to the `Source' class  within a maximum zenith angle of 90$^{\circ}$ to reduce contamination from the Earth limb $\gamma$ rays.
The spectral analysis was performed with the instrument response
functions \texttt{P8R2\_SOURCE\_V6} using a binned maximum-likelihood
method. 
Isotropic (`iso\_source\_v06.txt') and Galactic diffuse emission
(`gll\_iem\_v06.fit') components were used to model the background
\citep{acero16}\footnote{http://fermi.gsfc.nasa.gov/ssc/data/access/lat/\\BackgroundModels.html}. \\
We analysed a region of interest of $30^{\circ}$ radius centred at the location of S5\,0836$+$710. We evaluated the significance of the
$\gamma$-ray signal from the source by means of a maximum-likelihood test statistic (TS)\footnote{$\sqrt{TS}$ approximately corresponds to $\sigma$} defined as TS = 2$\times$(log$L_1$ - log$L_0$), where the likelihood $L$ is the probability of obtaining the data given the model with ($L_1$) or without ($L_0$) a point source at the position of S5\,0836$+$710 \citep[e.g.,][]{mattox96}. The source model used in \texttt{gtlike}
includes all the point sources from the 3FGL catalogue that fall
within $40^{\circ}$ of S5\,0836$+$710. The spectra of these sources
were parametrized by a power-law (PL), a log-parabola (LP), or a super
exponential cut-off, as in 
the 3FGL catalogue. We also included new candidates within $7^{\circ}$
of S5\,0836$+$710 from the LAT 8-year point source list
(FL8Y\footnote{https://fermi.gsfc.nasa.gov/ssc/data/access/lat/fl8y/}). \\
We used an iterative procedure to remove sources having TS $< 25$ from the model.
In the fitting procedure, the normalization factors and the spectral parameters of the sources within 10$^{\circ}$ of
S5\,0836$+$710 were left as free parameters. \\ 
\indent Integrating over the entire period the fit with an LP model, $dN/dE
\propto$ $(E/E_{0})^{-\alpha + \beta log (E/E_{0})}$, where $E_{0}$ is
fixed to 236 MeV as in the 3FGL catalogue, results in a TS
= 15743 in the 0.1--300 GeV energy range, with $\alpha$ = 2.59 $\pm$
0.02, $\beta$ = 0.19 $\pm$ 0.01, and a flux of (27.7 $\pm$
0.2)$\times$10$^{-8}$ ph cm$^{-2}$ s$^{-1}$. The corresponding
apparent isotropic $\gamma$-ray luminosity is
(9.1$\times$0.1)$\times$10$^{48}$ erg 
s$^{-1}$. As a reference, in the 3FGL catalogue the spectrum of the
source is described by an LP with $\alpha$ = 2.62 $\pm$ 0.05, and
$\beta$ = 0.19 $\pm$ 0.03, indicating no significant changes in the
average spectrum between the first four years of LAT operation (i.e.,
2008 August--2012 July) and the period studied here (i.e., 2014
January--2016 July). \\
\indent Fitting the entire dataset with a PL model, $dN/dE \propto$
$(E/E_{0})^{-\Gamma_{\gamma}}$, results in a TS = 15725 in the
0.1--300 GeV energy range, with an integrated average flux of (28.8
$\pm$ 1.6)$\times$10$^{-8}$ ph cm$^{-2}$ s$^{-1}$ and a photon index
of $\Gamma_\gamma$ =  2.78 $\pm$ 0.01. 
We used a likelihood ratio test to check the PL model (null
hypothesis) against the LP model (alternative hypothesis). These
values may be compared by defining the curvature test statistic
TS$_{\rm curve}$=TS$_{\rm LP}$--TS$_{\rm PL}$=18 ($\sim$
4.2-$\sigma$), meaning that we have statistical evidence of a curved
spectral shape.
Fig. \ref{LAT} shows the $\gamma$-ray flux evolution for the period 2014 January 1 -- 2016
July 31 (MJD 56658--57600) using an LP model and 1-month time bins with the spectral parameters
fixed to values obtained over the entire period.
Leaving the spectral parameters free to vary on a monthly time-scale during
the high activity period, at the peak of the activity (2015 November), the fit with an LP results in a TS = 6491 with $\alpha$ = 2.38 $\pm$ 0.05, $\beta$ = 0.29 $\pm$ 0.04, and a flux of (130.5 $\pm$ 3.3)$\times$10$^{-8}$ ph cm$^{-2}$ s$^{-1}$. 

\begin{figure}
\begin{center}
{\resizebox{!}{80mm}{\includegraphics{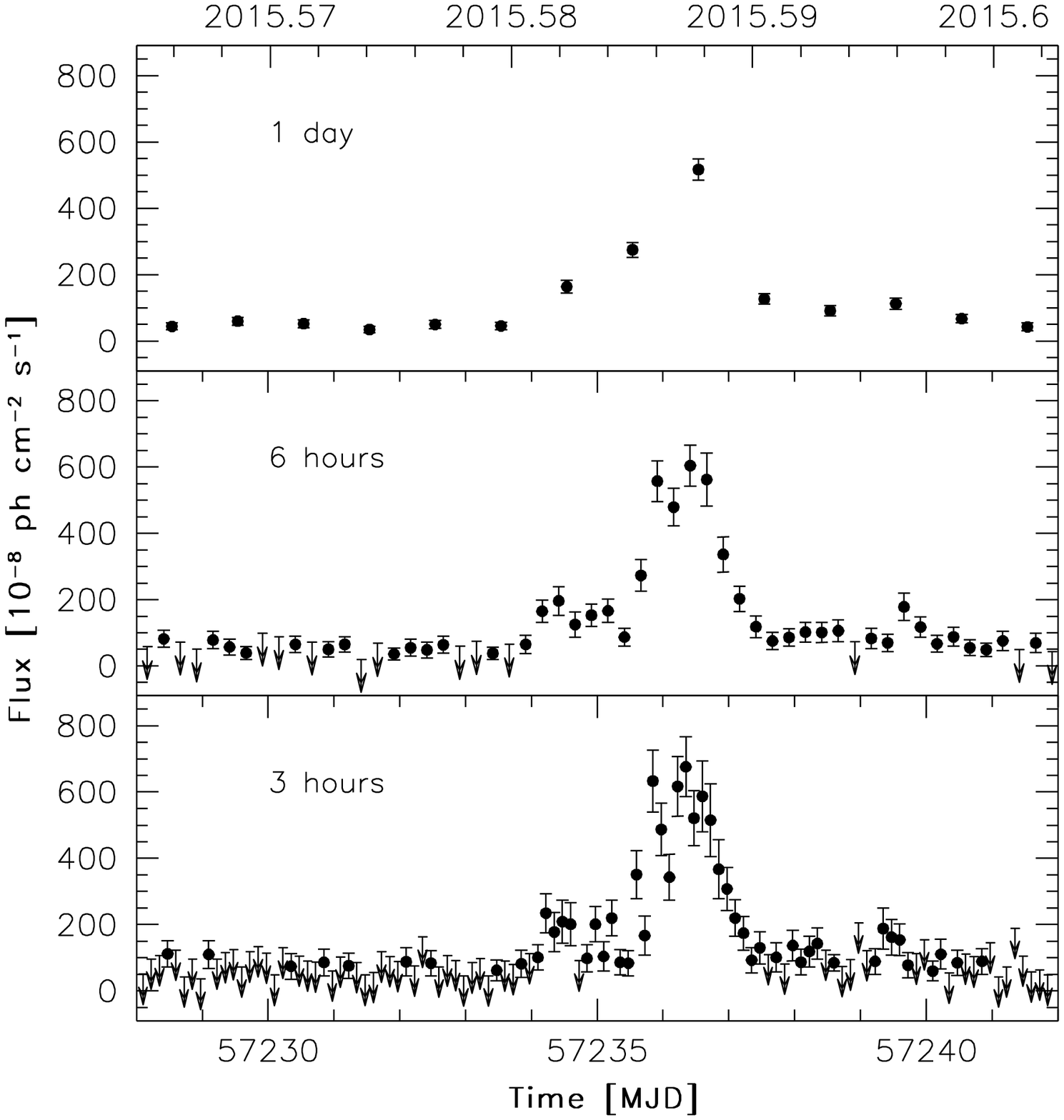}}}
{\resizebox{!}{80mm}{\includegraphics{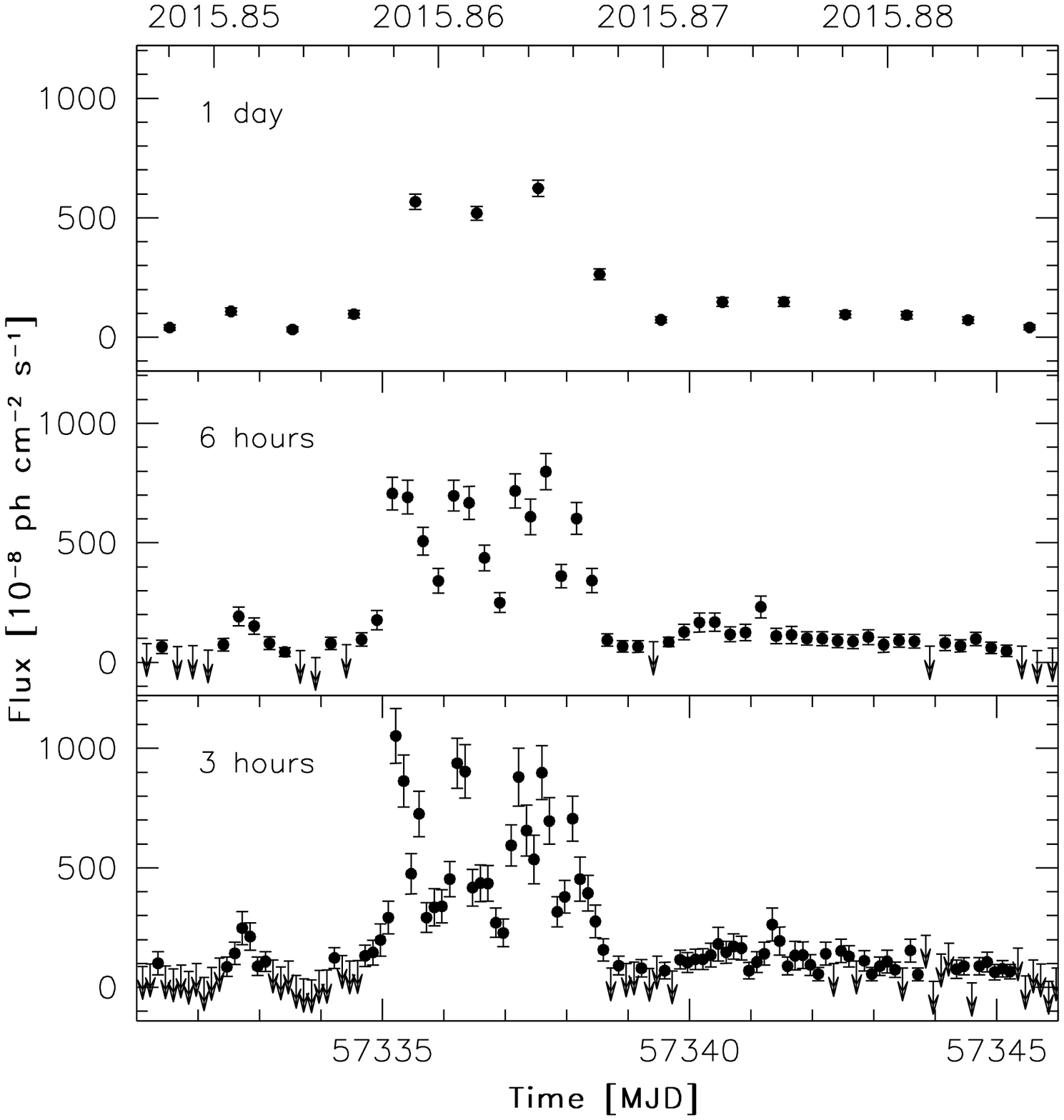}}}
\caption{Integrated {\it Fermi}-LAT flux light curve of S5\,0836$+$710
  obtained using an LP in the 0.1--300 GeV energy range during 2015 July
  25--August 7 (upper plot) and 2015 November 5--19 (bottom plot), with, from
  top to bottom, 1-d time bins, 6-h time bins, and 3-h time bins. Arrows refer to 2$\sigma$ upper limits on the source flux. Upper limits are computed when TS $<$ 10.}
\label{flare12}
\end{center} 
\end{figure}

On a monthly time-scale, the source is always detected and shows an
increase of activity starting from 2015 May with a first peak on 2015
August and the maximum reached on 2015 November. We investigated rapid flux variations during these two high activity periods by producing light curves with different time bins. Fig.~\ref{flare12} presents the light curve for the period 2015 July 25--August 7 (MJD 57228--57241; upper plot) and 2015 November 5--19 (MJD 57331-57345; lower plot), with 1-d (top panel), 6-h (middle panel), and 3-h (bottom panel) time bins using an LP. In the analysis of the sub-daily light curves,
we fixed the flux 
of the diffuse emission components at the value obtained by fitting
the data over the entire period analyzed in this paper. For each time
bin, the spectral parameters of S5\,0836$+$710 and all sources within
10$^{\circ}$ of it were frozen to the values resulting from the
likelihood analysis in the monthly time-bins. In both flares flux
  variations by a factor of 2 or more occurring on a 6-h time-scale are
  clearly visible. On the other hand, a possible double-peak structure is
  recognizable only in the 3-h light curve for the first flare. In the second
  flare, the 3-h light curve shows a different behaviour between sub-flares: a
  rising time shorter than the decaying time in the first sub-flare, and a
  comparable rising and decaying time in the other sub-flares. The rough
  symmetry of the sub-flares suggests that the relevant time-scale should not
  be too different from the light crossing time of the emitting region
  \citep[e.g.,][]{tavecchio10}. However, due to the low-statistics, we cannot make a definitive statement about the shape of the flares. \\

\indent By means of the \texttt{gtsrcprob} tool, we have estimated that the highest
energy photon emitted by S5 0836$+$710 (with probability $>$ 90 per
cent of being 
associated with the source) was observed on 2016 January 27 at a
distance of 0\fdg10 from the target with an energy of 15.3
GeV\footnote{At 15 GeV the LAT point spread function (68 per cent
  containment angle, front$+$back events) is $\sim$0.15$^{\circ}$.}.  \\

\subsection{{\it Swift} data}
\label{SwiftData}

The {\it Neil Gehrels Swift Observatory} \citep{gehrels04} carried out
forty-three observations of S5\,0836$+$710 between 2014 January 18 (MJD 56675) and
2016 July 3 (MJD 57572). The observations were performed with all three
instruments on board: the X-ray Telescope \citep[XRT; 0.2--10.0 keV;][]{burrows05}, the Ultraviolet/Optical Telescope \citep[UVOT; 170--600 nm;][]{roming05} and the Burst Alert Telescope \citep[BAT; 15--150 keV][]{barthelmy05}. \\
The hard X-ray flux of this source is below the
sensitivity of the BAT instrument for such short exposures and
therefore the data from this instrument collected during single
observations will not be used. However, the source is included in the
{\em Swift} BAT 105-month hard X-ray catalogue \citep{oh18}.\\
\indent The XRT data were processed with standard procedures
(\texttt{xrtpipeline v0.13.3}), filtering, and screening criteria by
using the \texttt{HEAsoft} package (v6.22). The data were collected in
photon counting mode in all the observations.  
The source count rate in some observations is higher than 0.5 counts s$^{-1}$:
these observations are checked for pile-up and a correction was applied
following standard procedures \citep[e.g.,][]{moretti05}. To correct for
pile-up we excluded from the source extraction region the inner circle of 3
pixel radius by considering an annulus region with outer radius of 30 pixels
(1 pixel $\sim$ 2.36 arcsec). For the other observations, source events were
extracted from a circular region with a radius of 20 pixels. Background events
were extracted from a circular region with radius of 50 pixels far away from
the source region. Ancillary response files were generated with
\texttt{xrtmkarf}, and account for different extraction regions, vignetting
and point spread function corrections. We used the spectral redistribution
matrices v014 in the Calibration data base maintained by \texttt{HEASARC}\footnote{https://heasarc.gsfc.nasa.gov/docs/heasarc/caldb/}. Data were grouped into a minimum of 20 counts per bin in order to apply $\chi$$^2$ spectrum fitting. Bad channels, including zero-count bins, were ignored in the fit.
We fitted the spectrum with an absorbed PL using the
photoelectric absorption model \texttt{tbabs} \citep{wilms00}, with a
neutral hydrogen column density fixed to its Galactic value
\citep[$N_{\rm H}$ = 2.83$\times$10$^{20}$
  cm$^{-2}$;][]{kalberla05}. The results of 
the fit are reported in Table~\ref{XRT_table}. The unabsorbed fluxes
in the 0.3--10 keV energy range are reported in Fig.~\ref{mwl}. \\
\indent For the longest {\em Swift} observation ($\sim$ 9.7 ks) carried out on 2014
October 17, we tested additional absorption at the redshift of the source
leaving $N_{\rm H}$ free to vary. The fit does not improve ($\chi^{2}$ /
d.o.f. = 154/168) with respect to a PL with the absorption fixed to the
Galactic value ($\chi^{2}$ / d.o.f. = 161/170), obtaining $N_{\rm H}$ = 5.1$^{+1.7}_{-1.6}$ $\times$10$^{20}$
cm$^{-2}$ and $\Gamma$ = 1.25 $\pm$ 0.06. We also tested an LP
model for checking spectral curvature of the X-ray spectrum. No improvement of
the fit is achieved using an LP ($\chi^{2}$ / d.o.f. = 154/169), with a slope
$\alpha$ = 1.10 $\pm$ 0.07 and a curvature parameter $\beta$ =
0.16$^{+0.11}_{-0.10}$. No substantial absorption is seen in
addition to Galactic and no spectral curvature is
observed in the X-ray spectrum of the source, in agreement with the
results obtained with {\em XMM-Newton} data \citep{vercellone19}. \\
\indent During the {\em Swift} pointings, the UVOT instrument observed S5
0836$+$710 in all its optical ($v$, $b$, and $u$) and UV ($w1$, $m2$,
and $w2$) photometric bands \citep{poole08,breeveld10}. We analysed
the data using the \texttt{uvotsource} task included in the
\texttt{HEAsoft} package (v6.22). Source counts were extracted from a
circular region of 5 arcsec radius centred on the source, while
background counts were derived from a circular region of 10 arcsec
radius in a nearby source-free region. The observed magnitudes are
reported in Table~\ref{UVOT}. The UVOT flux densities, 
corrected for extinction using the E(B--V) value of 0.026 from
\citet{schlafly11} and the extinction laws from \citet{cardelli89},
are reported in Fig.~\ref{mwl}. \\

\subsection{Radio data}

\subsubsection{VLBA observations and data reduction}
\label{vlba_obs}

Multi-frequency VLBA observations (project code BO051) of
S5\,0836$+$710 triggered by the 2015 August $\gamma$-ray flare were 
carried out at 15, 24 and 43 GHz during six observing epochs between
2015 August and 2016 July (Table \ref{log-vlba}),
with a recording bandwidth of 128 MHz at 2048
Mbps data rate. 
During each observing epoch, the source was observed for 50 min at 15
and 24 GHz, and for 90 min at 43 GHz, spread into 17 scans at 15
  and 24 GHz and 31 scans at 43 GHz to improve
the {\it uv}-coverage\footnote{The {\it uv}-coverage indicates how well
    the visibility plane is sampled. The visibility plane is the
    Fourier Transform of the brightness distribution of the sky as
    observed by an interferometer. For more details on radio astronomy
    see \citet{rohlfs86}.}. The duration of each scan is about 3
min.
For this reason the flux density
measurements of the 
pc-scale emission at the various frequencies can be considered roughly
simultaneous during each epoch.  
The observing epochs are separated by about 2 months.\\
\indent The initial data reduction and calibration were performed following
the standard procedures described in the NRAO's Astronomical Image
Processing System (\texttt{AIPS}) cookbook. The pulse calibration
signals were used in all the experiments to align the phases across
the intermediate frequencies (IFs). J0927$+$3902 was
used to generate the bandpass correction. The amplitudes were
calibrated using the 
antenna system temperatures and antenna gains
and applying an atmospheric opacity correction.
The uncertainties on the amplitude calibration were found 
to be approximately 
7 per cent at 15 and  24 GHz, and about 10 per cent at 43 GHz.
The target source S5\,0836$+$710 is
strong enough at all frequencies to allow the fringe fitting with a
solution interval of 
one/two minutes to preserve the phase coherence. \\
\indent For each frequency and epoch we determined the amplitudes and phases
of the complex feed leakage terms for  
each IF and antenna by using the \texttt{AIPS} task \texttt{LPCAL}.
The absolute EVPA was calibrated by using a knot in the jet at about 2.9 mas\footnote{This knot corresponds to component J in Fig.~\ref{full_res}.}, whose EVPA is relatively stable (EVPA $\sim$ $-$85$^{\circ}$ to $\sim$ $-$89$^{\circ}$) between 2015 September and 2016 June \citep[see Fig.~4 in][]{lister18}.
Furthermore, we confirmed the stability of the EVPA by performing two
epochs of Jansky 
Very Large Array (JVLA) observations of
S5\,0836$+$710 close in time with the VLBA observations of 2016 May and
July. Values of the VLA polarization and integrated polarization parameters of the VLBA images are reported in Table~\ref{integrated_pol}. 
The absolute error on the EVPA is about
5$^{\circ}$ $-$ 8$^{\circ}$ at all frequencies.\\ 
At 43 GHz we complemented our VLBA data with additional observations
from the VLBA Boston University (BU) blazar (VLBA-BU-BLAZAR) programme
with the aim of 
investigating the proper motion of jet components and the long-term
variability. Information on the 
monitoring programme and on data calibration can be found in
\citet{jorstad17}.\\

\begin{table*}
\caption{Integrated image parameters. Column 1: observation epoch; column 2: telescope; column 3; frequency band; column 4: full width half maximum (FWHM) major and minor axis of the restoring beam; Column 5: FWHM minor axis of the restoring beam; column 6: position angle of major axis of restoring beam; column 7: stokes I total flux density measured on the full resolution image; column 8: polarized flux density measured on the full resolution image; column 9: integrated EVPA.}
\begin{center}
\begin{tabular}{ccccccccc}
Epoch & Telescope & Band & $\theta_{\rm maj}$ & $\theta_{\rm min}$ & p.a. &
S$_{\rm I}$ & S$_{\rm P}$ & EVPA \\
  & & & mas & mas & $^{\circ}$ & Jy & mJy& $^{\circ}$ \\
\hline
2015-08-21 & VLBA & U & 0.87 & 0.42 & -4  & 2.19$\pm$0.15 & 17.3$\pm$1.2 & 60 \\
           &      & K & 0.53 & 0.27 & -5  & 2.08$\pm$0.14 & 44.3$\pm$3.1 & 76 \\
	   &      & Q & 0.30 & 0.14 &  0  & 1.66$\pm$0.25 & 34.6$\pm$5.2 & 128 \\
2015-10-23 & VLBA & U & 1.02 & 0.61 & -25 & 2.36$\pm$0.16 & 62.7$\pm$4.4 & 57 \\
           &      & K & 0.66 & 0.40 & -34 & 2.26$\pm$0.16 & 42.3$\pm$3.0 & 78 \\
	   &      & Q & 0.40 & 0.21 & -15 & 1.58$\pm$0.24 & 28.6$\pm$4.3 & 129 \\
2016-01-02 & VLBA & U & 0.88 & 0.41 &  13 & 2.51$\pm$0.17 & 13.3$\pm$0.9 & 68 \\
           &      & K & 0.53 & 0.26 &   3 & 2.04$\pm$0.14 & 38.0$\pm$2.7 & 90 \\
	   &      & Q & 0.35 & 0.15 &  14 & 1.27$\pm$0.19 & 32.8$\pm$4.9 & 132 \\
2016-03-15 & VLBA & U & 0.93 & 0.45 &   1 & 2.25$\pm$0.16 & 19.2$\pm$1.3 & 88 \\
           &      & K & 0.58 & 0.30 &  -3 & 1.99$\pm$0.14 & 41.7$\pm$2.9 & 128 \\ 
	   &      & Q & 0.35 & 0.15 &  14 & 1.28$\pm$0.19 & 33.0$\pm$4.9 & 135 \\
2016-05-14 & VLBA & U & 1.07 & 0.56 &  0 & 2.28$\pm$0.16 & 24.5$\pm$1.7 & 105 \\
           &      & K & 0.84 & 0.54 & -6 & 1.81$\pm$0.13 & 38.2$\pm$2.7 & 122 \\
	   &      & Q & 0.37 & 0.20 &  1 & 1.37$\pm$0.20 & 22.3$\pm$3.3 & 102 \\
2016-07-07 & VLBA & U & 0.86 & 0.40 &-10 & 2.17$\pm$0.15 & 41.2$\pm$2.9 & 116 \\
           &      & K & 0.52 & 0.26 &-13 & 1.59$\pm$0.11 & 32.2$\pm$2.2 & 78 \\
	   &      & Q & 0.33 & 0.14 & -5 & 0.97$\pm$0.14 & 21.1$\pm$3.1 & 133 \\
\hline
2016-05-10 & VLA & K & 0.34\tablenotemark{1} & 0.25\tablenotemark{1} &  32 & 2.60$\pm$0.13 & 52$\pm$3 & 118 \\
           &     & Q & 0.20\tablenotemark{1} & 0.14\tablenotemark{1} &  50 & 1.50$\pm$0.15 & 64$\pm$8 & 120 \\
2016-09-03 & VLA & U & 0.79\tablenotemark{1} & 0.40\tablenotemark{1} & -56 & 2.80$\pm$0.14 & 40$\pm$5 & 120 \\
           &     & K & 0.45\tablenotemark{1} & 0.26\tablenotemark{1} & -82 & 1.95$\pm$0.10 & 35$\pm$5 & 120 \\ 
          &      & Q & 0.21\tablenotemark{1} & 0.13\tablenotemark{1} & -77 & 1.40$\pm$0.07 & 25$\pm$5 & 165 \\
\hline
\end{tabular}
\end{center}
\tablenotetext{1}{The VLA FWHM major and minor axes are in arcseconds.}
\label{integrated_pol}
\end{table*}

\subsubsection{Radio images}
\label{radioim_section}

The calibrated data were edited and normal
  self-calibration and imaging techniques were then used within
  AIPS. Data were self-calibrated against the model in phase only
  and in both phase and amplitude on a 30 second timescale.
  Final images were 
  produced in Stokes I, Q, and U. 
The 1-$\sigma$ noise
level of the full-resolution images measured on the image plane is
about 0.1--0.3 mJy/beam.
Images at the same frequency were reconstructed with the same
restoring beam, that is 0.9$\times$0.5 mas$^2$ at 15 GHz,
0.6$\times$0.3 mas$^2$ at 24 GHz, and 0.38$\times$0.16 mas$^2$ at 43 GHz.
With the aim of producing spectral-index and rotation measure images,
for each frequency and at each epoch we produced another set of images 
in Stokes I, Q and U with the same {\it uv}-range between 29.3 and 280
M$\lambda$. Furthermore, the images were produced at 
the different frequencies with the same image sampling, natural grid weighting
and, in the case of 24 and 43 GHz, by forcing the beam major and minor
axes, and position angle to be equal to that of the 15-GHz image
(i.e. 0.9$\times$0.5 mas$^2$). Spectral-index images between 15 and 24
GHz and between 24 and 43 GHz plus the
associated statistical error images were produced by means of the
\texttt{AIPS} task \texttt{COMB}. Blanking was done clipping the
pixels of the input images with 
values below five times the rms measured on the off-source image
plane at each frequency. For each epoch we checked the image alignment
at the different frequencies by comparing the position of the bright optically-thin jet component that we have also used to calibrate the EVPA (see Section~\ref{vlba_obs}), and whose position should not depend on the observing frequency
\citep[e.g.][]{lobanov98}. The 
absolute shift between the 15 GHz and the other frequencies is
between 0 mas and 0.7 mas. If necessary, we shifted the Stokes I, Q,
and U images of the same amount using the \texttt{AIPS} task \texttt{LGEOM}. \\
\indent Images in Stokes Q and U
were then used to produce the polarization intensity and polarization
angle images, as well as the associated statistical error
images. Blanking was done clipping the 
pixels of the input images with 
values below five times the rms measured on the off-source image
plane at each frequency. For
each epoch, the polarization angle images and the associated
statistical error images at the three frequencies were combined with
the \texttt{AIPS} task \texttt{RM} to produce the RM
images, RM-corrected magnetic field images and the associated
statistical error images. Blanking was done clipping the
pixels of the input images with 
values on the polarization angle error image larger than the uncertainties
determined following the formulae reported in \citet{hovatta12}.\\   

\begin{table}
\caption{Log of VLBA observations. Column 1: date of
    observations; Column 2: epoch code; Column 3: frequency band; Column 4:
    notes. Mk and Pt refer to Mauna Kea antenna and Pie Town antenna, respectively.}
\begin{center}
\begin{tabular}{cccc}
\hline
Date& Code& Band& Notes\\
\hline
2015-08-21& A& U K Q& \\
2015-10-23& B& U K Q& No Mk \\
2016-01-02& C& U K Q&  \\
2016-03-15& D& U K Q&Pt\tablenotemark{a}\\\
2016-05-14& E& U K Q&Pt\tablenotemark{a}\\
2016-07-07& F& U K Q&Pt\tablenotemark{a,b}\\
\hline
\end{tabular}
\end{center}
\tablenotetext{a}{High K-band R/L cross polarization due to receiver swap.}
\tablenotetext{b}{Warm U-band receiver.}
\label{log-vlba}
\end{table}

\section{Results}

\begin{figure*}
\centering
\includegraphics[width=11cm]{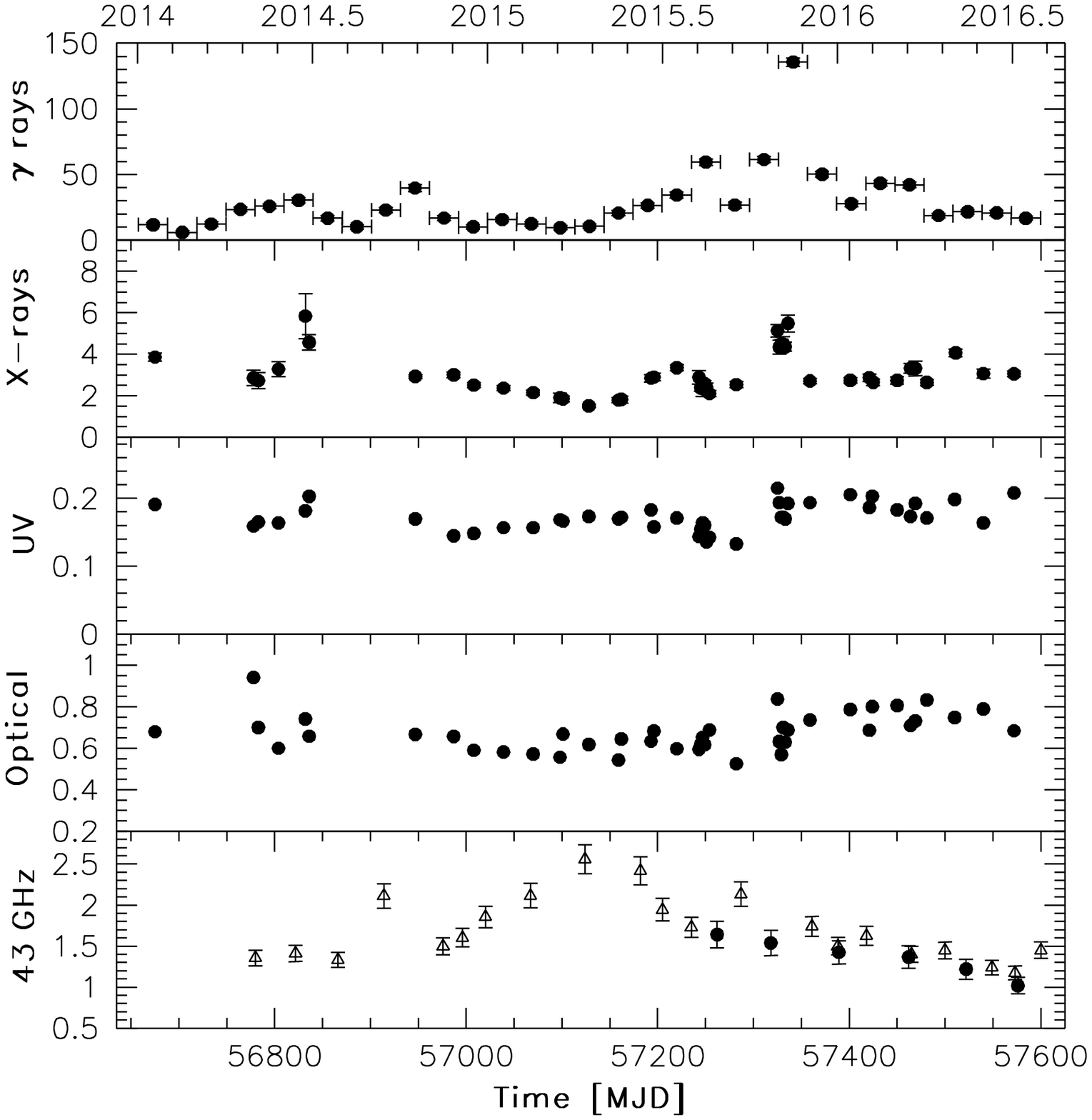}
\caption{Multiwavelength light curves of S5\,0836$+$710. From top to bottom:
  {\it Fermi}-LAT $\gamma$-ray flux, in units of 10$^{-8}$ ph cm$^{-2}$
  s$^{-1}$; {\it Swift}-XRT X-ray flux, in units of 10$^{-11}$ erg cm$^{-2}$
  s$^{-1}$; {\it Swift}-UVOT UV {\it w1}-band flux, in units of mJy;  {\it Swift}-UVOT
  optical {\it v}-band flux, in units of mJy; VLBA 43-GHz radio flux density,
  in units of Jy. In the
  bottom panel filled circles refer to our 6-epoch observations, while
  empty triangles refer to BU-blazar programme \citep{jorstad17}.}
\label{mwl}
\end{figure*}

\subsection{Variability}

The long-term light curves of S5\,0836$+$710 show low activity periods
interleaved with high activity phases in all energy bands 
(Fig.~\ref{mwl}). 
On average there seems to be an agreement between
$\gamma$-ray and X-rays/UV/optical light curves.
At the end of 2014 the flux density at 43 GHz
starts to increase and reaches about 2.55 Jy in 2015 April. During
this period the source is in a low activity state in $\gamma$ rays and
X-rays, while a hint of flux increase is observed in UV and marginally 
in optical. 
The amplitude variability (calculated as the ratio of maximum to minimum flux)
observed in $\gamma$ rays ($\sim$18 during the first flare, $\sim$22 during
the second flare) is significantly larger than the value estimated
in X-rays ($\sim$4). The small variability in X-rays could be an indication
that the X-ray emission is produced by the low-energy tail of the same
electron distribution that is also responsible for the $\gamma$-ray
emission.

The amplitude variability during the UVOT observations is
1.7, 1.6, 1.5, 1.6, 1.8 and 1.7 in the $v$, $b$, $u$, $w1$, $m2$, and
$w2$ bands, respectively. This is slightly larger than the variability
observed in the UVOT filters during 2006--2012 \citep[$<$ 50 per
  cent;][]{akyuz13}.

At 43 GHz the variability amplitude observed for the radio core is
$\sim$1.7, with a peak flux density significantly lower than the value
observed during 2006--2012 by Effelsberg \citep[$\sim$ 4
  Jy;][]{akyuz13}. \\ 
\indent The {\em
  Swift}-BAT spectrum is fitted in the 14--195 keV energy range by a
power law with a photon index 1.70 $\pm$ 0.08 and a corresponding flux
of 6.98$^{+0.24}_{-0.25}$ $\times$10$^{-11}$ erg cm$^{-2}$
s$^{-1}$. The source was detected in hard X-rays also by {\em
  BeppoSAX} \citep{tavecchio00}, {\em INTEGRAL} \citep{beckmann09},
and {\em NuSTAR} \citep{tagliaferri15}. In particular, two {\em NuSTAR} observations were performed on 2013 
December
15 and 2014 January 18 simultaneously to {\em Swift}-XRT observations. 
The
0.3--79 keV spectra of the source is well described by a broken 
power-law
model with photon indices 1.03$^{+0.20}_{-0.32}$ 
(1.18$^{+0.08}_{-0.10}$) and
1.66$\pm$0.02 (1.66$^{+0.02}_{-0.01}$) above and below the energy break 
of
1.73$^{+1.27}_{-0.48}$ keV (2.84$^{+1.03}_{-0.62}$ keV) for the first 
(second)
observation. The photon index obtained by 
{\em
NuSTAR} above 2--3 keV is compatible with the value obtained by {\em 
  Swift}-BAT in the 14--195 keV energy range. In the same way, as reported in the second INTEGRAL AGN catalogue \citep{beckmann09}, the photon index obtained by analyzing IBIS-ISGRI data in the 18--60 keV band collected between 2002 December 30 and 2007 February 17 for a total of 754 ks is 1.5$^{+0.2}_{-0.1}$, in agreement with the BAT and {\em NuSTAR} values. \\
\indent An increase of the flux by a factor of $\sim$ 1.5 was observed between 
the two {\em NuSTAR} observations (F$_{10-40\,keV}$ =
  2.3$\times$10$^{-11}$ erg cm$^{-2}$ s$^{-1}$ and 
3.6$\times$10$^{-11}$ erg cm$^{-2}$ s$^{-1}$). Extrapolating the 10--40 
keV flux to the {\it Swift}-BAT energy range 14--195 keV, we obtain a
value of 6.2$\times$10$^{-11}$ erg 
cm$^{-2}$ s$^{-1}$ and 9.7$\times$10$^{-11}$ erg cm$^{-2}$ s$^{-1}$, 
respectively, confirming a moderate variability of the hard X-ray flux.\\

\indent During the second half of 2015, S5\,0836$+$710 entered a high
activity phase observed from optical to high energies and culminating
in two major flares detected by {\it Fermi}-LAT. The daily peak of the
emission during the first flare was observed on 2015 August 2 (MJD
57236) with a flux of (517 $\pm$ 32)$\times$10$^{-8}$ ph cm$^{-2}$
s$^{-1}$ in the 0.1--300 GeV energy range, 18 times higher than the
average flux over the whole period of {\em Fermi}-LAT
observations. The corresponding apparent isotropic $\gamma$-ray
luminosity peak is (2.0$\pm$0.1)$\times$10$^{50}$ erg s$^{-1}$. 
The sub-daily analysis shows a clear flux rise followed by a sharp decay. The flare
is characterized by a rapid variability, with flux-doubling time scale
of about 3 hours. The flare lasted for approximately 48 hours (MJD
57235$-$57237) reaching a maximum value on a 3-h time scale of (676 $\pm$
90)$\times$10$^{-8}$ ph cm$^{-2}$ s$^{-1}$, corresponding to an
apparent isotropic $\gamma$-ray luminosity of
  (2.6$\pm$0.3)$\times$10$^{50}$  
erg s$^{-1}$, on 2015 August 2 (MJD 57236), followed by a sharp 24-hr time scale decay
(Fig.~\ref{flare12}, upper plot).  \\
\indent An increase of X-ray activity is observed by {\it Swift} on 2015 July 17 (MJD 57220) when the flux is 3.35$\times$10$^{-11}$ erg cm$^{-2}$
s$^{-1}$, together with a hardening of the X-ray photon index.
Unfortunately there are no {\it Swift}
observations during 
the peak of the first $\gamma$-ray 
flare. Observations performed a few days after the first $\gamma$-ray flare
indicate a decrease of the flux from about 2.9$\times$10$^{-11}$ to 2.1$\times$10$^{-11}$ erg
cm$^{-2}$ s$^{-1}$ between August 8 and August 20, suggesting that we are
observing the decreasing part of the flaring activity. In the optical $v$-band
the peak is observed on August 20 (MJD 57254), while in UV it seems to occur earlier, on June 20 (MJD 57193). A hint of flux density increase is observed at 43-GHz radio
frequency about 40 days after the X-ray flare. However, the poor time
sampling does not allow us to set stringent constraints on the
radio-to-X-rays light curve behaviour close in time with the first
$\gamma$-ray flare.\\

\begin{figure}
\centering
\includegraphics[width=7.5cm]{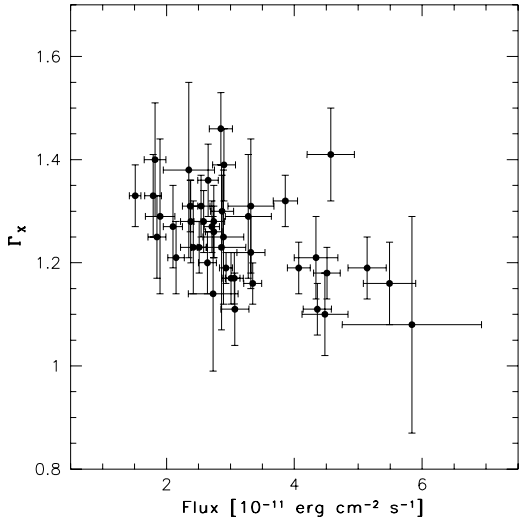}
\caption{{\em Swift}-XRT photon index as a function of the 0.3--10 keV flux.}
\label{XRT}
\end{figure}

The second $\gamma$-ray flare took place a few months later and
reached the maximum 
daily flux on 2015 November 11 (MJD 57337), with a value of (624 $\pm$ 
34)$\times$10$^{-8}$ ph cm$^{-2}$ s$^{-1}$, 22 times higher than the
average flux and corresponding to an apparent isotropic luminosity of
  (2.3$\pm$0.1)$\times 10^{50}$ erg s$^{-1}$. 
The flaring period lasts for about 6 days (from MJD 57332 to 57338) and shows several peaks with
flux doubling time scales of about 3 hours. 
The highest flux on a 3-h time-scale, (1052
$\pm$ 114)$\times$10$^{-8}$ ph cm$^{-2}$ s$^{-1}$, was
observed on November 9 (MJD 57335), and corresponds to an apparent isotropic
$\gamma$-ray luminosity of (3.7$\pm$0.4)$\times$10$^{50}$ erg s$^{-1}$
(Fig.~\ref{flare12}, bottom panel).\\ 
\indent {\it Swift} monitored S5\,0836$+$710 every two days between October 30
and November 10. The X-ray flux is high during the whole period, and 
reaches a peak of about 5.5$\times$10$^{-11}$ erg cm$^{-2}$ s$^{-1}$
on November 10 (MJD 57336). After this period a drop of the X-ray flux is 
observed. A hardening of the X-ray photon index is also observed during the
whole high activity period, suggesting a 'harder-when-brighter' effect
(Fig. \ref{XRT}). This behaviour is quite typical during flares in FSRQ \citep[e.g.,][]{vercellone10,dammando11}, and can be due to changes of the electron energy distribution in an acceleration and cooling scenario \citep[e.g.,][]{kirk98}.
The UV and optical fluxes reach their maximum  on October 30 (MJD 57325),
i.e. before the X-ray flux peak, and remain above the average value until
the end of the period considered here. No radio outburst is observed
close in time with this flare.\\  

In addition to S5\,0836$+$710, four other high-redshift blazars
  have been
studied in detail during a $\gamma$-ray flaring activity: TXS\,0536$+$145
\citep{mo14}, PKS 2149$-$306 \citep{dammando16}, PKS\,1830$-$211
\citep{abdo15}, and DA\,193 \citep{paliya19}. For the first two sources a significant curvature of
the $\gamma$-ray spectrum was observed at the peak of the $\gamma$-ray
activity, as seen in 
S5\,0836$+$710, while a curved model was not tested for
PKS\,1830$-$211 and no clear evidence of a hardening of the spectrum
was noted during high states. On the other hand, a hardening of the $\gamma$-ray spectrum, well described by a simple power-law, was observed during the flaring activity of DA\,193. Variability on a daily time-scale was
detected in all four blazars, down to sub-daily time-scales for
PKS\,1830$-$211 (12 hours) and PKS\,2149$-$306 (6 hours). As a
comparison, the 
doubling time-scale of 3 hours seen in the light curve of S5\,0836$+$710 is
the shortest variability time-scale observed for a high-redshift blazar in
$\gamma$ rays so far. A similar rapid variability has been observed from the
same source in 2011 November \citep{paliya15}. Moreover, the peak $\gamma$-ray luminosity
reached by S5\,0836$+$710 on a 3-hour time-scale in 2015 November puts the source among the brightest $\gamma$-ray sources ever observed
so far.\\

\subsection{Radio structure and spectral index distribution}

\begin{table*}
\caption{Flux density and spectral index of the source
  components. Column 1: epoch of observations from Table \ref{log-vlba}. Columns 2, 3, 4, 5 and 6: core peak flux density (mJy) at 15, 24, and 43 GHz, and core spectral index between 15 and 24 GHz ($\alpha_{15}^{24}$) and between 24 and 43 GHz ($\alpha_{24}^{43}$), respectively. Columns 7, 8, 9, 10 and 11: component B3 peak flux density (mJy) at 15, 24, and 43 GHz, and spectral index between 15 and 24 GHz ($\alpha_{15}^{24}$) and between 24 and 43 GHz ($\alpha_{24}^{43}$), respectively. Columns 12, 13, 14, 15 and 16: component J peak flux density (mJy) at 15, 24, and 43 GHz, and spectral index between 15 and 24 GHz ($\alpha_{15}^{24}$) and between 24 and 43 GHz ($\alpha_{24}^{43}$), respectively. Column 17: total flux density (mJy) at 15 GHz of the jet extended structure.} 
\centering
\resizebox{\textwidth}{!}{
\begin{tabular}{lcccccccccccccccc}
\hline
Epoch& \multicolumn{5}{c}{Core}& \multicolumn{5}{c}{B3}&
\multicolumn{5}{c}{J}& Extended \\
\cmidrule(l{10pt}r{5pt}){2-6} \cmidrule(l{10pt}r{5pt}){7-11} \cmidrule(l{10pt}r{5pt}){12-16}
 &S$_{15}$&S$_{24}$&S$_{43}$&$\alpha_{15}^{24}$&$\alpha_{24}^{43}$&S$_{15}$&S$_{24}$&S$_{43}$&$\alpha_{15}^{24}$&$\alpha_{24}^{43}$&S$_{15}$&S$_{24}$&S$_{43}$&$\alpha_{15}^{24}$&$\alpha_{24}^{43}$&S$_{15}$\\
\hline
A & 1220$\pm$85& 1468$\pm$102& 1365$\pm$136& 0.4$\pm$0.2& $-$0.1$\pm$0.2& 507$\pm$35& 325$\pm$23& 153$\pm$15& $-$1.0$\pm$0.2& $-$1.2$\pm$0.2& 185$\pm$13& 120$\pm$8& 51$\pm$5& $-$1.0$\pm$0.2& $-$1.4$\pm$0.2& 14$\pm$1\\
B & 1407$\pm$98& 1730$\pm$121& 1247$\pm$125& 0.5$\pm$0.2& $-$0.5$\pm$0.2& 488$\pm$34& 296$\pm$21&  72$\pm$7& $-$1.1$\pm$0.2& $-$2.4$\pm$0.2& 184$\pm$13& 127$\pm$9& 55$\pm$5& $-$0.9$\pm$0.2& $-$1.3$\pm$0.2& 17$\pm$1 \\
C & 1414$\pm$99& 1548$\pm$108& 1201$\pm$120& 0.2$\pm$0.2& $-$0.4$\pm$0.2& 408$\pm$29& 253$\pm$18& 108$\pm$11& $-$1.1$\pm$0.2& $-$1.4$\pm$0.2& 169$\pm$12& 104$\pm$7& 45$\pm$4& $-$1.1$\pm$0.2& $-$1.3$\pm$0.2& 13$\pm$1\\
D & 1368$\pm$96& 1473$\pm$103& 1074$\pm$107& 0.2$\pm$0.2& $-$0.5$\pm$0.2& 331$\pm$23& 239$\pm$17& 108$\pm$11& $-$0.8$\pm$0.2& $-$1.3$\pm$0.2& 156$\pm$11& 101$\pm$7& 48$\pm$5& $-$1.0$\pm$0.2& $-$1.2$\pm$0.2& 20$\pm$2 \\
E & 1449$\pm$101& 1404$\pm$98& 1000$\pm$100& $-$0.1$\pm$0.2& $-$0.5$\pm$0.2& 323$\pm$23& 179$\pm$12&  86$\pm$9& $-$1.4$\pm$0.2& $-$1.2$\pm$0.2& 138$\pm$10&  80$\pm$6& 34$\pm$3& $-$1.3$\pm$0.2& $-$1.4$\pm$0.2& 18$\pm$1 \\
F & 1391$\pm$97& 1164$\pm$81&  799$\pm$80& $-$0.4$\pm$0.2& $-$0.6$\pm$0.2& 298$\pm$21& 160$\pm$11&  70$\pm$7& $-$1.4$\pm$0.2& $-$1.3$\pm$0.2& 120$\pm$8&  70$\pm$5& 32$\pm$3& $-$1.3$\pm$0.2& $-$1.3$\pm$0.2& 13$\pm$1\\ 
\hline
\end{tabular}}
\label{fluxvlba}
\end{table*}

\begin{figure}
\begin{center}
\includegraphics{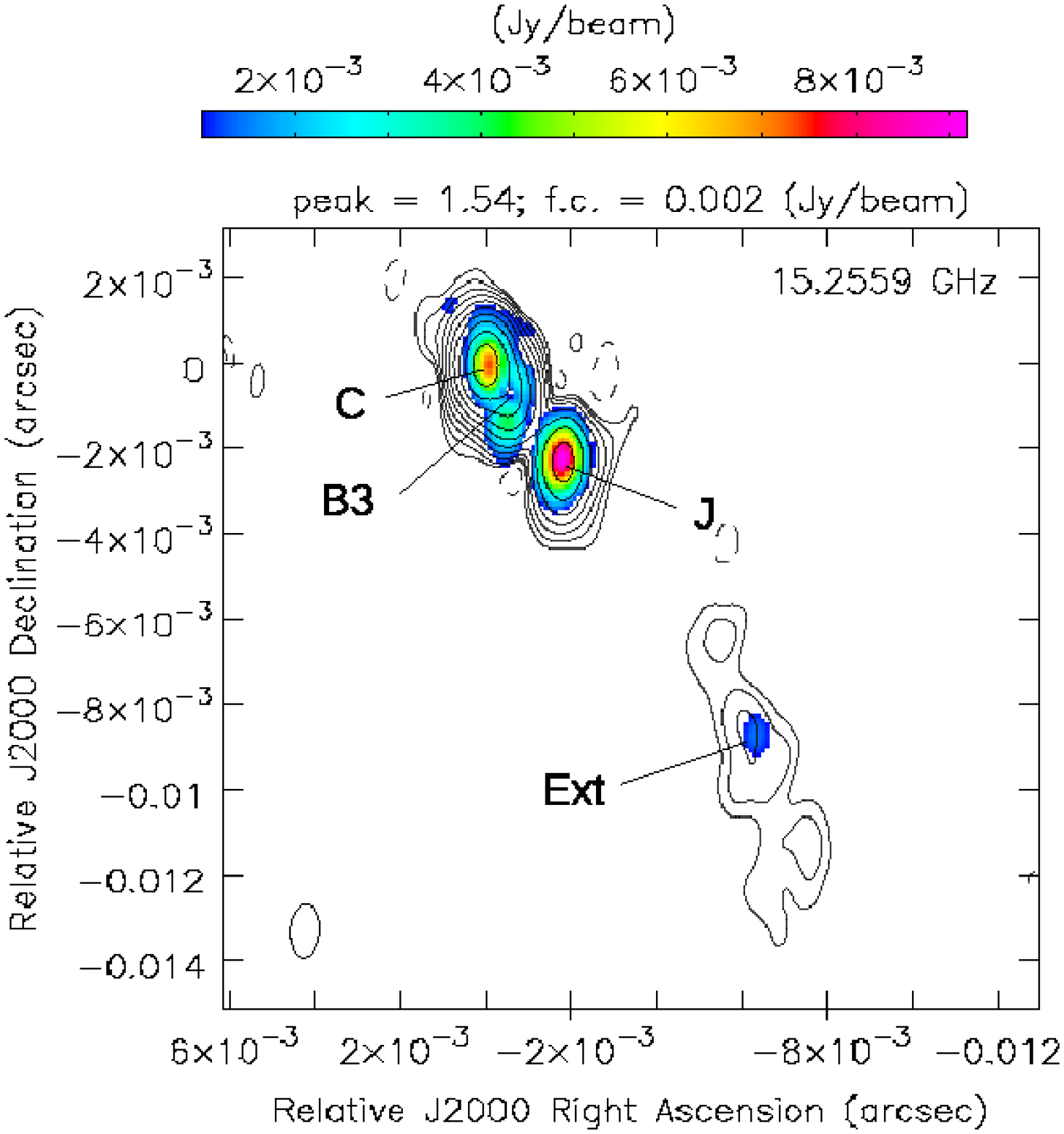}
\includegraphics{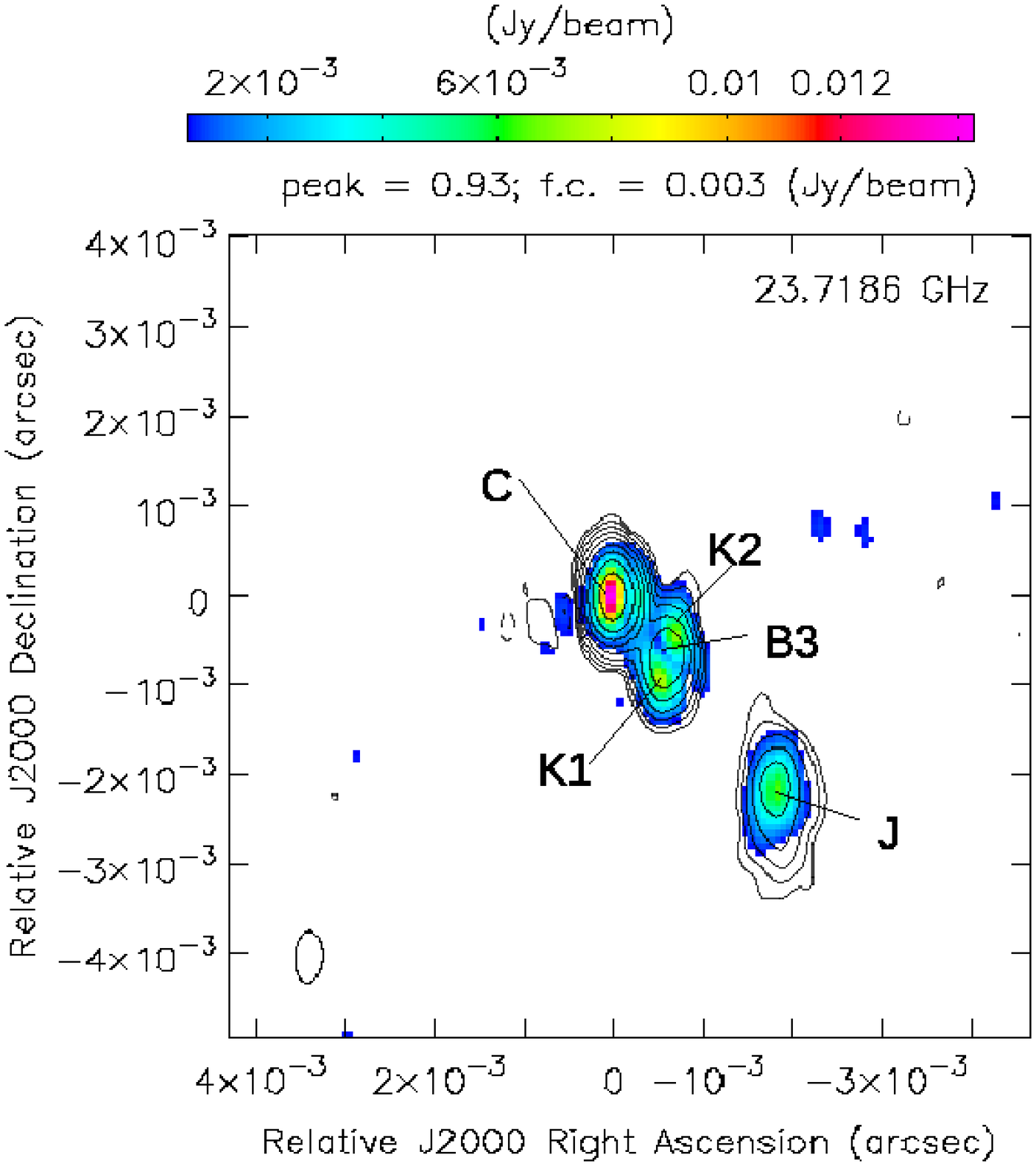}
\includegraphics{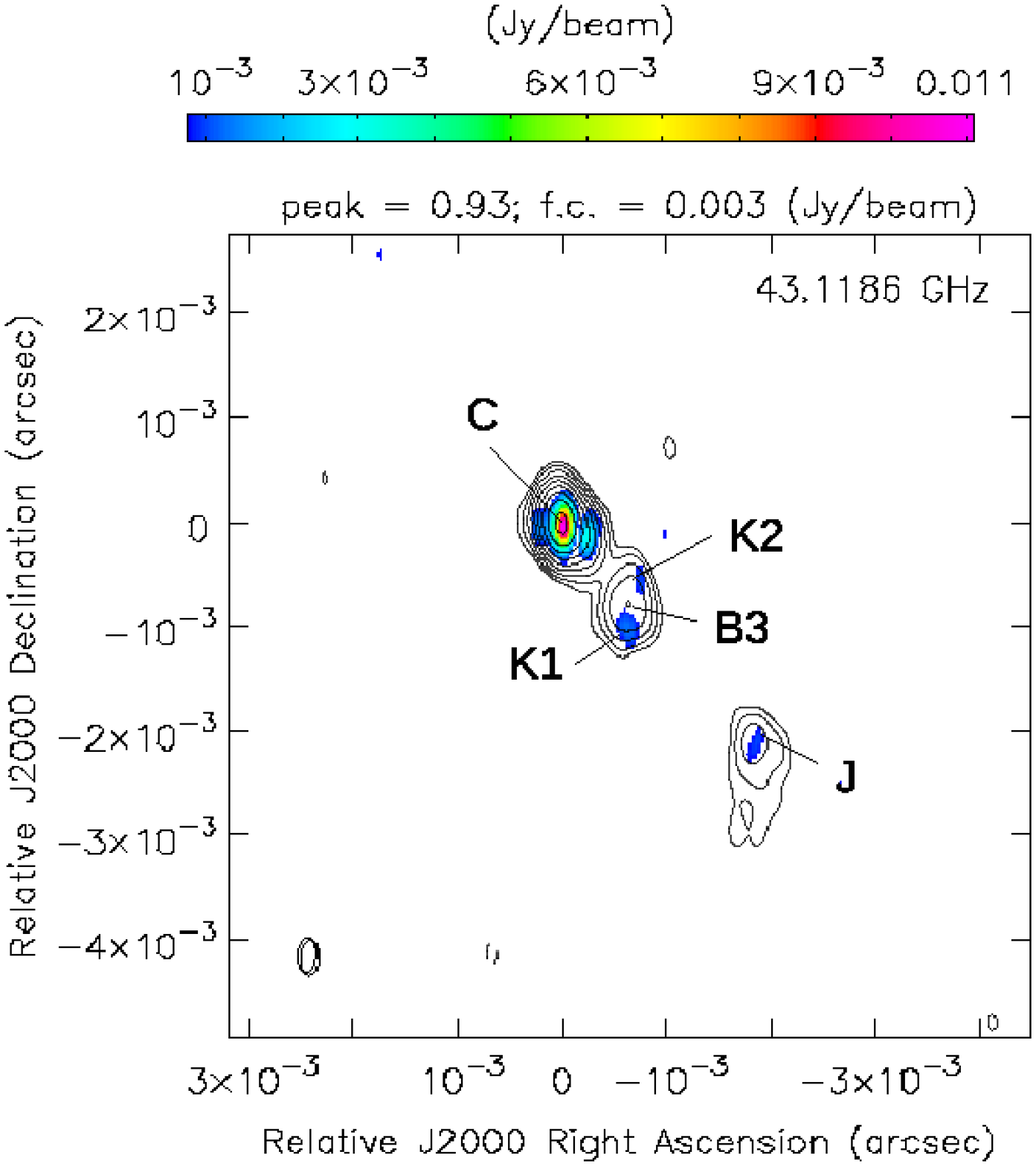}
\vspace{21.2cm}
\caption{An example of full resolution images of S5\,0836$+$710 at 15 {\it (top)},
  24 {\it (center)}, and 43 GHz {\it (bottom)} from the observations performed on 2016 May 14. On each image, we
  provide the peak flux density in Jy/beam and the first contour (f.c.)
  intensity in Jy/beam, which corresponds to three times the
  off-source noise level. Contour levels increase by a factor of
  2. The restoring beam is plotted in the bottom left-hand corner. The
  colour-scale represents the polarization intensity.} 
\label{full_res}
\end{center}
\end{figure}

\begin{figure}
\centering
\includegraphics[width=9.0cm]{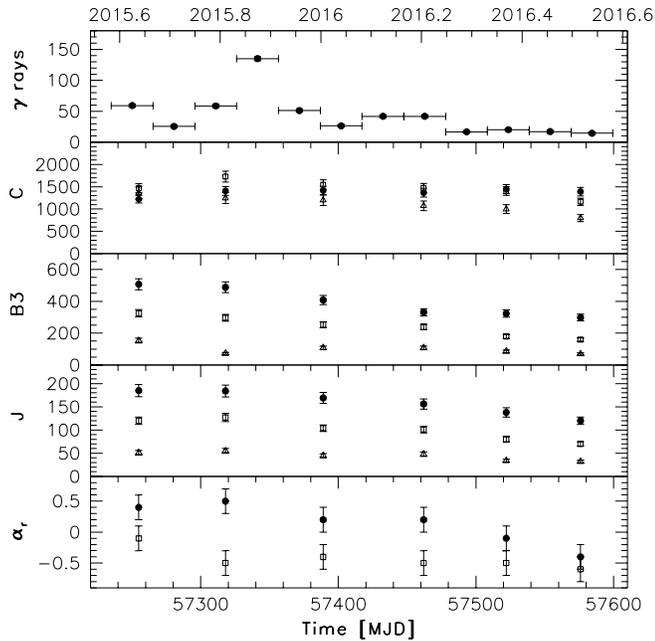}
\caption{Flux density of the component C (second panel from the top), component B3 (third panel from the top), component J (forth panel from the
  top), is compared to the $\gamma$-ray light curve in the 0.1--300 GeV energy
  range with 1-month time bins (top panel). Filled circles refer to 15 GHz
  data, open squares to 24 GHz data, open triangles to 43 GHz data. In the
  bottom panel the spectral index of the core $\alpha_r$ between 15 and 24 GHz (filled
  circle) and 24 and 43 GHz (open square) are shown.}
\label{radio_multipanel}
\end{figure}

At parsec scale the radio source S5\,0836$+$710 is characterized by a
compact core and a jet that emerges from the core with a position 
angle (PA) of $-$130$^{\circ}$ up to $\sim$10 mas and then changes to
PA $-$155$^{\circ}$ in agreement with previous studies 
\citep[e.g.][]{krichbaum90,lobanov98}. The radio emission originates
mainly in the radio core (component C in Fig. \ref{full_res}),
which accounts for more than 65 per 
cent of the 
total flux density measured on our VLBA images. Two compact features
are observed along the jet at $\sim$ 1 mas (component B3 in Fig.
\ref{full_res}) and at $\sim$ 3 mas (component J in
Fig. \ref{full_res}) from the core. Component B3 is resolved into
  two sub-components visible only in polarization intensity and
  labeled K1 and K2 in Fig. \ref{full_res} (see Section \ref{sect_polla}). 
The low dynamic range of our
observations prevents us from producing detailed images of the jet
structure, and the region in which the jet changes the position angle
is visible in some 15-GHz images only
(Fig. \ref{full_res}). Multi-frequency VLBA flux
densities are reported in Table \ref{fluxvlba}. For a reliable comparison of
flux density at different epochs for the main components we
prefer to report the peak flux density measured on images obtained
with the same beam. In fact our images are dynamic range limited and a
variation of the total flux density may be not related to intrinsic variability
of the component, but it may be due to the presence of low-surface
brightness diffuse jet emission that is not detectable in all the
observing epochs.\\
\indent Fig. \ref{radio_multipanel} shows the evolution
  of the peak flux density of component C, B3 and J. Between 2015
  August and 2016 July the peak flux density at 43 GHz of component C shows a decreasing trend, whereas at 24 GHz an increase of
  the flux density is observed during the first two observing epochs,
  followed by a decreasing trend. At 15 GHz the variability is less
  evident with respect to the trend observed at higher frequencies. 
  We observe a
decrease of the flux density at each frequency for both components B3 and J, as expected
in presence of adiabatic expansion. \\
\indent To derive structural changes we complemented our observations with
those from the VLBA-BU-BLAZAR program
performed between 2014 September and 2018 May. To this aim we fitted
the visibility data with circular Gaussian components at each epoch
using the model-fit option in \texttt{DIFMAP}.
This approach is used in order to derive small structure variation and
provide an accurate fit of unresolved components close to the core component.

\begin{figure}
\begin{center}
\includegraphics{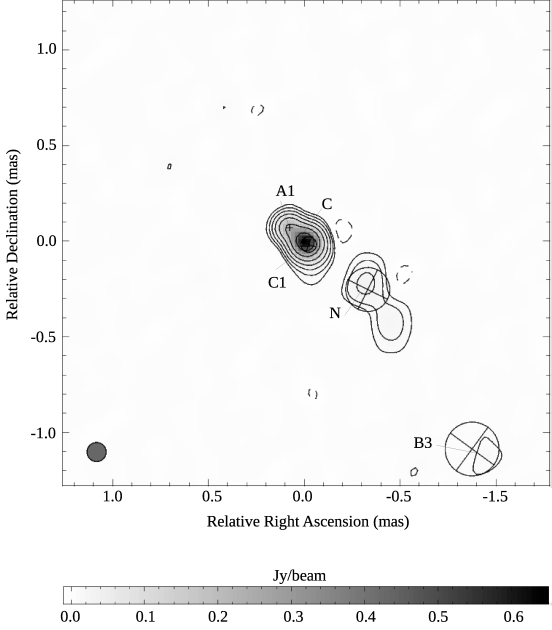}
\vspace{8.5cm}
\caption{43-GHz VLBA image from the observations performed on 2017 November 6 in the framework of the VLBA-BU-BLAZAR program. The image has been reconstructed with a beam of 0.1 $\times$ 0.1 mas$^{2}$. 
  The peak flux density is 0.80 Jy/beam and the
  first contour is 1 per cent of the peak. Contour levels increase by a factor
  of 2. The restoring beam is plotted in the bottom left-hand
  corner. The grey-scale is shown by the wedge at the bottom of the
  figure and represents the total intensity flux density in Jy/beam. }
\label{modelfit-image}
\end{center}
\end{figure}

\begin{figure}
\begin{center}
    \includegraphics{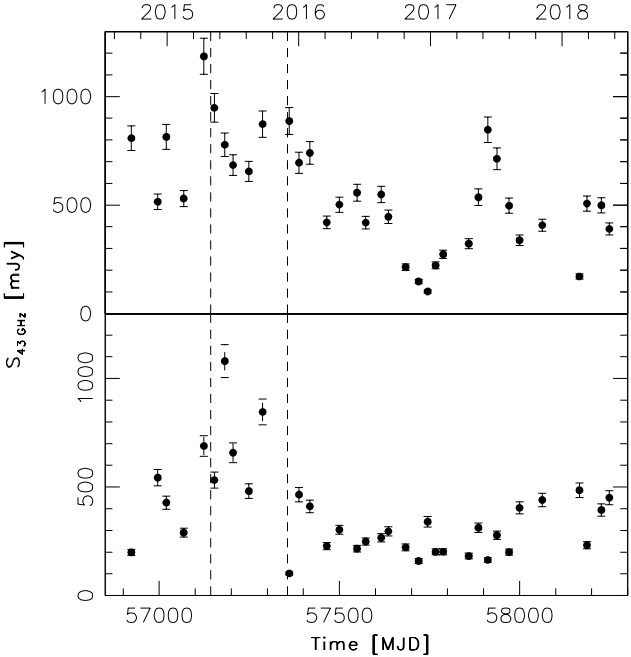}
\vspace{8cm}
\caption{Flux density at 43 GHz of the component C (top panel), and the component C1 (bottom panel) in the period 2014 September--2018 May. Vertical dashed lines mark the high-activity period in $\gamma$ rays (i.e. 2015 May--November).}
\label{43GHz_component}
\end{center}
\end{figure}

This analysis points out the presence of one (quasi-)stationary feature at about 0.03--0.1 mas from the core, labeled C1 in
Fig. \ref{modelfit-image} with a position angle that ranges between
$-$110$^{\circ}$ and $-$140$^{\circ}$, and two superluminal components, N and
B3 with position angle of about $-$125$^{\circ}$ and $-$140$^{\circ}$,
respectively. Component N is first detected by the visibility
  model-fit analysis at 43 GHz. Its presence on the image plane could
  be resolved at 43 GHz 
  only after 2016 October. Fig. \ref{43GHz_component} shows the
    evolution of the flux density at 43 GHz of components C and C1 between 2014 September and 2018 May. The core component shows variability throughout the period, reaching a flux peak in 2015 April when the flux density doubled with respect to the value observed in 2015 February. The core peak flux density occurred close in time with the ejection of the new component N. In the same way, during the second half of 2015, when the $\gamma$-ray activity of the source was higher, the radio flux density of component C1 was higher than the values observed between 2014 September and 2015 February, reaching peak values in 2015 June and in September. After the high $\gamma$-ray activity period the flux density of C1 significantly decreased.\\

\indent We derive the proper motion of these components by
means of a linear fit. We find that
component N is moving with an apparent velocity v$_{\rm app}$ = (14.8 $\pm$ 0.6)$c$ and the estimated epoch of passage through the VLBI core is
2015.28$\pm$0.07 (i.e. 2015 April), in good
agreement with the increase of the flux density at 43 GHz (Fig. \ref{mwl}) and the beginning of the high-activity period observed in $\gamma$ rays. Component B3 is moving with an apparent velocity v$_{\rm app}$ = (21.0 $\pm$ 0.4)$c$, and corresponds to the component that emerged after the
$\gamma$-ray flare in 2011 
\citep{akyuz13},
discussed by \citet{jorstad17} and \citet{jorstad13}. Results on the model-fit analysis of the 
  visibility data are
reported in Fig. \ref{proper_motion} and in Appendix
  \ref{section-modelfit}. A
stationary feature, labeled A1 in Fig. \ref{modelfit-image}, at 0.1 mas
on the opposite side of the core is 
present during the entire
period monitored by our VLBA campaign. This feature was already
reported by \citet{jorstad17}.\\

Between 15 and 24 GHz the spectrum of the core is inverted after the
$\gamma$-ray flare, with a spectral index $\alpha_{15}^{24}$ = 0.5
$\pm$ 0.3\footnote{The spectral index $\alpha$ is defined as S($\nu$)
  $\propto$ $\nu^{\alpha}$} (Fig. \ref{spix}, right
panel). 
Errors on the spectral index are computed in two steps. First,
we determine the errors associated with the flux density scale
uncertainty $\sigma_c$ such
that:\\

\begin{displaymath}
\sigma_c = \sqrt{ \left(\frac{\sigma_{\rm S1}}{S_1}\right)^2 + \left(\frac{\sigma_{\rm S2}}{S_2}\right)^2} \frac{1}{ {\rm ln(\nu_{2}) - ln(\nu_1)}}
  \end{displaymath}

\noindent where $S_i$ and $\sigma_{{\rm S}i}$ are the flux density and the flux density uncertainty, respectively, at the
frequency $i$ (see Section 2). Then $\sigma_c$ is combined with the
value from the spectral index error maps, $\sigma_{\alpha}$, obtained
by error propagation theory. $\sigma_c$ is about 0.2, while
$\sigma_{\alpha}$ is generally below 0.05 with the
exception of the edges of the radio structures. The resulting error is
$\sigma_{\rm tot} = \sqrt{ \sigma_c^2 + \sigma_{\alpha}^2}$, and is
usually dominated by $\sigma_c$.\\ 
\indent Fig. \ref{spix} shows how the 
ridge line spectral index values change across the source in 2015
August and 2016 July. In the former, the spectrum  
is inverted up to 2 mas from the core and then steepens smoothly,
whereas in the latter the spectrum is steeper and a flattening is
present at the position of B3 and corresponds to a peak in
polarization (labelled K1 in Fig. \ref{full_res}). The
gradients that are highlighted by a shaded area in Fig. \ref{spix} are
likely due to ({\it
  u,v})-coverage effects \citep[see e.g.][]{hovatta14}. In these
regions the values measured on the spectral index error images are
$\sigma_{\alpha} >$ 0.3, i.e. more than an order of magnitude larger
than in the other regions. In
the last epochs
the spectrum of the core flattens up to reaching
$\alpha_{15}^{24}$ = $-$0.4 $\pm$ 0.3 in 2016 July
(Fig. \ref{spix}, bottom panel). Between 24 and 43 GHz the variation of
the spectral shape is smoother and the spectral index
$\alpha_{24}^{43}$ ranges between $-$0.1 $\pm$ 0.3 and $-$0.6 $\pm$
0.3 (Fig. \ref{radio_multipanel}, bottom panel). Jet components have a 
steep spectrum ($-1.1 < \alpha < -0.8$).\\

\begin{figure}
\begin{center}
    \includegraphics{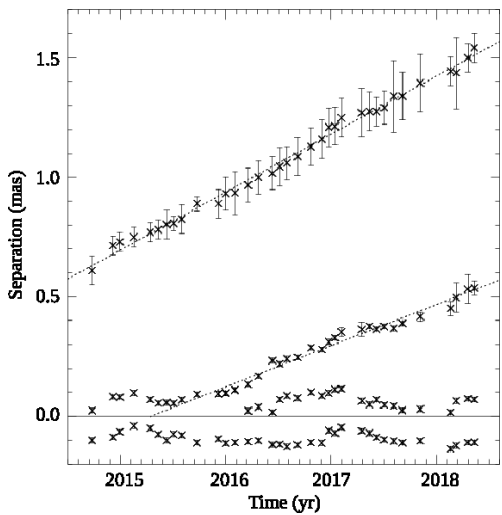}
\vspace{7cm}
\caption{Separation of components from the core (dashed line) vs
  time. Labels refer to the source components found during the
    visibility model-fit analysis and reported in Fig. \ref{modelfit-image}.}
\label{proper_motion}
\end{center}
\end{figure}

\subsection{Polarization and Rotation Measure}
\label{sect_polla}

\begin{table*}
\caption{Polarized flux density measured on the full-resolution
  images. Images at the same frequency were reconstructed with the
  same restoring beam (see Section \ref{radioim_section}). Column 1: epoch of
  observations from Table \ref{log-vlba}. Columns 2, 3, and 4: core polarized flux density (mJy) at 15, 24 and 43 GHz, respectively. Columns 5, 6, and 7: component
  K2 polarized flux density (mJy) at 15, 24 and 43 GHz, respectively. Columns 8, 9, and 10:
  component K1 polarized flux density (mJy) at 15, 24 and 43 GHz, respectively. Columns 11,
  12, and 13: component J polarized flux density (mJy) at 15, 24 and 43 GHz, respectively.}
\centering
\resizebox{\textwidth}{!}{ 
\begin{tabular}{lcccccccccccc}  
\hline
  Epoch & \multicolumn{3}{c}{C} & \multicolumn{3}{c}{K2} &
\multicolumn{3}{c}{K1} & \multicolumn{3}{c}{J}\\
 \cmidrule(l{10pt}r{5pt}){2-4} \cmidrule(l{10pt}r{5pt}){5-7} \cmidrule(l{10pt}r{5pt}){8-10} \cmidrule(l{10pt}r{5pt}){11-13} 
&P$_{15}$&P$_{24}$&P$_{43}$&P$_{15}$&P$_{24}$&P$_{43}$&P$_{15}$&P$_{24}$&P$_{43}$&P$_{15}$&P$_{24}$&P$_{43}$\\
\hline
A & - & 8.0$\pm$0.6 & 10.8$\pm$1.1 & 11.7$\pm$0.8\tablenotemark{1} & 18.5$\pm$1.3\tablenotemark{1} &
11.1$\pm$1.1 & - & - & - & 7.7$\pm$0.6 & 13.2$\pm$0.9 &
4.7$\pm$0.5 \\
B & 0.7$\pm$0.1 & 11.5$\pm$0.8 & 14.1$\pm$1.4 & 30.7$\pm$2.1\tablenotemark{1}& 12.1$\pm$0.9\tablenotemark{1} &
4.0$\pm$0.5\tablenotemark{1} & - & - & - & 23.5$\pm$1.6 & 13.3$\pm$0.9 & 1.6$\pm$0.3 \\
C & 1.3$\pm$0.2 & 7.6$\pm$0.5 & 24.3$\pm$2.4 & 18.1$\pm$1.3 \tablenotemark{1}& 13.7$\pm$1.0\tablenotemark{1} &
2.7$\pm$0.3 & - & - & 5.5$\pm$0.6 & 16.4$\pm$1.2 & 11.5$\pm$0.8 & 2.1$\pm$0.3\\ 
D & 1.3$\pm$0.2 & 15.0$\pm$1.0 & 17.8$\pm$1.8 & 3.0$\pm$0.2 & 4.6$\pm$0.3 & 2.4$\pm$0.3 & 4.0$\pm$0.3 & 6.5$\pm$0.5 & 3.0$\pm$0.3 & 10.3$\pm$0.9 & 11.2$\pm$0.8 &
1.5$\pm$0.2 \\
E & 7.3$\pm$0.5 & 24.1$\pm$1.7\tablenotemark{2}& 16.7$\pm$1.7 & 1.1$\pm$0.2 & - & 2.0$\pm$0.2 & 3.2$\pm$0.3 & 4.5$\pm$0.3 & 2.0$\pm$0.2 & 11.4$\pm$0.8 & 9.1$\pm$0.6 &
1.6$\pm$0.2 \\
F & 12.5$\pm$0.9 & 11.9$\pm$0.8 & 15.5$\pm$1.6& 2.9$\pm$0.2 & 2.0$\pm$0.2 & 0.4$\pm$0.2 & 7.4$\pm$0.5 & 5.0$\pm$0.4 & 1.2$\pm$0.2 & 14.7$\pm$1.0 & 7.5$\pm$0.5
& 1.5$\pm$0.2\\
\hline
\end{tabular}}
\tablenotetext{1}{K1+K2 flux density.}
\tablenotetext{2}{C+K2 flux density.}
\label{tab_polla}
\end{table*}

At 43 GHz and 24 GHz the core region is polarized during the entire monitoring campaign. No significant polarization ($<$1 mJy) is observed at 15 GHz in 2015
August, then the polarized flux density increases from 0.7 mJy in
2015 October up to $\sim$13 mJy in 2016 July. The polarized flux density reaches a maximum at 43 GHz in 2016 January followed with some time delay at 24 GHz and then at 15 GHz (see Table  \ref{tab_polla}). This may be related to the change in opacity with time, suggested by the spectral index behaviour (see Fig. \ref{radio_multipanel}). \\
\indent Significant
polarization is observed for component J at each frequency
during the whole monitoring period (Table \ref{tab_polla}). The
polarization angle  
is stable at about 90$^{\circ}$ at all frequencies, consistent with other VLBA observations at 15 GHz
\citep{lister18}. As a consequence, no significant RM is 
observed in component J, and values are 
consistent with the errors.\\ 
\indent Polarized emission from component B3 is detected during all
epochs. At 24 GHz the polarized emission is resolved into two
components, one to the north, K2 (with position angle $-$125$^{\circ}$
 with respect to the core component), and
one to the south, K1 (with position angle $-$145$^{\circ}$
 with respect to the core component), of the peak 
of the B3 component as observed in total intensity images. The
polarization morphology resembles a limb-brightened structure. 
At 15 GHz the two
polarized components are resolved from 2016 March, whereas in the
first three epochs they are blended together, in agreement with what
is found by \citet{lister18}. At 43 GHz polarized
emission from component K1 is detected during all epochs with the
exception of 2015 October, whereas significant polarized emission from
component K2 is detected sporadically. Polarized flux density of the
sub-components of S5\,0836$+$710 are reported in Table
\ref{tab_polla}, while the full set of polarization images are
presented in Appendix \ref{section_polla} (Fig. \ref{appendix_polla}).\\

\indent In the core component we observe very high values of RM, that may
exceed $|$5000$|$ rad m$^{-2}$. The RM is highly variable with sign
changes and its structure is patchy,
indicating either opacity gradients and/or different components that
evolve with time, expected in the case of a perturbed flow which is moving along the jet. In
the last epoch, when the radiation is optically thin at all three
frequencies, we observe an RM of $-$1400
$\pm$ 500 rad m$^{-2}$, and an RM-corrected magnetic field direction of 58 $\pm$
5$^{\circ}$, roughly parallel to the jet direction. RM-images and the
associated error images at
  each observing epoch are presented in Fig. \ref{rm-image}.\\

During the epochs
in which component K2 has significant polarized emission at the three
frequencies the RM is about $-$1950 $\pm$ 150 rad m$^{-2}$ in 2015 August, and 
$-$1200 $\pm$ 400 rad m$^{-2}$ in 2015 October and 2016 March. The
RM-corrected magnetic field direction ranges between 160$^{\circ}$ and
175$^{\circ}$. Component K1 has 
RM values between 850 $\pm$ 120 and $-$180 $\pm$ 120 rad m$^{-2}$, with a
tentative sign change 
observed in the last epoch, while the RM-corrected
magnetic field direction ranges between 60$^{\circ}$ and
75$^{\circ}$, roughly 
parallel to the jet direction.\\
We perform the analysis of the jet transverse structure making use of
the data taken in March 2016, i.e. when polarized emission from both K1 and
K2 components is clearly visible at all frequencies. Fig. \ref{rm-slice}
indicates the RM image and the slice considered for the analysis. A
transverse RM gradient is clearly visible with K1 component showing positive
RM values, while K2 component has negative RM values
(Fig. \ref{rm-slice}). The shaded area marks regions with low polarization levels consistent with noise, where no reliable RM could be estimated. Total intensity and polarization profiles
on the same transverse slice show a ridge-brightened profile and a
limb-brightened profile, respectively. The transverse spectral index
profile between 15 and 24 GHz indicates a smooth flattening of
the spectrum towards the ridge of the jet, with the spectral index
values moving from about 1.0 at the borders to about 0.7 at the centre.\\

\section{Discussion}

\subsection{Localization of the $\gamma$-ray emitting region and energetics}

One of the main characteristics of blazars is the high variability in all
bands, with a high fraction of energy in $\gamma$ rays. Information about the
variability time-scale and the highest energy photons observed from a source
may provide stringent constraints on the location of the $\gamma$-ray
emitting region. Since most of the luminosity of blazars is often released at
extreme energies, coverage of the $\gamma$-ray band is necessary to
properly infer the energetic budget of these sources. \\
\indent In FSRQ the $\gamma$-$\gamma$ collision between photons produced in the jet
and broad line region (BLR) photons may produce a strong cut-off in
the $\gamma$-ray spectrum 
above $\sim$ 20 GeV. In case of high-redshift blazars the $\gamma$-ray
emission above a few GeV should be suppressed also by the pair production due
to interaction of these $\gamma$-rays with the low-energy photons of the EBL.
S5\,0836$+$710 is not included in the Third Catalog of Hard Fermi-LAT sources
\citep[3FHL;][]{ajello17}, based on 7 years of LAT data analysed in the 10
GeV--2 TeV energy range, suggesting how difficult is to detect photons with energy higher than 10 GeV from S5\,0836$+$710. 
The maximum photon energy observed from the source during 2014--2016 is 15.3
GeV, consistent with current EBL models for a source at redshift 2.2 \citep[e.g.,][]{finke10,dominguez11}. No evidence of cut-off in the
$\gamma$-ray spectrum of the source due to $\gamma$-$\gamma$ interaction with BLR photons have been reported in
\citet{costamante18}. This suggests that the $\gamma$-ray emission from this
source is due to inverse Compton (IC) scattering off infrared photons from the dusty torus and
the spectrum above a few dozen GeV is significantly attenuated by the $\gamma$-$\gamma$ interaction with the EBL photons. \\  

During the 2015 November flaring activity of S5\,0836$+$710
significant $\gamma$-ray flux variation by a factor of 2 or more is
clearly visible on 3-h time-scales. This short time variability
observed in $\gamma$ rays constrains the size of the emitting region
to $R < c t_{var} \delta / (1+z)$. 
Assuming a bulk Lorentz factor $\Gamma =$ 16 \citep{tagliaferri15}, we
find that the size of the emitting region responsible for 3-h
variability is R $\sim$ 2$\times$10$^{15}$ cm.
The inferred size is comparable to the gravitational radius
(r$_{\rm\,g}$/$c$ = G\,M /$c^2$) for a black hole with mass
5$\times$10${^9}$ M$_{\odot}$, as the one estimated for S5\,0836$+$710
\citep{tagliaferri15}.

Although the high activity observed in the radio band starting at the beginning of 2015 does not seem associated with any significant increase of flux in the other bands, after the emergence of the new superluminal component we observe the beginning of high activity in $\gamma$ rays, X-rays, UV, and then in optical. The high activity in $\gamma$ rays reaches two peaks, in 2015 August and November, i.e., about 80 and 210 days after the ejection of the new component from the radio core. During this period the C1 component shows high variability, roughly doubling its flux density in one month, and its centroid moves from about 0.05 to 0.09 mas, which corresponds to a deprojected distance from the core of about 7--15 pc, assuming a viewing angle of 3.2 degrees \citep{pushkarev09}. These pieces of evidence suggest that a perturbed flow is moving along the jet and crosses the C1 component that may represent several standing shocks. Observations at higher resolution are necessary to confirm the presence of multiple shocks by resolving C1 into sub-components. In this scenario, the $\gamma$-ray activity should be produced at about 6 pc and 15 pc from the radio core, and the short-term $\gamma$-ray variability might be explained by the turbulent, extreme multi-zone model proposed by \citet{marscher14}, although magnetic reconnection cannot be excluded \citep[e.g.,][]{petropoulou16}. However, the sparse radio light curve does not allow us to claim a clear connection between the radio and optical-to-$\gamma$-ray variability. A similar conclusion was suggested for the flare observed in 2012 from the same source \citep{akyuz13,jorstad13}.
 
The interaction between a superluminal jet component and a standing shock as the origin of
$\gamma$-ray flares has been proposed for several blazars, like the
case of CTA\,102 \citep{casadio19}, PKS\,1510$-$089
\citep{marscher10,orienti13}, and BL Lacertae \citep{marscher08}. The
lack of any evidence of $\gamma$-$\gamma$ absorption from the BLR
during the high-activity period in S5\,0836$+$710 supports the location of the $\gamma$-ray flaring region far away from
the central region.\\  
 
If we consider the $\gamma$-ray luminosity of S5\,0836$+$710 at the daily peak
($\sim 2.6 \times 10^{50}$ erg s$^{-1}$) as the total luminosity emitted during
the major flare ($L_{\gamma,iso}$), after the beaming correction, we obtain the jet power spent to produce the observed radiation as $P_{\rm\,rad}$ $\simeq$ $L_{\gamma,iso}$ / 2$\Gamma^{2}$ = 5.0 $\times$ 10$^{47}$ erg s$^{-1}$ (assuming $\Gamma$ = 16). For a comparison, the radiative jet power is about 65 per cent of the Eddington luminosity (L$_{\rm\,Edd}$ =
6.9 $\times$ 10$^{47}$ erg s$^{-1}$) and a factor of two higher than
the accretion disc luminosity. 
Assuming that the radiative power is about 10 per cent of the total jet power \citep[e.g.,][]{celotti08}, we have $P_{\rm jet,tot}$ = 5.0 $\times$
10$^{48}$ erg s$^{-1}$. The total jet power can be compared to the
accretion power, P$_{\rm\,acc}$ = L$_{\rm\,disc}$ / $\eta_{\rm disc}$
= 2.3$\times$10$^{48}$ erg s$^{-1}$ (assuming $\eta_{disc}$ = 0.1),
indicating that the total jet power is larger than the accretion power
in active states 
of S5\,0836$+$710, as observed for other blazars \citep{ghisellini14}.\\

\subsection{Jet structure}

From the analysis of the multi-epoch polarimetry observations of
S5\,0836$+$710 we find that in the limb-brightened polarization 
structure that is observed at a projected distance of about 1 mas
from the core, RM values vary 
between 1000 and -2000 rad m$^{-2}$. These
values are much larger than 
those reported by \citet{hovatta12} for this
source. However, in their
work \citet{hovatta12} detected RM only from a component that is a few
parsecs away from the 
core, and is likely consistent with component J, which also does not
show any significant RM during our VLBA monitoring campaign. On the
contrary, we observe some variability in the RM values observed in the
limb-brightened polarization structure, as well as in the polarization
intensity, suggesting that the Faraday screen and the emitting jet
are closely connected. Furthermore, 
this structure shows a clear RM
gradient transverse to the jet direction. In 2016 March the RM values vary
from $\sim$800 to $\sim$ $-$1200 moving from the eastern to the
western edge, with the exception of the
central region where no significant 
polarization is detected. \citet{gabuzda17} found that a high fraction
of the sources that were found to have a `spine-sheath' polarization structure in \citet{gabuzda14} display transverse RM gradient with a high
incidence of sign change. Detection of sign changes indicates a change in the
direction of the 
line-of-sight magnetic field. Although we observe some RM variability
in the limb-brightened polarization
structure, the magnetic field direction in K1 is roughly
parallel to the jet 
axis during the three epochs in which polarized emission from this
component is clearly detected. These characteristics are consistent
with a scenario 
in which Faraday rotation is produced by a sheath or boundary layer of
thermal electrons with a toroidal magnetic
field that surrounds the emitting jet \citep[e.g.][]{broderick10}. \citet{gabuzda14} observed for this source  
a transverse RM gradient with a sign change at 5 mas from the
core. Interestingly, \citet{asada10} reported a similar result, but
with the gradient moving in the opposite direction, which may be
interpreted in terms of a change in domination between an inner and
outer region of helical magnetic field as suggested for
the jet in 1803+784 by \citet{mahmud09}.\\
\indent When polarization is
detected in the core, the RM is highly variable and may
exceed $|$5000$|$ rad 
m$^{-2}$. Such large values have been measured in the core region of
other blazars \citep[e.g.][]{jorstad07,hovatta12} and may indicate that in
this region the 
relation between the polarization vector
and lambda square is not linear. As pointed out by the model-fit of
visibility at 43 GHz, in the core region there are several components
that are unresolved with the beam at 15 GHz, and blending of 
components with different opacity and polarization properties may
cause spurious RM values \citep{hovatta12}. The variation of the
spectral index of the core, from inverted soon after the $\gamma$-ray
flare to slightly steep in the last observing epochs, suggests
changes in opacity of the core region. A similar steepening of the
core was observed in the VLBA monitoring of the high-z source
TXS\,0536+145 \citep{mo14}. However, the lack of multi-frequency VLBA
observations before the $\gamma$-ray flare precludes us to
unambiguously connect the high opacity of the core region to the
$\gamma$-ray flare.\\ 

\begin{figure*}
\begin{center}
\includegraphics{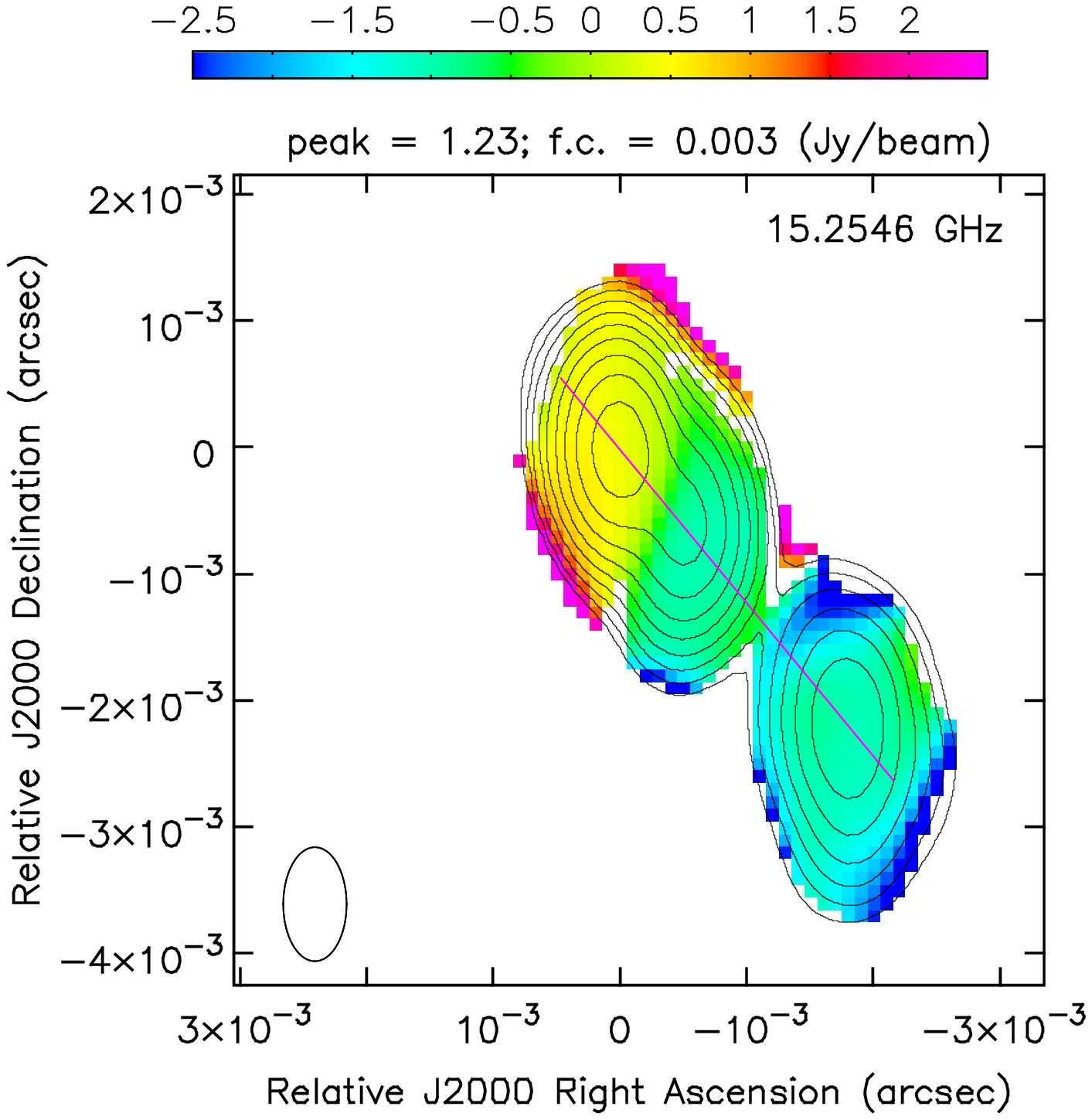}
\includegraphics{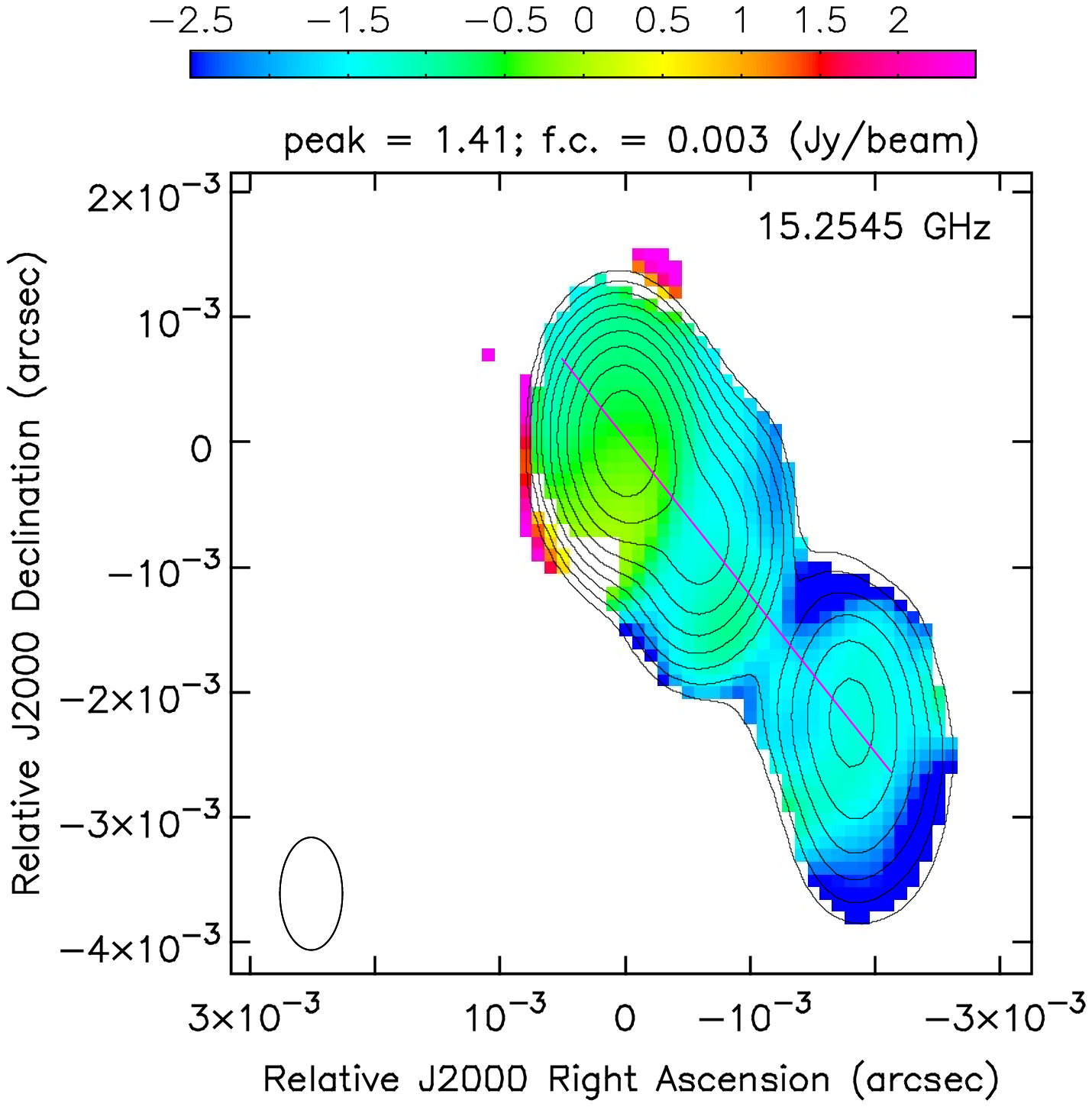}
\includegraphics{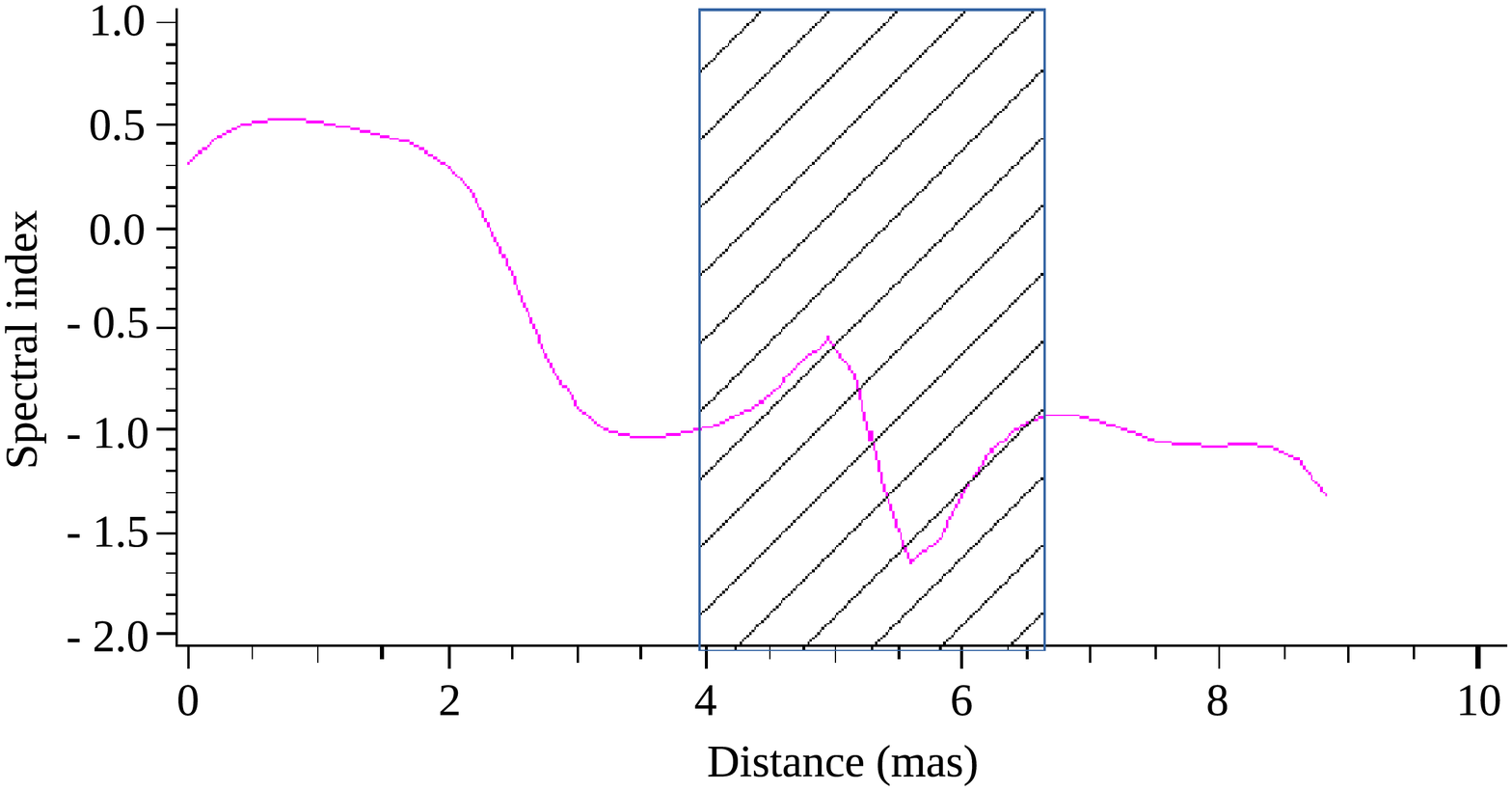}
\includegraphics{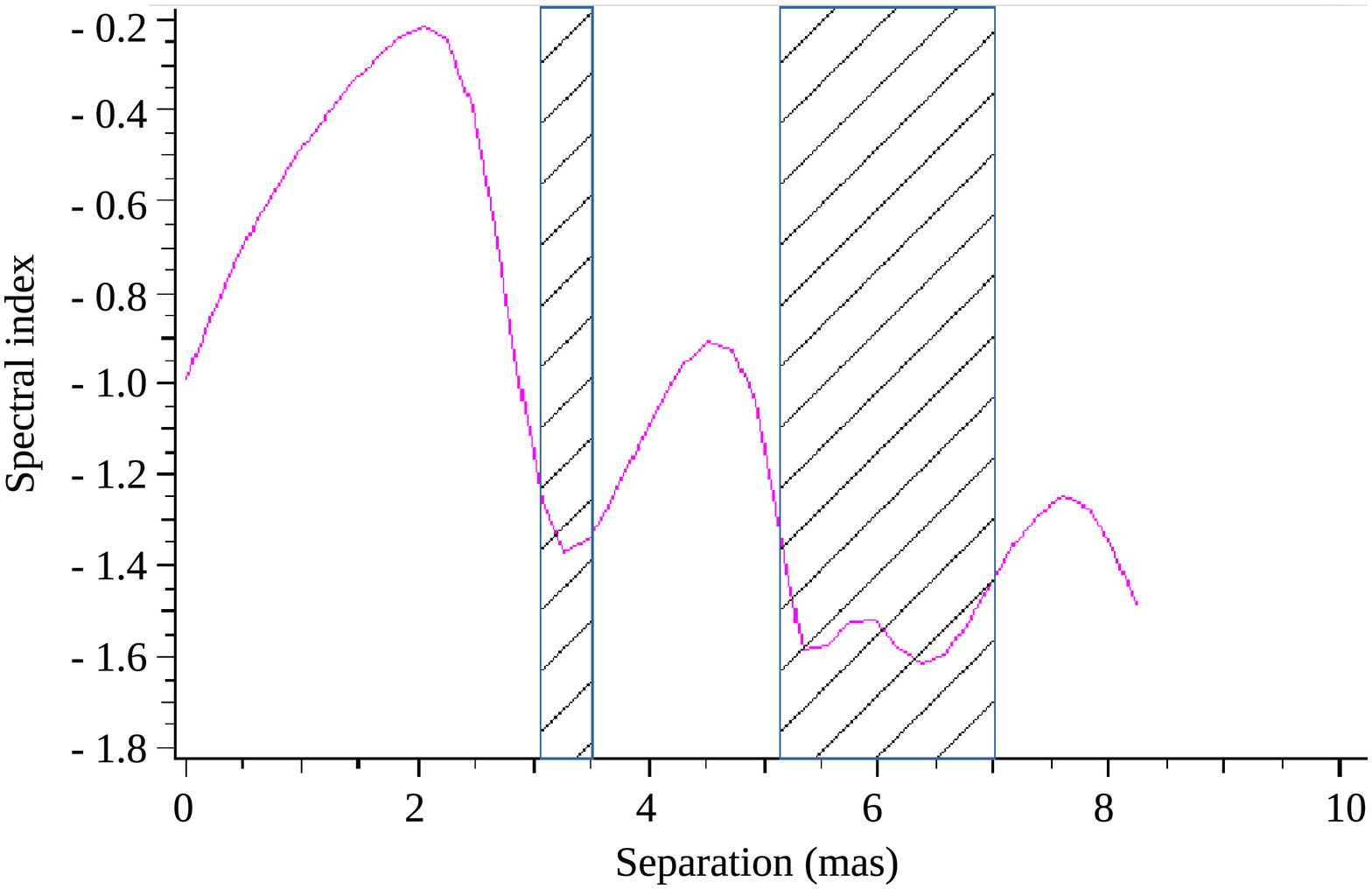}
\vspace{12.0cm}
\caption{Spectral index distribution between 15 and 24 GHz in 2015
  August ({\it top left panel}) and in 2016 July ({\it bottom left
    panel}); spectral index 
values along the ridge line in 2015 August ({\it top right panel}) and in 2016
July ({\it bottom right panel}). The line indicates the slice
used to derive 
the spectral profiles. Shaded areas represent regions of artificial
gradients with high errors ($\sigma_{\alpha} >$ 0.3),
likely caused by poor ({\it u,v})-coverage.}
\label{spix}
\end{center}
\end{figure*}

\begin{figure*}
\begin{center}
\includegraphics{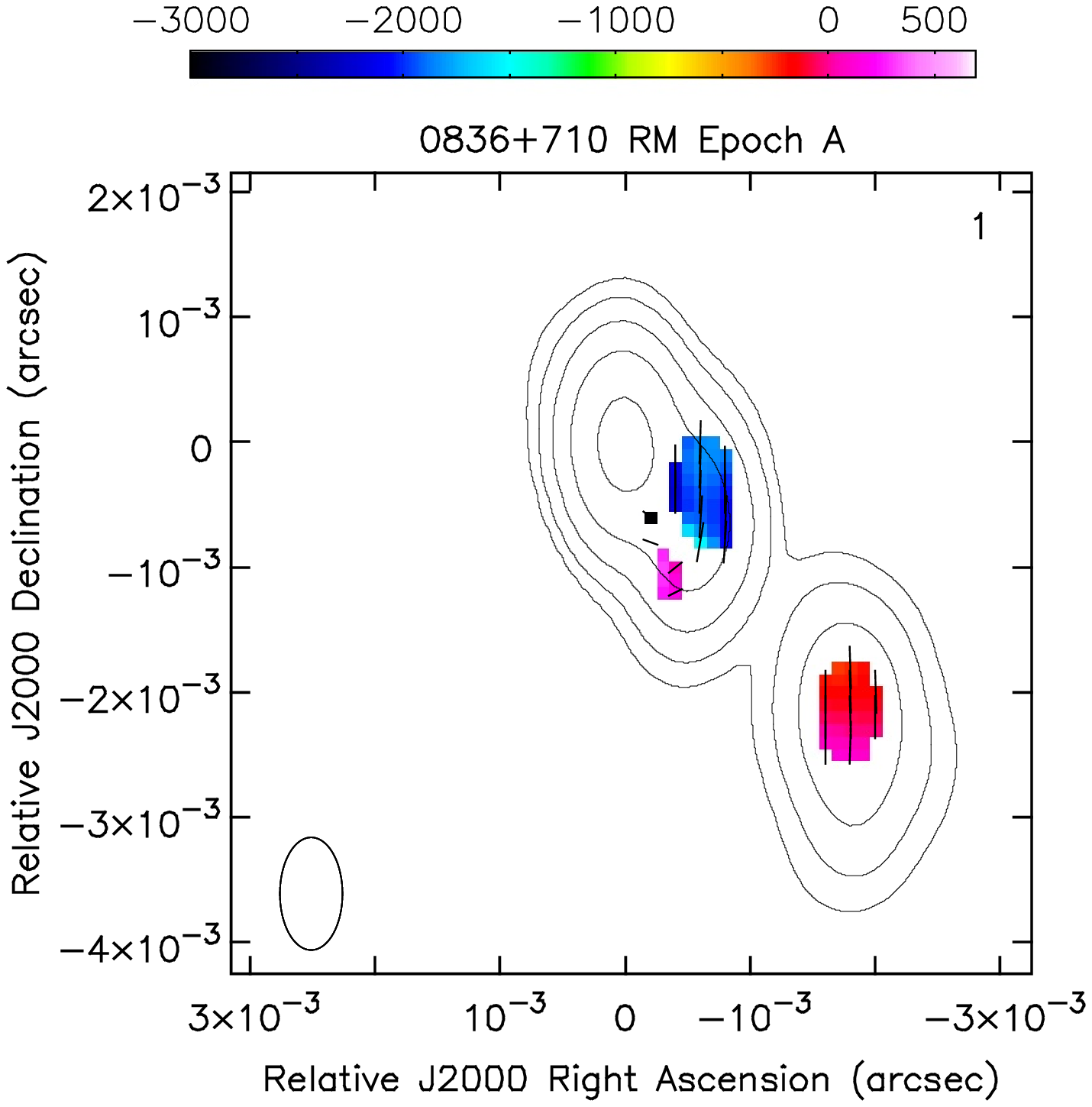}
\includegraphics{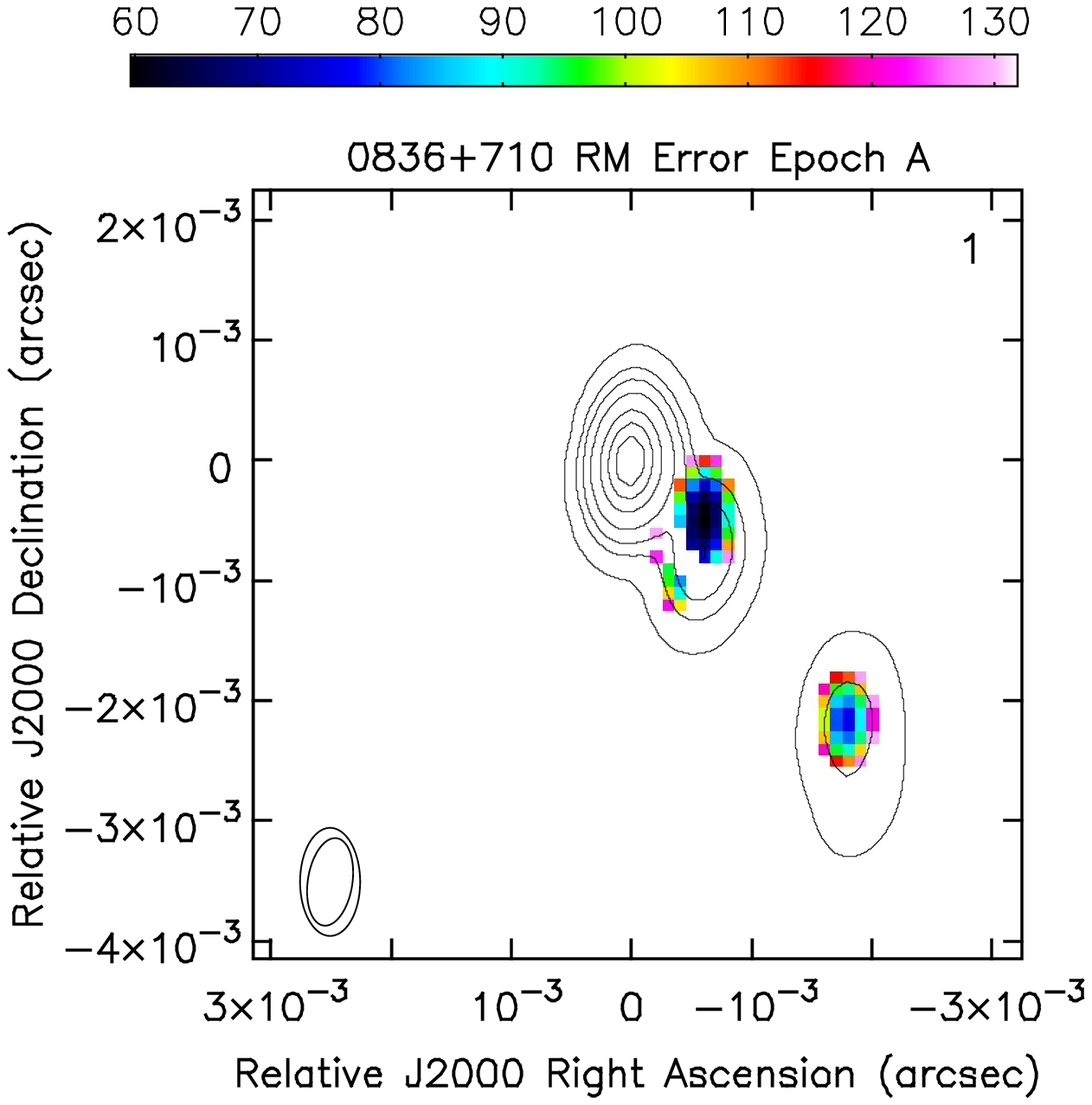}
\includegraphics{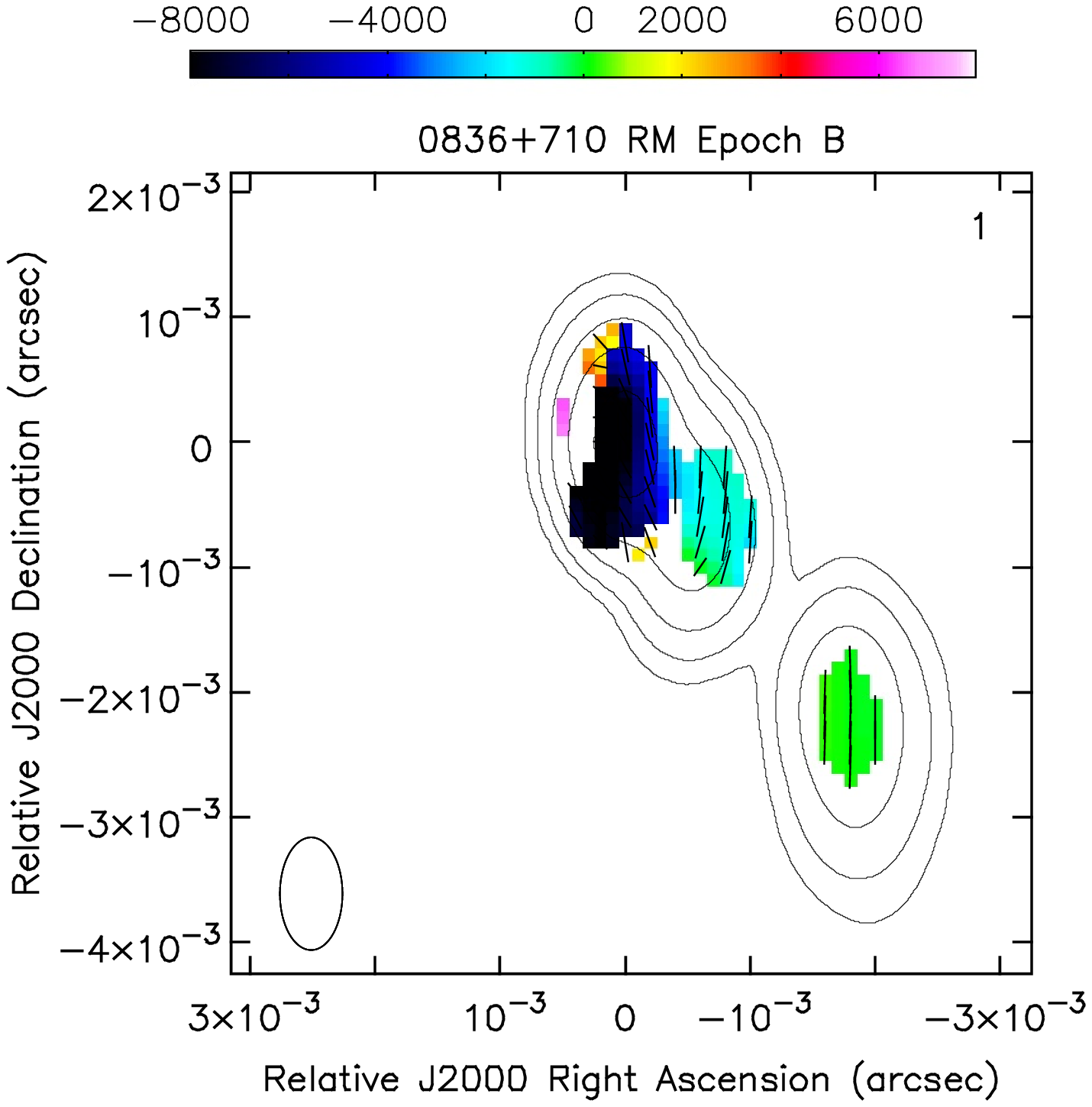}
\includegraphics{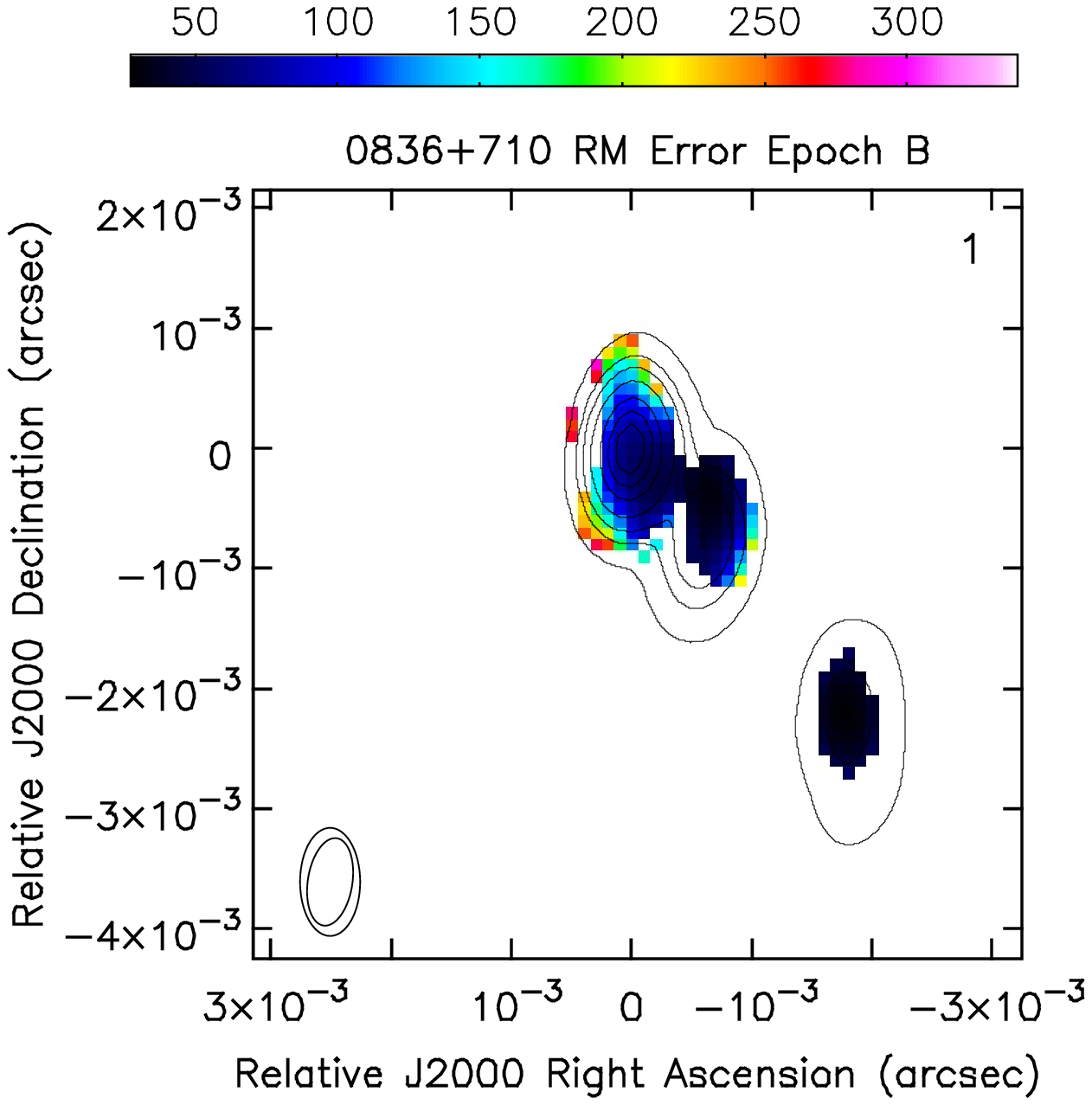}
\includegraphics{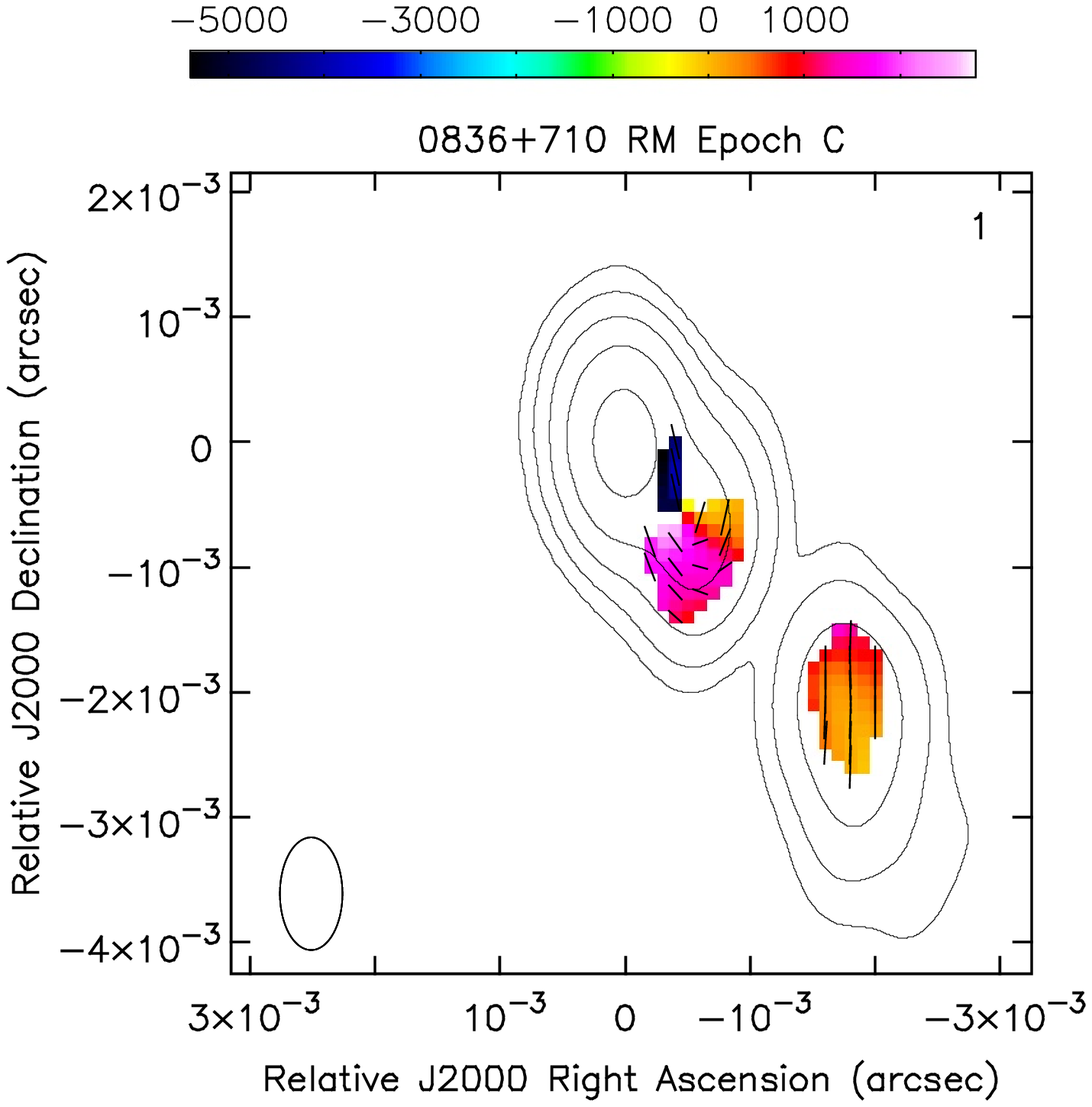}
\includegraphics{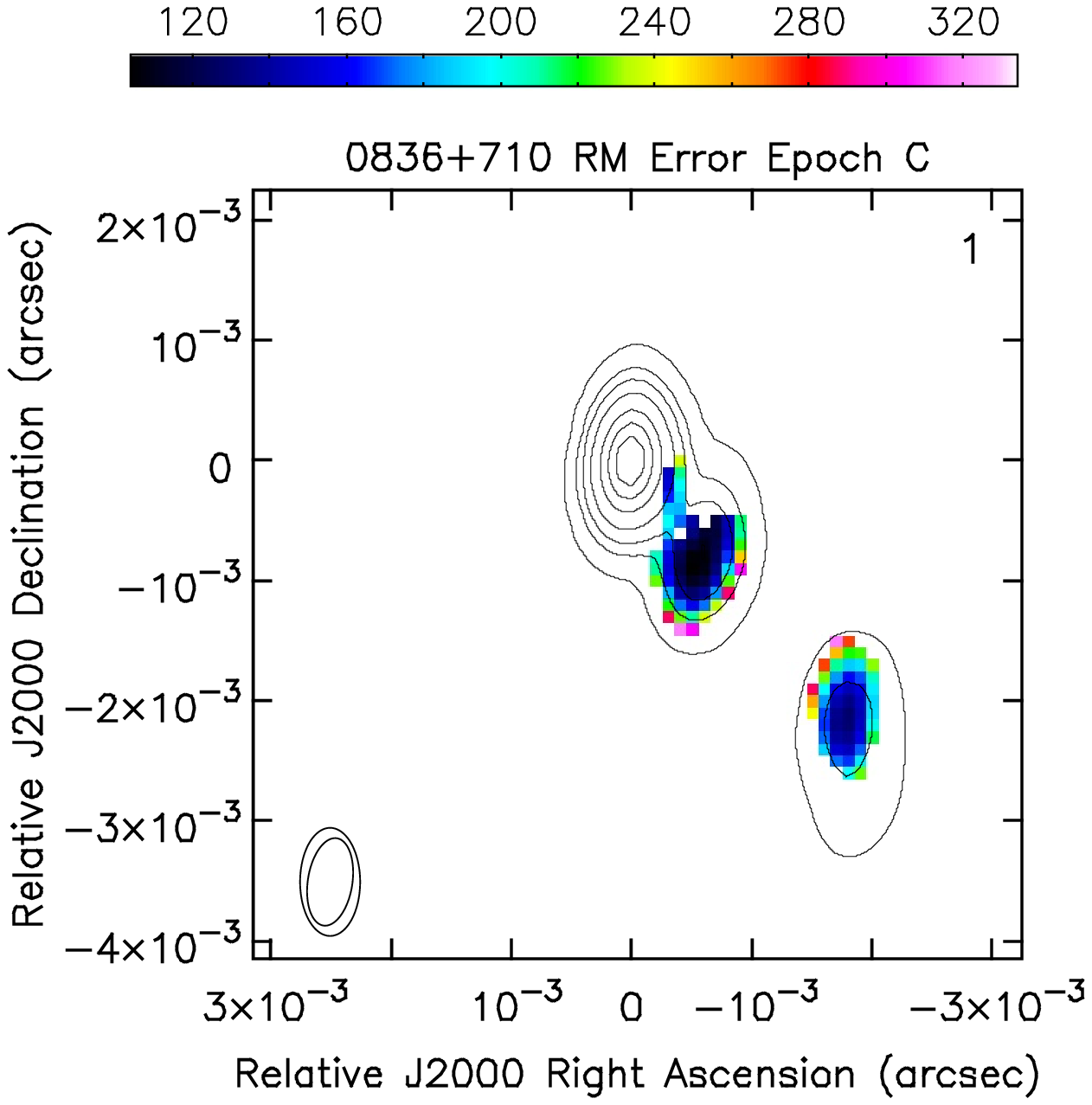}
\vspace{22cm}
\caption{Rotation measure images (colour-scale) for
  S5\,0836$+$710 overlaid with total intensity contours (left column), and the
  associated rotation measure error images (colour-scale; right column). 
Vectors represent the RM-corrected magnetic field (B) vectors.}
\label{rm-image}
\end{center}
\end{figure*}

\addtocounter{figure}{-1}
\begin{figure*}
\begin{center}
\includegraphics{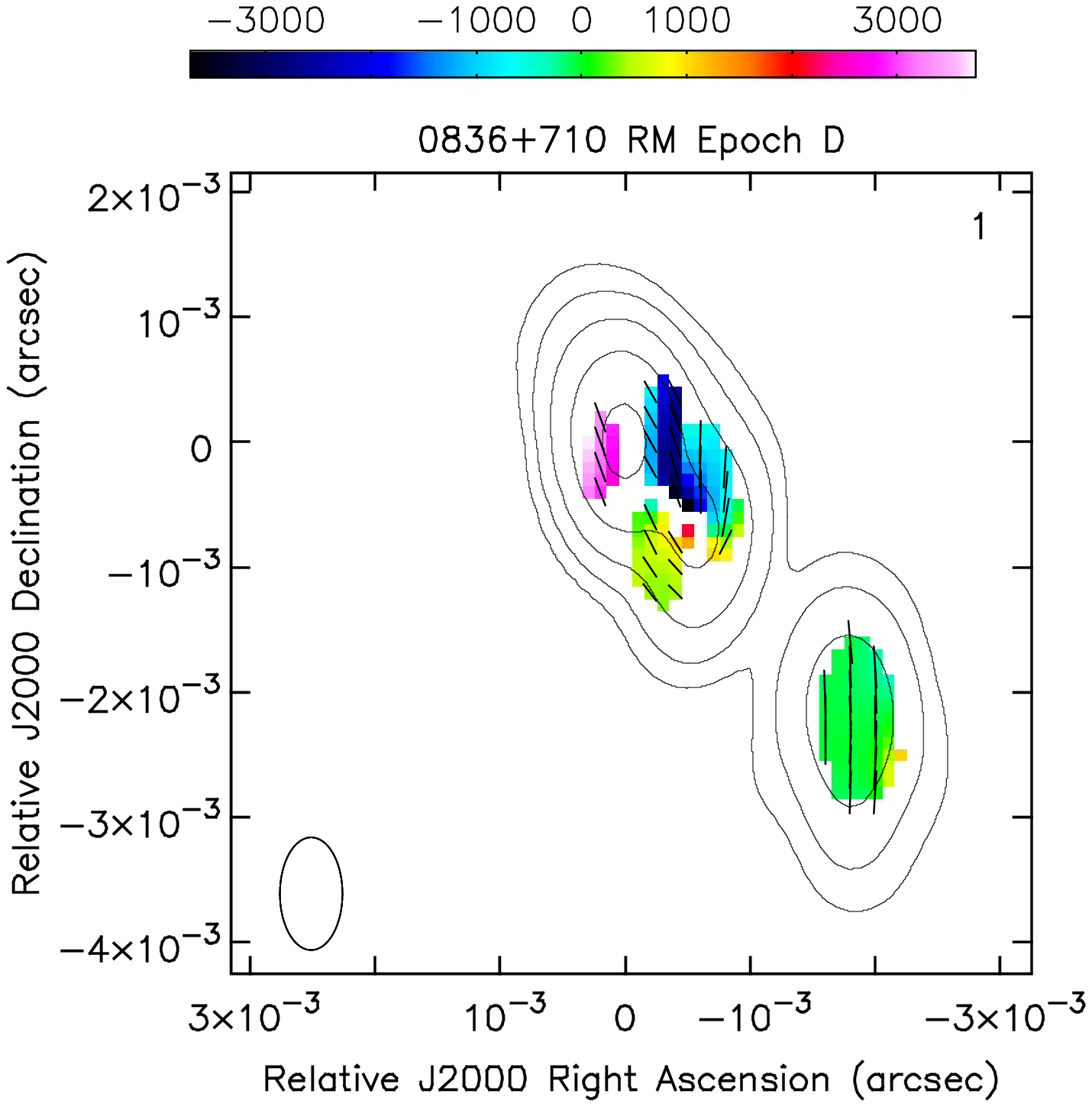}
\includegraphics{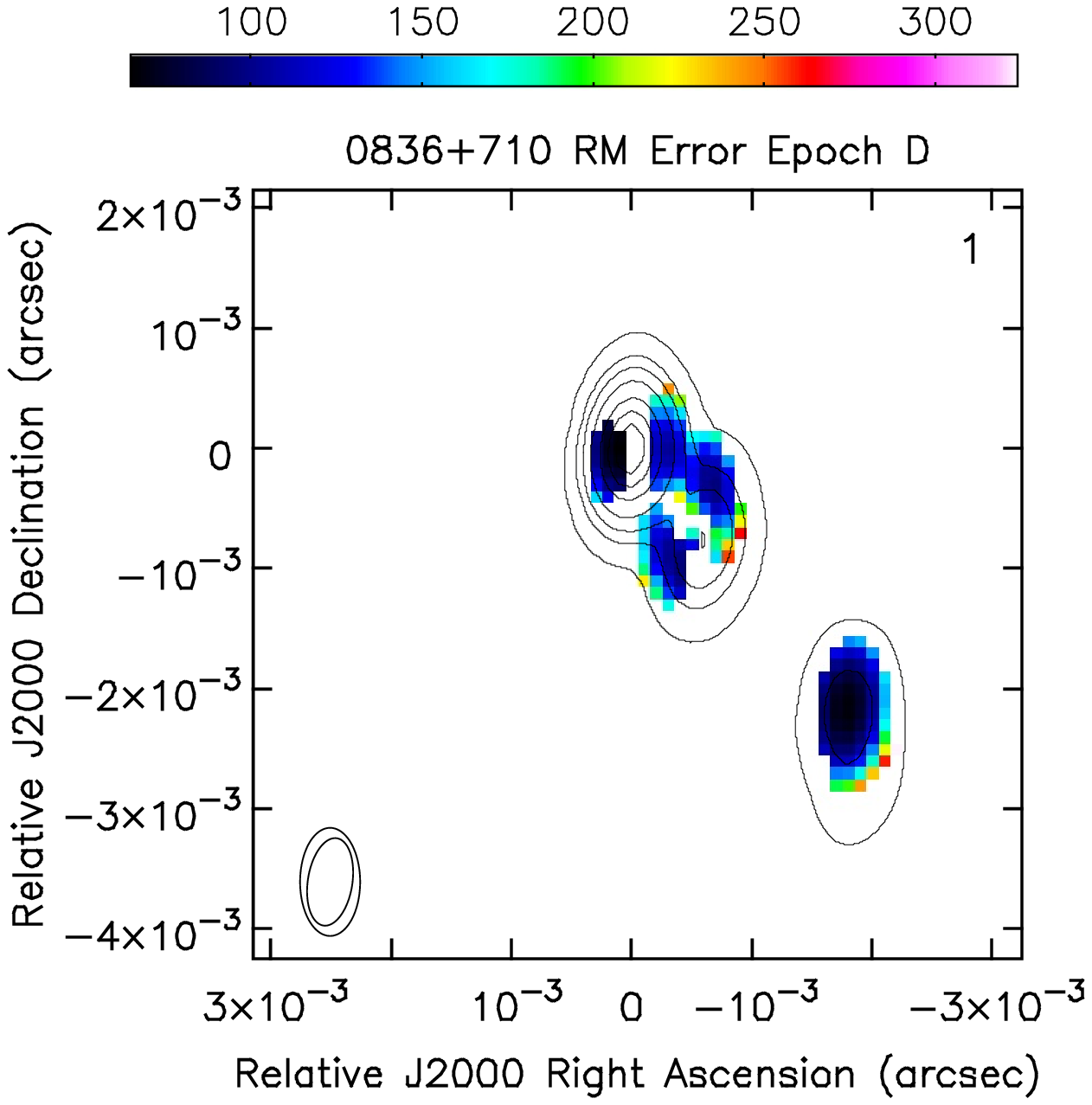}
\includegraphics{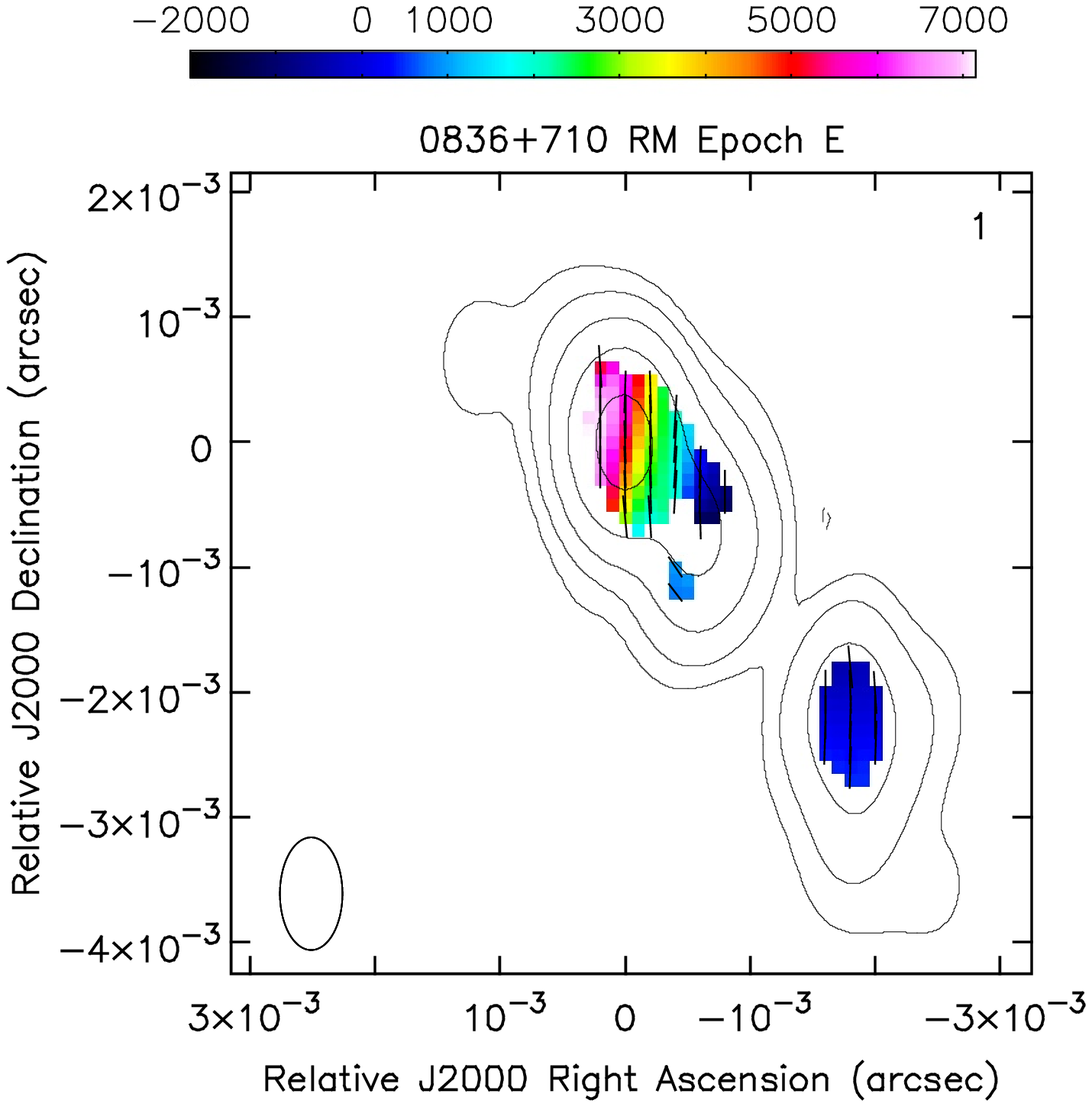}
\includegraphics{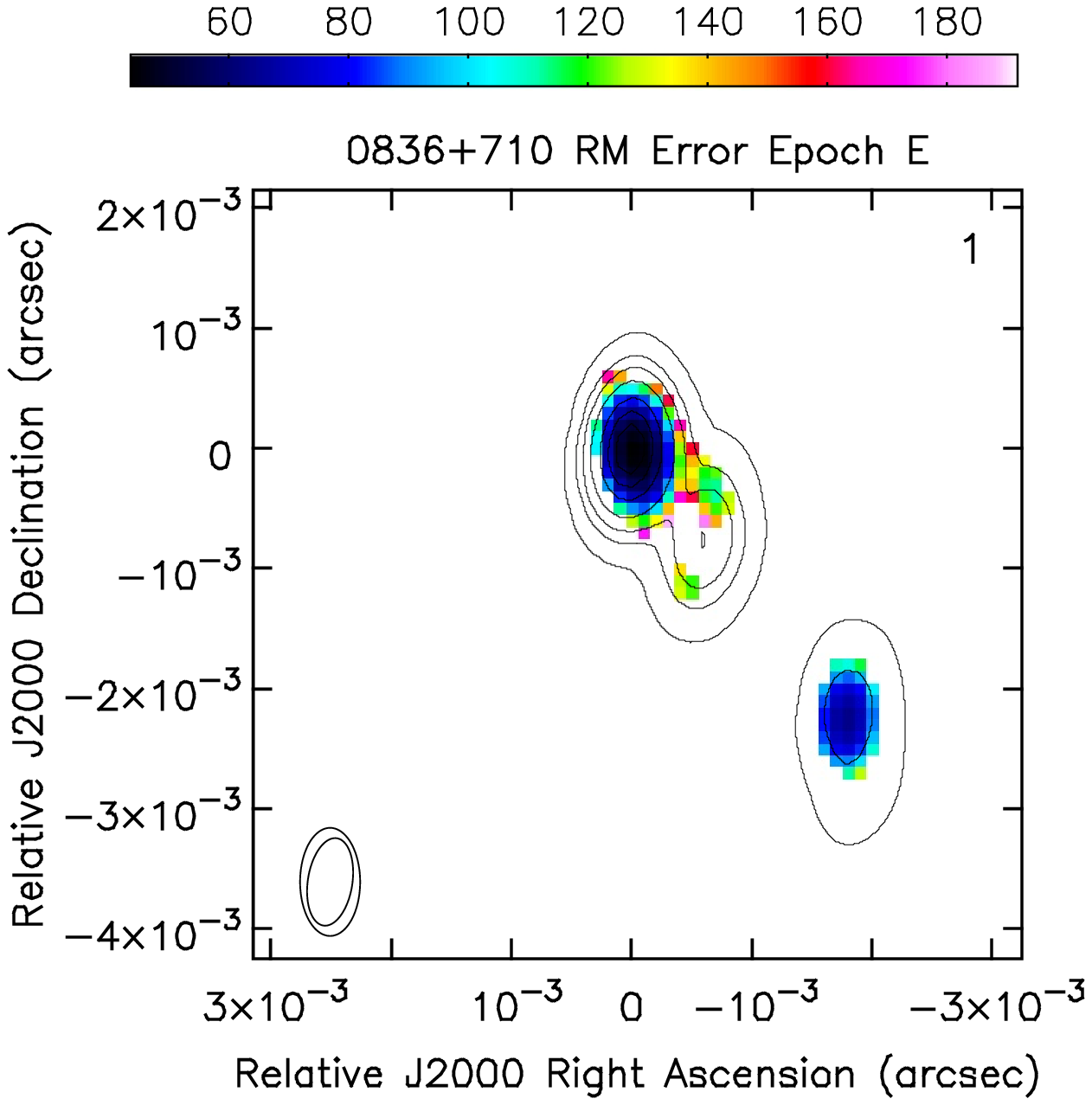}
\includegraphics{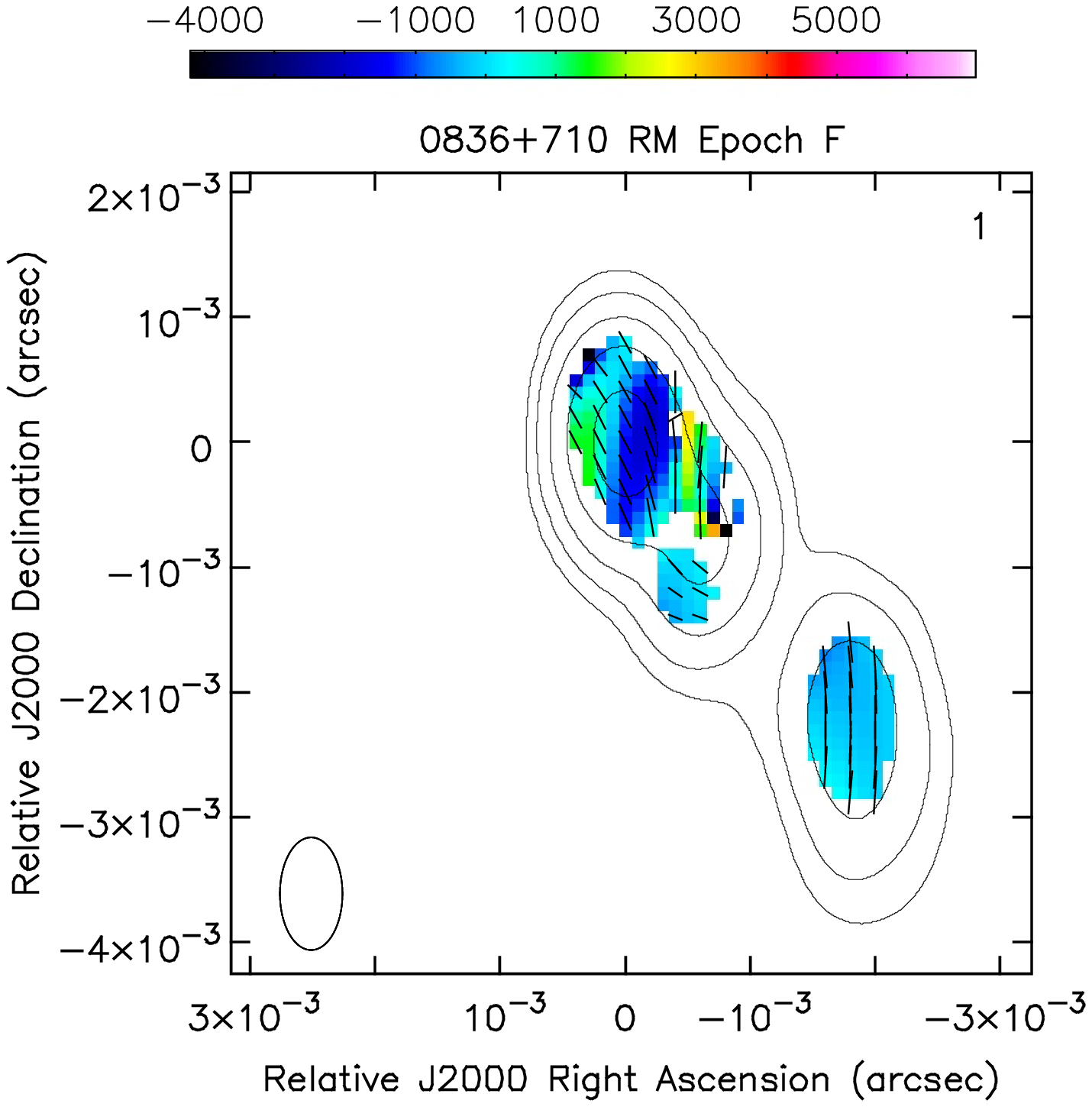}
\includegraphics{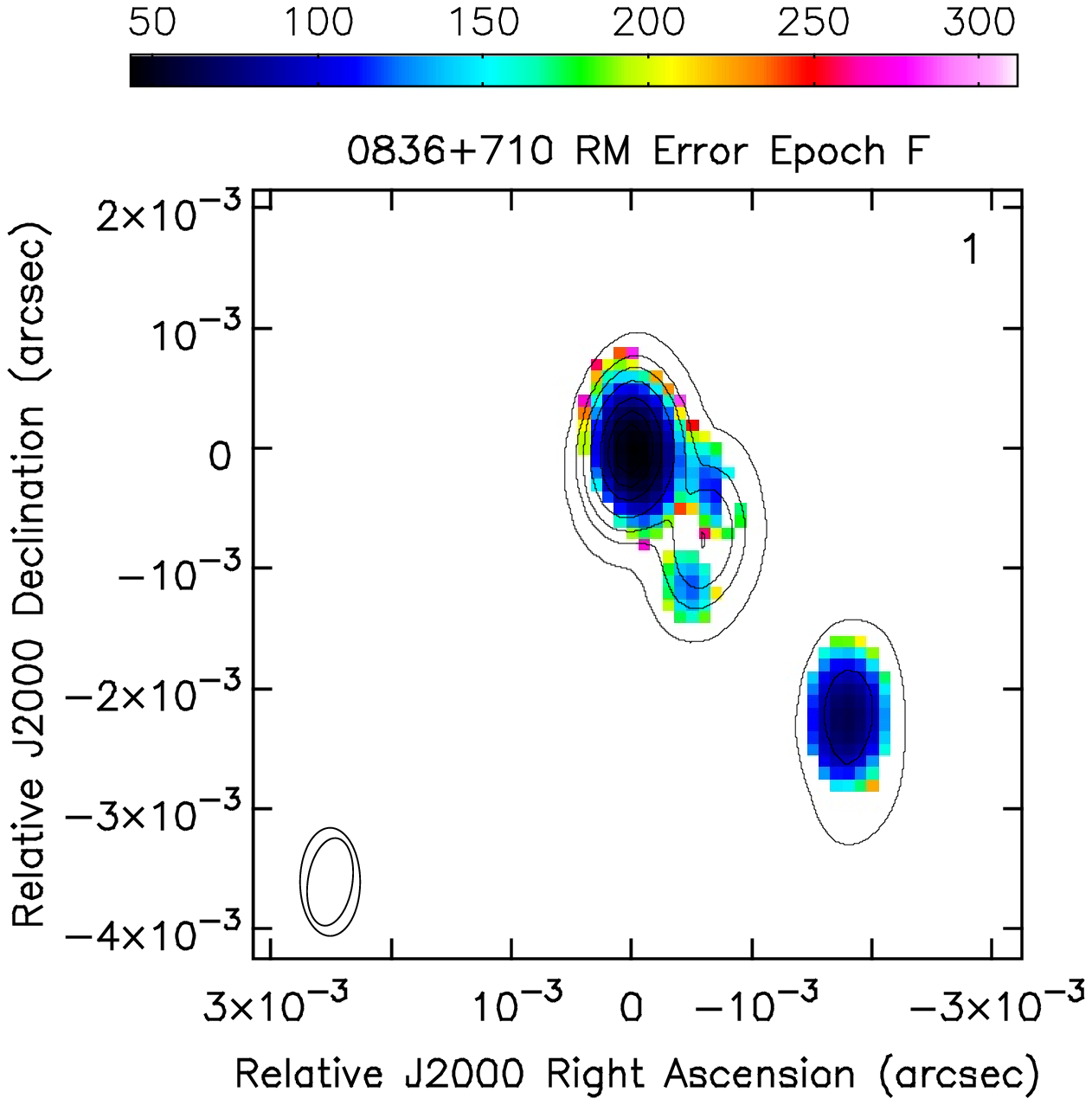}
\vspace{22cm}
\caption{continued}
\end{center}
\end{figure*}

\begin{figure*}
\begin{center}
\includegraphics{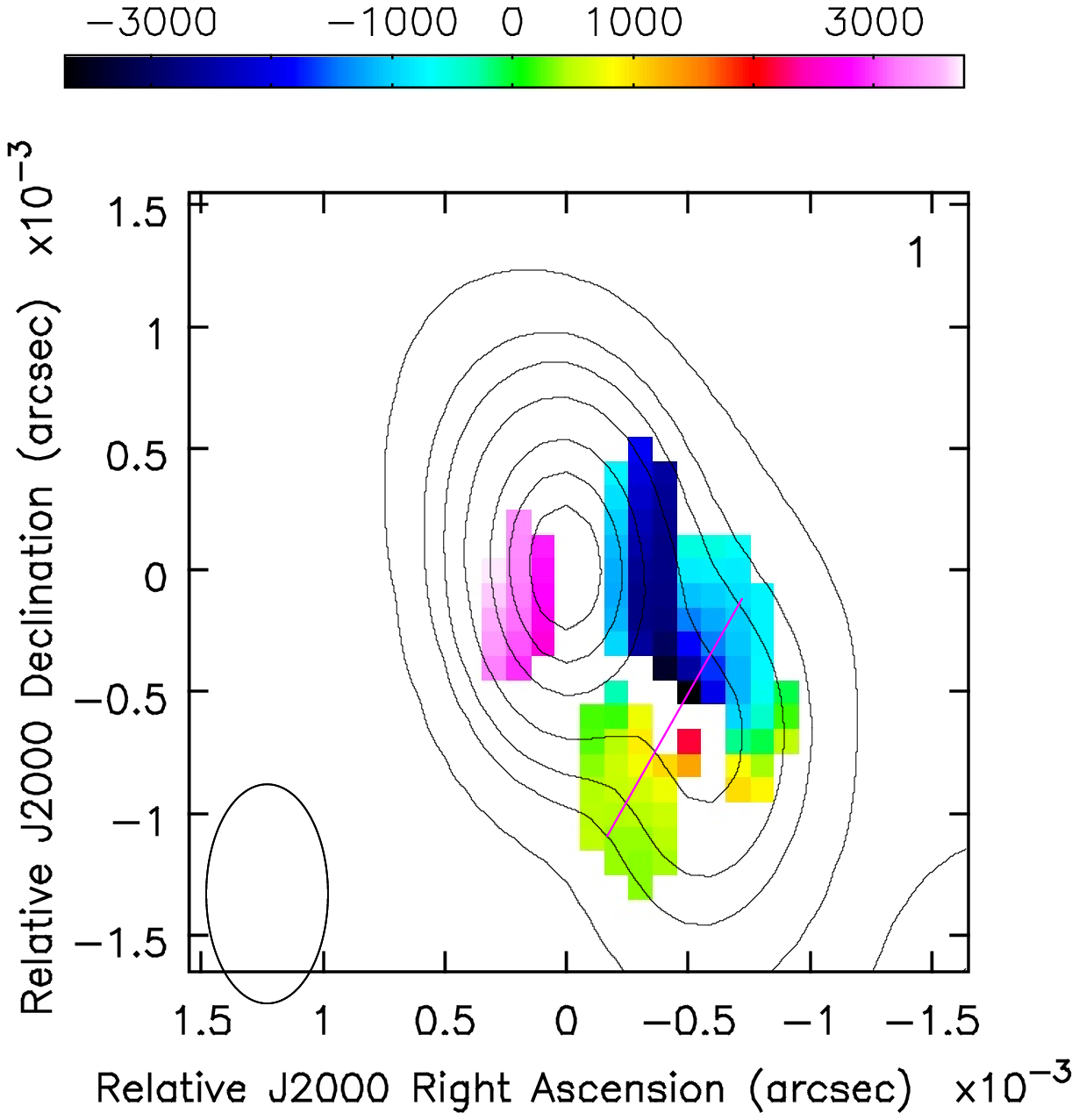}
\includegraphics{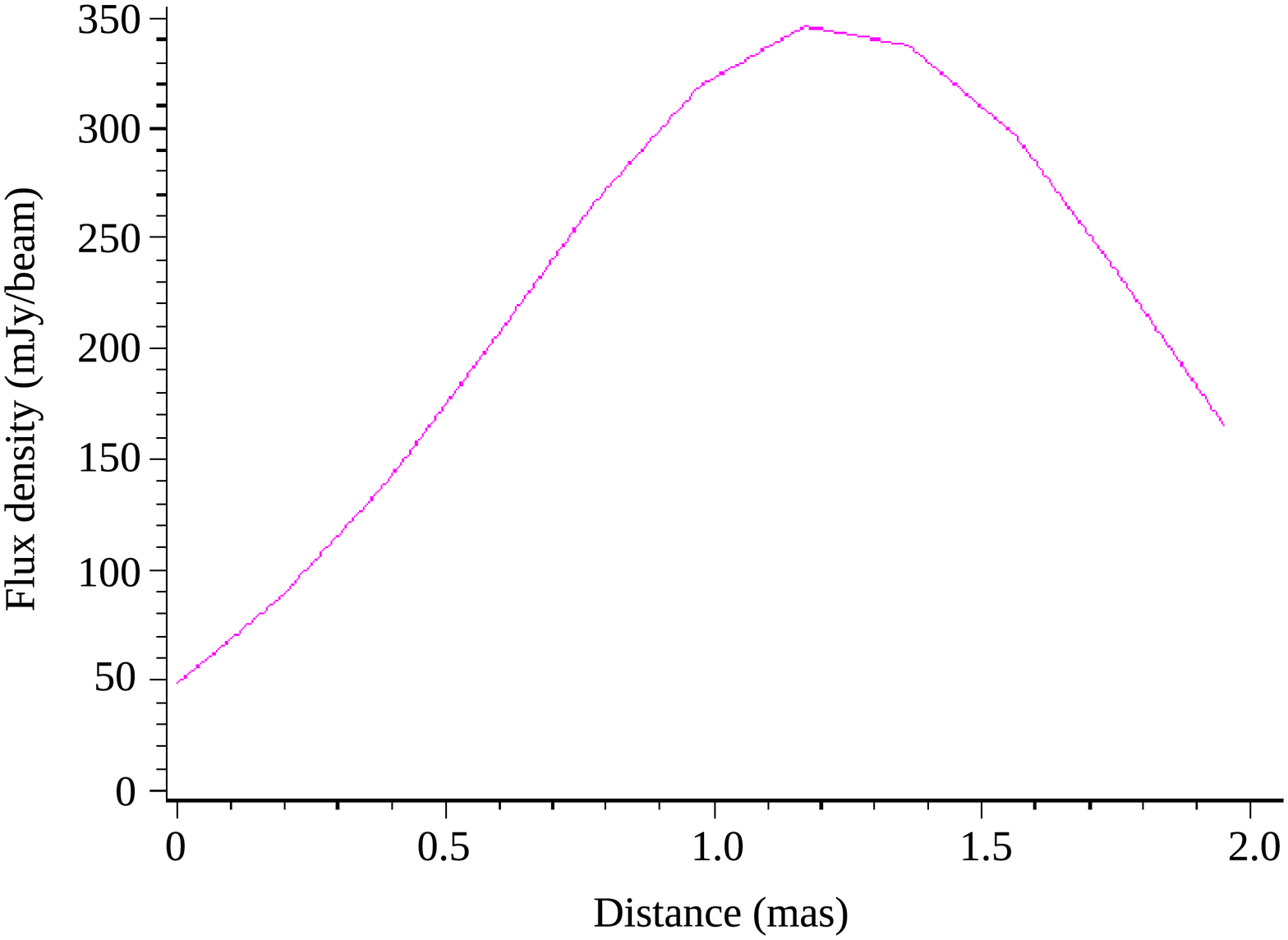}
\includegraphics{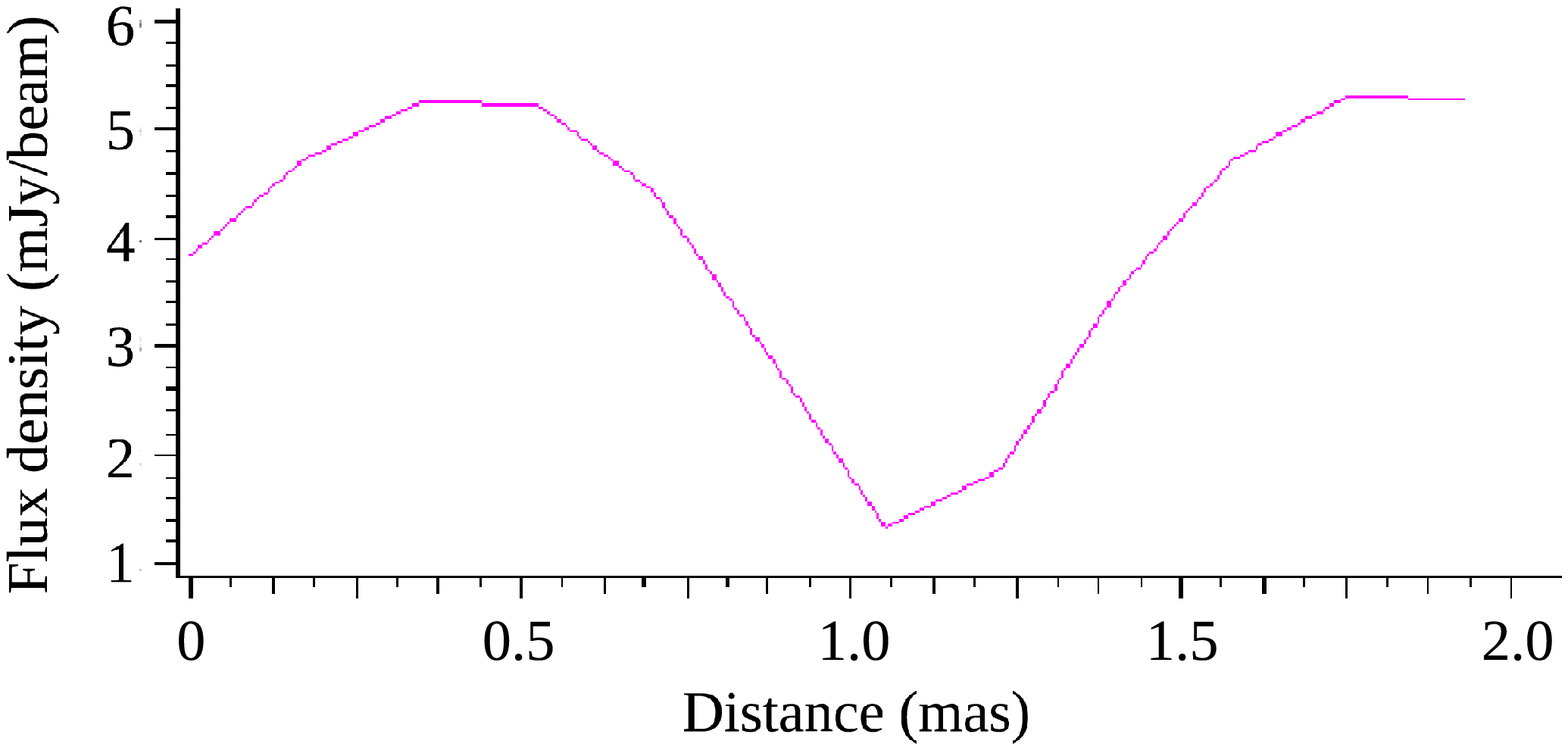}
\includegraphics{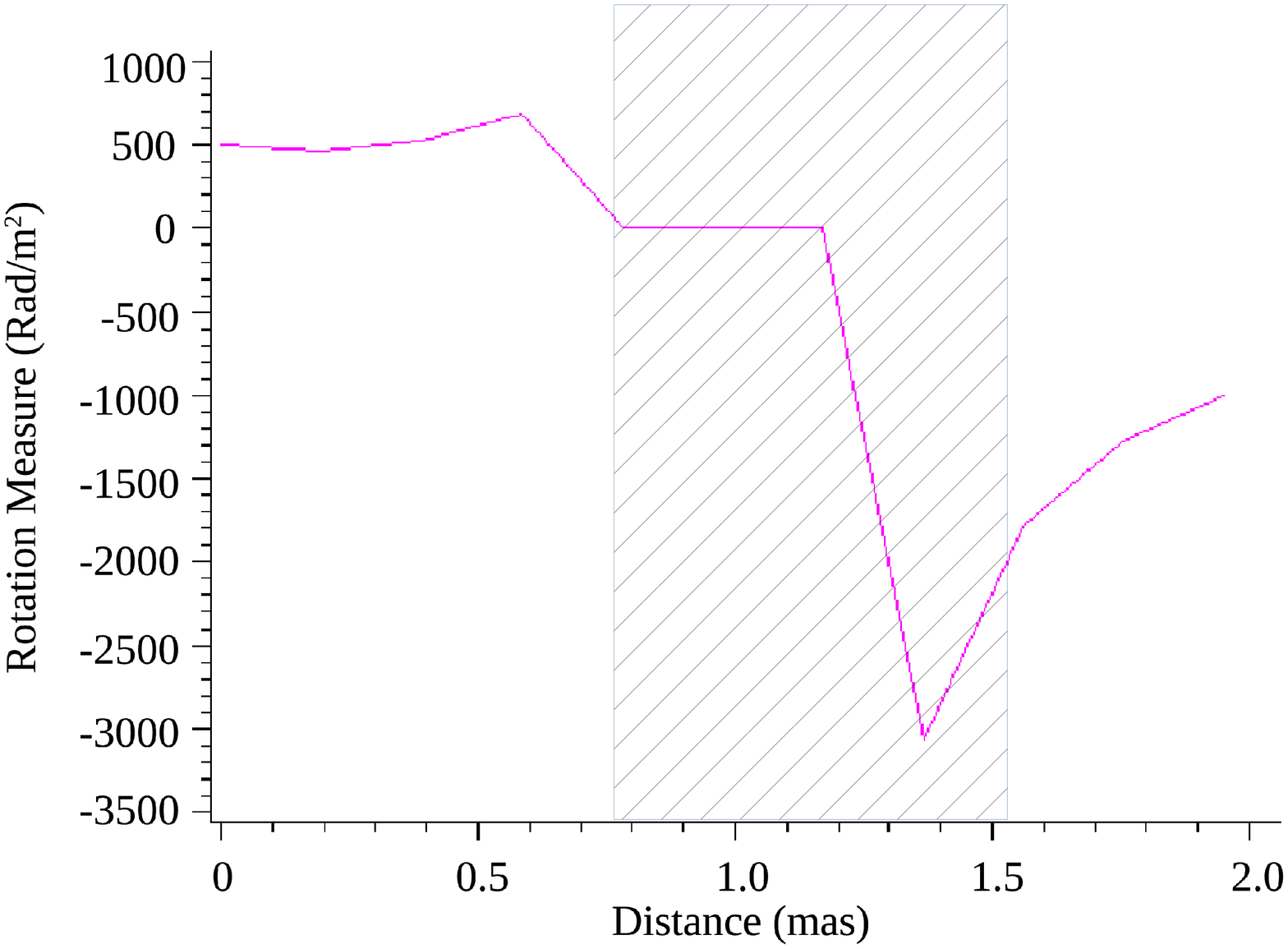}
\includegraphics{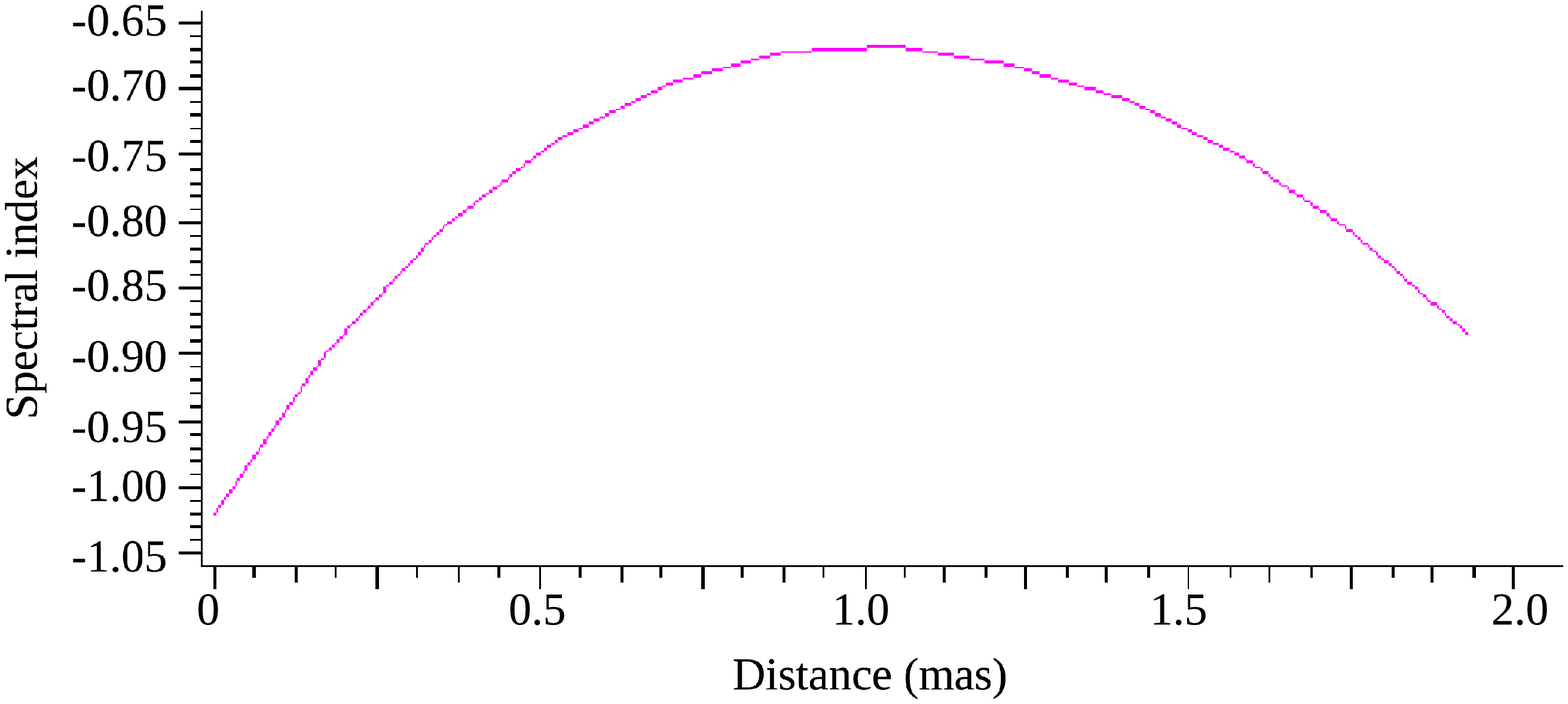}
\vspace{18cm}
\caption{Rotation measure map of S5\,0836$+$710 in 2016 March ({\it top panel}). The
  line represents the transverse slice used for the analysis. {\it Middle left panel}: total intensity
  profile; {\it middle right panel}: polarization profile;
  {\it bottom left panel}: RM
  profile; {\it bottom right panel}: spectral index profile between 15 and 24
  GHz.}
\label{rm-slice} 
\end{center}
\end{figure*}

\section{Summary}

In this paper we reported on results of a broad-band monitoring
campaign, from radio to $\gamma$ rays, of the high redshift FSRQ
S5\,0836$+$710 following a period of high activity detected by {\it
  Fermi}-LAT. During the $\gamma$-ray flares the apparent isotropic
$\gamma$-ray luminosity of the source exceeds 10$^{50}$ erg s$^{-1}$, similar
to other high-redshift objects detected in flares by {\it Fermi}-LAT. In
particular, on 2015 November 9 (MJD 57335) the source reached on 3-hour time-scale the
highest $\gamma$-ray luminosity observed by a blazar ($\sim$3.7
$\times$ 10$^{50}$ 
erg s$^{-1}$). The flux doubling time of 3 hours at the peak of the
$\gamma$-ray activity indicates that the 
size of the emitting region is comparable to the gravitational radius
for this source.

The high $\gamma$-ray activity observed in 2015 might be related to the new superluminal component that emerged from the core at the peak of the radio activity, with the short
variability explained by a strong turbulence in the jet plasma or magnetic reconnection. However, the available data cannot allow us to infer a clear connection between the radio and the $\gamma$-ray activity.

The smaller variability
observed in X-rays with respect to $\gamma$ rays may indicate that the X-ray
emission is produced by the low-energy tail of the same electron distribution
that produces the $\gamma$-ray emission through IC. The optical-UV part of the
spectrum of the source is dominated by the accretion disc emission also during
high activity states.
The small variability observed
in optical and UV bands during our monitoring campaign, suggests that
the optical-UV part of the spectrum has a large contribution from the
accretion disc.\\ 
 
\indent The analysis of multi-epoch full polarization radio observations suggests a
change in the opacity in the core component with time with a
steepening of the spectral index during the latest observing
epochs. Although in total intensity the jet has a
ridge-brightened structure, the polarized emission has a clear
limb-brightened structure in which a RM gradient is observed
transverse to the jet direction. Furthermore, some RM variability is
observed in the core and jet structures with the exception of a knot
in the jet with stable RM.
The polarization properties are consistent with a helical field in a
two-fluid jet model, consisting of an inner, emitting jet and a sheath
containing non-relativistic electrons. In addition, we observe a region
with highly ordered magnetic field in which strong shocks are likely
taking place. However the low dynamic range of these observations
could not allow us to study in detail the polarization structure at
large distances and deeper observations are needed for a better
characterization of the magnetic field along the jet. \\

\label{lastpage}

\section*{Acknowledgments}
We thank the anonymous referee for reading the manuscript carefully
and making valuable suggestions.
MO is grateful to S. Jorstad for fruitful discussion. This study
makes use of 43 GHz VLBA data from the VLBA-BU Blazar Monitoring
Program (VLBA-BU-BLAZAR; 
http://www.bu.edu/blazars/VLBAproject.html), funded by NASA through
the Fermi Guest Investigator Program. The Long Baseline Observatory is
a facility of the National Science Foundation operated by Associated
Universities, Inc.  \\  
The {\it Fermi} LAT Collaboration acknowledges generous ongoing support
from a number of agencies and institutes that have supported both the
development and the operation of the LAT as well as scientific data analysis.
These include the National Aeronautics and Space Administration and the
Department of Energy in the United States, the Commissariat \`a
l'Energie Atomique 
and the Centre National de la Recherche Scientifique / Institut
National de Physique 
Nucl\'eaire et de Physique des Particules in France, the Agenzia
Spaziale Italiana 
and the Istituto Nazionale di Fisica Nucleare in Italy, the Ministry
of Education, 
Culture, Sports, Science and Technology (MEXT), High Energy Accelerator Research
Organization (KEK) and Japan Aerospace Exploration Agency (JAXA) in Japan, and
the K.~A.~Wallenberg Foundation, the Swedish Research Council and the
Swedish National Space Board in Sweden. Additional support for science analysis during the operations phase is gratefully
acknowledged from the Istituto Nazionale di Astrofisica in Italy and the
Centre National d'\'Etudes Spatiales in France. This work performed in part under DOE Contract
  DE-AC02-76SF00515
\\
This research has made use of the
data from the MOJAVE database that is maintained by the MOJAVE team
(Lister et al. 2009, AJ, 137, 3718).\\
This research has made use of the NASA/IPAC
Extragalactic Database NED which is operated by the JPL, California
Institute of Technology, under contract with the National Aeronautics
and Space Administration. \\
This work was supported by the Korea's National Research Council of
Science \& Technology (NST) granted by the International joint
research project (EU-16-001). \\

\appendix

\section{Swift data results}

\begin{table*}
 \caption{Log and fitting results of {\em Swift}-XRT observations of
   S5\,0836$+$710 using a PL model with $N_{\rm H}$ fixed to Galactic
   absorption.} 
 \begin{center}
 \begin{tabular}{cccccc}
 \hline
 \multicolumn{1}{c}{MJD} &
 \multicolumn{1}{c}{Date (UT)} &
 \multicolumn{1}{c}{Net exposure time} &
 \multicolumn{1}{c}{Photon index} &
 \multicolumn{1}{c}{Flux 0.3--10 keV$^{\rm a}$} &
 \multicolumn{1}{c}{$\chi^{2}$ / d.o.f.} \\
 \multicolumn{1}{c}{} &
 \multicolumn{1}{c}{} &
 \multicolumn{1}{c}{(sec)} &
 \multicolumn{1}{c}{($\Gamma_{\rm\,X}$)} &
 \multicolumn{1}{c}{($\times$10$^{-11}$ erg cm$^{-2}$ s$^{-1}$)} &
 \multicolumn{1}{c}{} \\
 \hline
56675 & 2014-01-18 & 4735 &  1.32  $\pm$ 0.05 &  3.86 $\pm$ 0.19  & 67/80   \\   
56778 & 2014-05-01 & 649  &  1.23  $\pm$ 0.16 &  2.86 $\pm$ 0.38  &  9/12   \\
56783 & 2014-05-06 & 719  &  1.14  $\pm$ 0.15 &  2.73 $\pm$ 0.39  & 13/12   \\
56804 & 2014-05-27 & 1076 &  1.29  $\pm$ 0.12 &  3.28 $\pm$ 0.36  & 17/14   \\ 
56832 & 2014-06-24 & 415  &  1.08  $\pm$ 0.21 &  5.84 $\pm$ 1.09  &  6/7    \\  
56836 & 2014-06-28 & 824  &  1.41  $\pm$ 0.09 &  4.57 $\pm$ 0.37  & 33/29   \\
56947 & 2014-10-17 & 9709 &  1.19  $\pm$ 0.03 &  2.93 $\pm$ 0.10  & 161/170 \\   
56987 & 2014-11-26 & 4700 &  1.17  $\pm$ 0.05 &  3.01 $\pm$ 0.14  & 68/98   \\   
57008 & 2014-12-17 & 5017 &  1.23  $\pm$ 0.05 &  2.51 $\pm$ 0.12  & 90/86   \\   
57039 & 2015-01-17 & 4792 &  1.31  $\pm$ 0.05 &  2.37 $\pm$ 0.12  & 91/80   \\   
57070 & 2015-02-17 & 4755 &  1.21  $\pm$ 0.07 &  2.15 $\pm$ 0.13  & 58/58   \\   
57098 & 2015-03-17 & 1104 &  1.29  $\pm$ 0.15 &  1.90 $\pm$ 0.23  & 10/14   \\   
57101 & 2015-03-20 & 3718 &  1.25  $\pm$ 0.08 &  1.85 $\pm$ 0.14  & 33/42   \\   
57128 & 2015-04-16 & 4550 &  1.33  $\pm$ 0.06 &  1.51 $\pm$ 0.09  & 68/55   \\   
57159 & 2015-05-17 & 3174 &  1.33  $\pm$ 0.08 &  1.79 $\pm$ 0.13  & 50/42   \\   
57162 & 2015-05-20 & 1608 &  1.40  $\pm$ 0.11 &  1.82 $\pm$ 0.17  & 26/24   \\   
57193 & 2015-06-20 & 2023 &  1.46  $\pm$ 0.07 &  2.85 $\pm$ 0.18  & 45/47   \\   
57196 & 2015-06-23 & 2108 &  1.39  $\pm$ 0.07 &  2.90 $\pm$ 0.18  & 60/52   \\  
57220 & 2015-07-17 & 4915 &  1.16  $\pm$ 0.04 &  3.35 $\pm$ 0.14  & 119/115 \\  
57243 & 2015-08-09 & 954  &  1.25  $\pm$ 0.13 &  2.89 $\pm$ 0.32  & 21/19   \\   
57245 & 2015-08-11 & 2677 &  1.28  $\pm$ 0.08 &  2.38 $\pm$ 0.16  & 44/44   \\   
57247 & 2015-08-13 & 2103 &  1.38  $\pm$ 0.17 &  2.35 $\pm$ 0.40  & 30/39   \\   
57249 & 2015-08-15 & 2957 &  1.28  $\pm$ 0.06 &  2.58 $\pm$ 0.15  & 65/58   \\    
57251 & 2015-08-17 & 1915 &  1.23  $\pm$ 0.09 &  2.42 $\pm$ 0.20  & 46/37   \\   
57254 & 2015-08-20 & 3276 &  1.27  $\pm$ 0.08 &  2.10 $\pm$ 0.15  & 49/49   \\   
57282 & 2015-09-17 & 3971 &  1.31  $\pm$ 0.06 &  2.54 $\pm$ 0.14  & 77/62   \\   
57325 & 2015-10-30 & 2924 &  1.19  $\pm$ 0.06 &  5.14 $\pm$ 0.30  & 65/62   \\   
57327 & 2015-11-01 & 1915 &  1.21  $\pm$ 0.08 &  4.34 $\pm$ 0.34  & 33/37   \\   
57329 & 2015-11-03 & 5042 &  1.18  $\pm$ 0.05 &  4.51 $\pm$ 0.21  & 106/96  \\   
57331 & 2015-11-05 & 2320 &  1.10  $\pm$ 0.08 &  4.48 $\pm$ 0.36  & 29/39   \\  
57333 & 2015-11-07 & 2929 &  1.11  $\pm$ 0.05 &  4.36 $\pm$ 0.22  & 73/85   \\   
57336 & 2015-11-10 & 1968 &  1.16  $\pm$ 0.08 &  5.49 $\pm$ 0.41  & 39/44   \\   
57359 & 2015-12-03 & 4827 &  1.27  $\pm$ 0.05 &  2.71 $\pm$ 0.12  & 99/99   \\   
57401 & 2016-01-14 & 4665 &  1.26  $\pm$ 0.05 &  2.74 $\pm$ 0.13  & 96/91   \\   
57421 & 2016-02-03 & 2632 &  1.30  $\pm$ 0.07 &  2.87 $\pm$ 0.18  & 47/52   \\   
57425 & 2016-02-07 & 2602 &  1.36  $\pm$ 0.07 &  2.65 $\pm$ 0.16  & 46/55   \\   
57450 & 2016-03-03 & 3207 &  1.28  $\pm$ 0.07 &  2.74 $\pm$ 0.17  & 58/57   \\   
57464 & 2016-03-17 & 3359 &  1.22  $\pm$ 0.07 &  3.32 $\pm$ 0.22  & 50/48   \\   
57469 & 2016-03-22 & 1408 &  1.31  $\pm$ 0.13 &  3.32 $\pm$ 0.36  & 17/19   \\   
57481 & 2016-04-03 & 4420 &  1.20  $\pm$ 0.06 &  2.64 $\pm$ 0.14  & 80/80   \\   
57511 & 2016-05-03 & 4445 &  1.19  $\pm$ 0.05 &  4.07 $\pm$ 0.18  & 123/107 \\   
57540 & 2016-06-01 & 3526 &  1.11  $\pm$ 0.07 &  3.07 $\pm$ 0.22  & 80/68   \\   
57572 & 2016-07-03 & 4917 &  1.17  $\pm$ 0.05 &  3.06 $\pm$ 0.14  & 112/103 \\     
\hline
\end{tabular}
\end{center}
$^{\rm a}$Unabsorbed flux
\label{XRT_table}
\end{table*}

\begin{table*} 
\caption{Observed magnitudes obtained by {\em Swift}-UVOT for S5\,0836$+$710.} 
\begin{center} 
\begin{tabular}{cccccccc} 
\hline 
\multicolumn{1}{c}{MJD} & 
\multicolumn{1}{c}{Date (UT)} & 
\multicolumn{1}{c}{$v$} & 
\multicolumn{1}{c}{$b$} & 
\multicolumn{1}{c}{$u$} & 
\multicolumn{1}{c}{$w1$} & 
\multicolumn{1}{c}{$m2$} & 
\multicolumn{1}{c}{$w2$} \\ 
\hline 
56675 & 2014-01-18 & 16.91 $\pm$ 0.10 & 17.11 $\pm$ 0.07 & 16.23 $\pm$ 0.07 & 16.86 $\pm$ 0.10 & 17.54 $\pm$ 0.12  & 17.90 $\pm$ 0.11 \\
56778 & 2014-05-01 & 16.55 $\pm$ 0.31 & 17.06 $\pm$ 0.08 & 16.38 $\pm$ 0.08 & 17.06 $\pm$ 0.11 & -                 & 18.08 $\pm$ 0.13 \\
56783 & 2014-05-06 & 16.87 $\pm$ 0.14 & 17.26 $\pm$ 0.10 & 17.26 $\pm$ 0.09 & 17.02 $\pm$ 0.13 & 17.68 $\pm$ 0.16  & 17.86 $\pm$ 0.14 \\
56804 & 2014-05-27 & 17.04 $\pm$ 0.13 & 17.03 $\pm$ 0.09 & 16.18 $\pm$ 0.08 & 17.03 $\pm$ 0.11 & 17.44 $\pm$ 0.13  & 17.70 $\pm$ 0.11 \\
56832 & 2014-06-24 & 16.81 $\pm$ 0.19 & 17.10 $\pm$ 0.14 & 16.10 $\pm$ 0.11 & 16.91 $\pm$ 0.15 & 17.74 $\pm$ 0.23  & 17.78 $\pm$ 0.16 \\
56836 & 2014-06-28 & 16.94 $\pm$ 0.20 & 17.20 $\pm$ 0.14 & 16.23 $\pm$ 0.11 & 16.79 $\pm$ 0.14 & 17.47 $\pm$ 0.22  & 17.84 $\pm$ 0.17 \\
56947 & 2014-10-17 & 16.93 $\pm$ 0.17 & 17.13 $\pm$ 0.10 & 16.37 $\pm$ 0.10 & 16.99 $\pm$ 0.13 & 17.56 $\pm$ 0.15  & 17.64 $\pm$ 0.13 \\
56987 & 2014-11-26 & 16.94 $\pm$ 0.16 & 17.24 $\pm$ 0.12 & 16.24 $\pm$ 0.09 & 17.16 $\pm$ 0.14 & 17.50 $\pm$ 0.15  & 17.85 $\pm$ 0.14 \\
57008 & 2014-12-17 & 17.06 $\pm$ 0.10 & 17.15 $\pm$ 0.07 & 16.30 $\pm$ 0.07 & 17.13 $\pm$ 0.10 & 17.63 $\pm$ 0.12  & 17.93 $\pm$ 0.10 \\
57039 & 2015-01-17 & 17.07 $\pm$ 0.11 & 17.16 $\pm$ 0.08 & 16.34 $\pm$ 0.07 & 17.07 $\pm$ 0.11 & 17.68 $\pm$ 0.14  & 18.05 $\pm$ 0.12 \\
57070 & 2015-02-17 & 17.09 $\pm$ 0.10 & 17.13 $\pm$ 0.07 & 16.32 $\pm$ 0.07 & 17.08 $\pm$ 0.10 & 17.83 $\pm$ 0.13  & 17.98 $\pm$ 0.11 \\
57098 & 2015-03-17 & 17.12 $\pm$ 0.12 & 17.11 $\pm$ 0.08 & 16.32 $\pm$ 0.08 & 17.00 $\pm$ 0.11 & 17.43 $\pm$ 0.13  & 17.93 $\pm$ 0.12 \\
57101 & 2015-03-20 & 16.92 $\pm$ 0.10 & 17.12 $\pm$ 0.07 & 16.35 $\pm$ 0.07 & 17.01 $\pm$ 0.10 & 17.58 $\pm$ 0.12  & 18.07 $\pm$ 0.11 \\
57128 & 2015-04-16 & 17.01 $\pm$ 0.14 & 17.16 $\pm$ 0.10 & 16.26 $\pm$ 0.09 & 16.97 $\pm$ 0.11 & 17.37 $\pm$ 0.12  & 18.09 $\pm$ 0.14 \\
57159 & 2015-05-17 & 17.15 $\pm$ 0.13 & 17.11 $\pm$ 0.08 & 16.28 $\pm$ 0.08 & 16.99 $\pm$ 0.11 & 17.40 $\pm$ 0.13  & 17.75 $\pm$ 0.11 \\
57162 & 2015-05-20 & 16.96 $\pm$ 0.10 & 17.23 $\pm$ 0.08 & 16.23 $\pm$ 0.07 & 16.97 $\pm$ 0.10 & 17.38 $\pm$ 0.11  & 17.84 $\pm$ 0.11 \\
57193 & 2015-06-20 & 16.98 $\pm$ 0.15 & 17.07 $\pm$ 0.10 & 16.08 $\pm$ 0.08 & 16.91 $\pm$ 0.13 & 17.72 $\pm$ 0.21  & 17.80 $\pm$ 0.14 \\
57196 & 2015-06-23 & 16.90 $\pm$ 0.16 & 17.07 $\pm$ 0.09 & 16.23 $\pm$ 0.08 & 17.07 $\pm$ 0.12 & 17.51 $\pm$ 0.19  & 18.12 $\pm$ 0.14 \\
57220 & 2015-07-17 & 17.05 $\pm$ 0.16 & 17.22 $\pm$ 0.11 & 16.36 $\pm$ 0.09 & 16.98 $\pm$ 0.12 & 17.60 $\pm$ 0.15  & 17.90 $\pm$ 0.13 \\
57243 & 2015-08-09 & 17.05 $\pm$ 0.17 & 17.10 $\pm$ 0.11 & 16.37 $\pm$ 0.10 & 17.17 $\pm$ 0.16 & 17.38 $\pm$ 0.15  & 17.89 $\pm$ 0.14 \\
57245 & 2015-08-11 & 17.01 $\pm$ 0.13 & 17.05 $\pm$ 0.10 & 16.26 $\pm$ 0.09 & 17.09 $\pm$ 0.13 & 17.66 $\pm$ 0.14  & 17.92 $\pm$ 0.25 \\
57247 & 2015-08-13 & 16.95 $\pm$ 0.15 & 17.22 $\pm$ 0.11 & 16.39 $\pm$ 0.10 & 17.03 $\pm$ 0.13 & 17.59 $\pm$ 0.16  & 17.94 $\pm$ 0.15 \\
57249 & 2015-08-15 & 17.01 $\pm$ 0.12 & 17.25 $\pm$ 0.10 & 16.32 $\pm$ 0.08 & 17.05 $\pm$ 0.10 & 17.64 $\pm$ 0.12  & 17.95 $\pm$ 0.11 \\
57251 & 2015-08-17 & -                & 17.21 $\pm$ 0.08 & 16.38 $\pm$ 0.08 & 17.23 $\pm$ 0.11 & -                 & 17.80 $\pm$ 0.23 \\
57254 & 2015-08-20 & 16.89 $\pm$ 0.18 & 17.33 $\pm$ 0.14 & 16.47 $\pm$ 0.11 & 17.18 $\pm$ 0.16 & 17.60 $\pm$ 0.17  & 17.81 $\pm$ 0.16 \\
57282 & 2015-09-17 & 17.18 $\pm$ 0.22 & 16.95 $\pm$ 0.11 & 16.28 $\pm$ 0.11 & 17.25 $\pm$ 0.19 & 17.55 $\pm$ 0.27  & 17.87 $\pm$ 0.18 \\
57325 & 2015-10-30 & 16.68 $\pm$ 0.10 & 16.83 $\pm$ 0.08 & 16.11 $\pm$ 0.07 & 16.73 $\pm$ 0.11 & 17.39 $\pm$ 0.08  & 17.64 $\pm$ 0.12 \\
57327 & 2015-11-01 & 16.98 $\pm$ 0.12 & 17.14 $\pm$ 0.09 & 16.21 $\pm$ 0.08 & 16.85 $\pm$ 0.11 & 17.33 $\pm$ 0.13  & 17.81 $\pm$ 0.12 \\
57329 & 2015-11-03 & 17.10 $\pm$ 0.18 & 17.05 $\pm$ 0.09 & 16.38 $\pm$ 0.09 & 16.97 $\pm$ 0.13 & 17.26 $\pm$ 0.17  & 17.77 $\pm$ 0.14 \\
57331 & 2015-11-05 & 16.87 $\pm$ 0.17 & 17.13 $\pm$ 0.11 & 16.33 $\pm$ 0.10 & 16.97 $\pm$ 0.15 & 17.89 $\pm$ 0.20  & 17.59 $\pm$ 0.15 \\
57333 & 2015-11-07 & 16.99 $\pm$ 0.13 & 17.06 $\pm$ 0.09 & 16.24 $\pm$ 0.08 & 16.99 $\pm$ 0.12 & 17.38 $\pm$ 0.13  & 17.84 $\pm$ 0.13 \\
57336 & 2015-11-10 & 16.89 $\pm$ 0.11 & 17.12 $\pm$ 0.08 & 16.25 $\pm$ 0.08 & 16.86 $\pm$ 0.11 & 17.33 $\pm$ 0.10  & 17.63 $\pm$ 0.11 \\
57359 & 2015-12-03 & 16.82 $\pm$ 0.10 & 17.10 $\pm$ 0.07 & 16.18 $\pm$ 0.07 & 16.85 $\pm$ 0.09 & 17.37 $\pm$ 0.11  & 17.85 $\pm$ 0.11 \\
57401 & 2016-01-14 & 16.75 $\pm$ 0.10 & 16.98 $\pm$ 0.08 & 16.17 $\pm$ 0.07 & 16.78 $\pm$ 0.10 & 17.38 $\pm$ 0.13  & 17.58 $\pm$ 0.14 \\
57421 & 2016-02-03 & 16.90 $\pm$ 0.11 & 16.94 $\pm$ 0.08 & 16.17 $\pm$ 0.07 & 16.89 $\pm$ 0.11 & 17.46 $\pm$ 0.14  & 17.75 $\pm$ 0.12 \\
57425 & 2016-02-07 & 16.73 $\pm$ 0.11 & 16.99 $\pm$ 0.08 & 16.07 $\pm$ 0.07 & 16.79 $\pm$ 0.11 & 17.54 $\pm$ 0.14  & 17.58 $\pm$ 0.11 \\
57450 & 2016-03-03 & 16.72 $\pm$ 0.14 & 16.90 $\pm$ 0.10 & 16.21 $\pm$ 0.10 & 16.91 $\pm$ 0.14 & 17.36 $\pm$ 0.17  & 17.67 $\pm$ 0.15 \\
57464 & 2016-03-17 & 16.86 $\pm$ 0.17 & 16.88 $\pm$ 0.11 & 16.06 $\pm$ 0.10 & 16.96 $\pm$ 0.16 & 17.25 $\pm$ 0.15  & 17.57 $\pm$ 0.16 \\
57469 & 2016-03-22 & 16.83 $\pm$ 0.12 & 17.06 $\pm$ 0.09 & 16.10 $\pm$ 0.08 & 16.85 $\pm$ 0.12 & 17.43 $\pm$ 0.14  & 17.76 $\pm$ 0.14 \\
57481 & 2016-04-03 & 16.68 $\pm$ 0.13 & 17.00 $\pm$ 0.10 & 16.17 $\pm$ 0.09 & 16.98 $\pm$ 0.13 & 17.53 $\pm$ 0.16  & 17.56 $\pm$ 0.13 \\
57511 & 2016-05-03 & 16.80 $\pm$ 0.09 & 17.07 $\pm$ 0.07 & 16.17 $\pm$ 0.07 & 16.82 $\pm$ 0.09 & 17.35 $\pm$ 0.12  & 17.63 $\pm$ 0.10 \\
57540 & 2016-06-01 & 16.74 $\pm$ 0.15 & 16.98 $\pm$ 0.11 & 16.12 $\pm$ 0.09 & 17.03 $\pm$ 0.19 & 17.24 $\pm$ 0.13  & 17.71 $\pm$ 0.14 \\
57572 & 2016-07-03 & 16.90 $\pm$ 0.14 & 17.10 $\pm$ 0.10 & 16.15 $\pm$ 0.08 & 16.77 $\pm$ 0.10 & 17.36 $\pm$ 0.12  & 17.57 $\pm$ 0.11 \\
\hline
\end{tabular}
\end{center}
\label{UVOT}
\end{table*}

\clearpage

\section{Model-fit analysis}
\label{section-modelfit}
To derive structural changes we complemented our observations with
those from the VLBA Boston University blazar (VLBA-BU-BLAZAR) program
performed between 2015 May and 2018 May. To this aim we fitted
the visibility data with circular Gaussian components at each epoch
using the model-fitting option in \texttt{DIFMAP}.
This approach is used in order to derive small structure
  variation and provide an accurate fit of unresolved components close to the
core components. Direct comparison of models obtained independently at
each epoch is not the best approach to detect small changes
\citep{conway92}. For this reason, we produced a zero-order model
consisting 
of four circular Gaussian components, which was used as the initial
model in model-fitting the visibility data of each observing
epoch. The convergence of the fit is reached when both the core and A1
components are considered point-like sources. After 2016 March an
additional circular component was included in the model.  
Results are reported in Table \ref{modelfit_table}.\\
Errors on the component position are $\Delta r = a/(S_{\rm p}/{\rm
  rms})$, where $a$ is the component deconvolved major-axis, $S_{\rm
  p}$ is the component peak flux density and the rms is the 1$\sigma$
noise level measured on the image plane. In case
the errors are unreliably small, we assume a more conservative value
that corresponds to 10 per cent of the beam \citep[e.g.][]{mo11}.\\

\begin{table*}
\caption{Results of the visibility model-fit analysis at 43 GHz. Column 1: observing epoch; column 2: source component as from Fig. \ref{modelfit-image}; column 3: flux density at 43 GHz (in mJy); column 4: component angular size (in milliarcsecond); Column 5: distance from the core component (in milliarcsecond); column 6: position angle with respect to the core component (in degree).}
\begin{center}
\begin{tabular}{cccccc}
  \hline
Epoch & Comp. & $S_{\rm 43 GHz}$& $a$ & $r$ & $\Theta$ \\  
\hline
23/09/2014 & C  &  808 & -     & -     & - \\
           & C1 &  199 & 0.14  & 0.024 & $-$133  \\
           & B3 &  440 & 0.40  & 0.610 & $-$136 \\
           & A1 &  523 &  -    & 0.100 & 54 \\
05/12/2014 & C  &  515 & -     & -     & - \\
           & C1 &  543 & 0.05  & 0.082 & $-$106  \\
           & B3 &  328 & 0.41  & 0.715 & $-$136 \\
           & A1 &  107 &  -    & 0.087 & 24 \\
29/12/2014 & C  &  814 & -     & -     & - \\
           & C1 &  428 & 0.01  & 0.080 & $-$102  \\
           & B3 &  330 & 0.41  & 0.730 & $-$136 \\
           & A1 &  109 &  -    & 0.065 & 9 \\
15/02/2015 & C  &  530 & -     & -     & - \\
           & C1 &  290 & 0.07  & 0.098 & $-$133  \\
           & B3 &  312 & 0.40  & 0.750 & $-$136 \\
           & A1 &  856 &  -    & 0.040 & 165 \\
12/04/2015 & C  & 1185 & -     & -     & - \\
           & C1 &  689 & 0.03  & 0.070 & $-$97  \\
           & B3 &  315 & 0.41  & 0.770 & $-$136 \\
           & A1 &  220 &  -    & 0.050 & 19 \\
11/05/2015 & C  &  948 & - & -              & - \\
           & C1 &  532 & 0.094 & 0.055 & $-$101  \\
           & B3 &  211 & 0.240 & 0.781 & $-$137 \\
           & A1 &  271 &  -    & 0.075 & 57 \\
09/06/2015 & C  &  778 & -     &  -    & -  \\
           & C1 & 1080 & 0.094 & 0.057 & $-$104 \\
           & B3 &  305 & 0.360 & 0.802 & $-$136 \\
           & A1 &  141 &   -   & 0.101 &  36 \\
02/07/2015 & C  &  684 &   -   &  -    &  -  \\
           & C1 &  658 & 0.094 & 0.054 & $-$122 \\
           & B3 &  193 & 0.240 & 0.807 & $-$137 \\
           & A1 &  286 &  -    & 0.074 &  57 \\
15/08/2015 & C  &  655 &  -    &   -   &  - \\
           & C1 &  481 & 0.094 & 0.070 & $-$111 \\
           & B3 &  172 & 0.250 & 0.825 & $-$138 \\
           & A1 &  306 &   -   & 0.079 &  52 \\
22/09/2015 & C  &  873 &   -   &   -   &  -  \\
           & C1 &  846 & 0.094 & 0.092 & $-$117 \\
           & B3 &  190 & 0.250 & 0.890 & $-$138 \\
           & A1 &   95 &   -   & 0.111 &  37 \\
05/12/2015 & C  &  887 &   -   &   -   &  -  \\
           & C1 &  102 & 0.094 & 0.094 & $-$120 \\
           & B3 &  148 & 0.290 & 0.888 & $-$138 \\
           & A1 &  240 &   -   & 0.095 &  45 \\
01/01/2016 & C  &  695 &   -   &   -   &  - \\
           & C1 &  465 & 0.093 & 0.097 & $-$120 \\
           & B3 &  133 & 0.280 & 0.932 & $-$139 \\
           & A1 &  110 &  -    & 0.113 &  48 \\
31/01/2016 & C  &  740 &  -    &   -   &  - \\
           & C1 &  411 & 0.093 & 0.110 & $-$117 \\
           & B3 &  180 & 0.370 & 0.935 & $-$138 \\
           & A1 &  180 &   -   & 0.110 & 38 \\
18/03/2016 & C  &  420 &   -   &   -   &  -  \\
           & C1 &  228 & 0.093 & 0.023 & 177 \\
           & N  &  308 & 0.130 & 0.134 & $-$123 \\
           & B3 &  133 & 0.310 & 0.967 & $-$139 \\
           & A1 &  221 &   -   & 0.106 &  37 \\
22/04/2016 & C  &  502 &   -   &   -   &  - \\
           & C1 &  303 & 0.093 & 0.039 & $-$155 \\
           & N  &  232 & 0.130 & 0.168 & $-$124 \\
           & B3 &  123 & 0.320 & 1.000 & $-$140 \\
           & A1 &  195 &  -    & 0.102 &  41 \\
\hline
\end{tabular}
\end{center}
\label{modelfit_table}
\end{table*}

\addtocounter{table}{-1}
\begin{table*}
\caption{Continued.}
\begin{center}
\begin{tabular}{cccccc}
  \hline
  Epoch & Comp. & $S_{\rm 43 GHz}$& $a$ & $r$ & $Theta$ \\  
  \hline
 10/06/2016 & C  &  557 &   -   &   -   &  -  \\
           & C1 &  216 & 0.094 & 0.090 & $-$131 \\
           & N  &  122 & 0.130 & 0.233 & $-$122 \\
           & B3 &  105 & 0.330 & 1.017 & $-$140 \\
           & A1 &  171 &   -   & 0.119 &  38 \\
04/07/2016 & C  &  419 &   -   &   -   &  -  \\
           & C1 &  249 & 0.091 & 0.070 & $-$134 \\
           & N  &  148 & 0.130 & 0.220 & $-$123 \\
           & B3 &  105 & 0.340 & 1.045 & $-$140 \\
           & A1 &  178 &   -   & 0.118 &  41 \\
16/08/2016 & C  &  549 &   -   &   -   &  -  \\
           & C1 &  266 & 0.110 & 0.086 & $-$121 \\
           & N  &  157 & 0.120 & 0.240 & $-$123 \\
           & B3 &  150 & 0.340 & 1.060 & $-$140 \\
           & A1 &  258 &   -   & 0.127 &  36 \\
05/09/2016 & C  &  446 &   -   &   -   &  - \\ 
           & C1 &  296 & 0.092 & 0.076 & $-$137 \\
           & N  &  193 & 0.130 & 0.246 & $-$124 \\
           & B3 &  127 & 0.335 & 1.087 & $-$141 \\
           & A1 &  235 &   -   & 0.112 &  41 \\
23/10/2016 & C  &  214 &   -   &  -    &  - \\ 
           & C1 &  222 & 0.090 & 0.101 & $-$135 \\
           & N  &  156 & 0.140 & 0.288 & $-$126 \\
           & B3 &   94 & 0.360 & 1.128 & $-$141 \\
28/11/2016 & C  &  148 &   -   &  -    &  -  \\
           & C1 &  159 & 0.080 & 0.086 & $-$135 \\
           & N  &  138 & 0.160 & 0.280 & $-$126 \\
           & B3 &   81 & 0.360 & 1.159 & $-$142 \\
           & A1 &  157 &   -   & 0.112 &   39 \\
23/12/2016 & C  &  102 &   -   &   -   &  - \\
           & C1 &  340 & 0.090 & 0.098 & $-$153 \\
           & N  &  254 & 0.200 & 0.311 & $-$131  \\
           & B3 &  120 & 0.350 & 1.210 & $-$142 \\
           & A1 &  255 &   -   & 0.060 &  61 \\
14/01/2017 & C  &  222 &   -   &   -   &  - \\
           & C1 &  201 & 0.090 & 0.112 & $-$139 \\
           & N  &  193 & 0.170 & 0.328 & $-$131 \\
           & B3 &  105 & 0.370 & 1.214 & $-$142 \\
           & A1 &  228 &   -   & 0.070 &  40 \\
04/02/2017 & C  &  273 &   -   &   -   &  - \\
           & C1 &  202 & 0.070 & 0.115 & $-$142 \\
           & N  &  180 & 0.200 & 0.353 & $-$128 \\
           & B3 &   90 & 0.350 & 1.249 & $-$142 \\
           & A1 &  191 &  -    & 0.047 &  36 \\
16/04/2017 & C  &  322 &  -    &   -   &  - \\
           & C1 &  182 & 0.080 & 0.064 & $-$130 \\
           & N  &  156 & 0.240 & 0.364 & $-$128 \\
           & B3 &   66 & 0.360 & 1.270 & $-$141 \\
           & A1 &  122 &   -   & 0.061 &  29 \\
13/05/2017 & C  &  536 &   -   &   -   &  -  \\
           & C1 &  312 & 0.060 & 0.050 & $-$124 \\
           & N  &  216 & 0.240 & 0.377 & $-$128 \\
           & B3 &   95 & 0.360 & 1.276 & $-$141 \\
           & A1 &  280 &  -    & 0.070 &   33 \\
08/06/2017 & C  &  847 &  -    &   -   &  -  \\
           & C1 &  164 & 0.050 & 0.069 & $-$135 \\ 
           & N  &  226 & 0.240 & 0.365 & $-$127 \\
           & B3 &  101 & 0.360 & 1.275 & $-$142 \\
           & A1 &  194 &   -   & 0.088 &  44 \\
03/07/2017 & C  &  713 &   -   &   -   &  -  \\
           & C1 &  278 & 0.090 & 0.047 & $-$138 \\
           & N  &  189 & 0.180 & 0.376 & $-$128 \\
           & B3 &   96 & 0.410 & 1.291 & $-$141 \\ 
           & A1 &  176 &   -   & 0.098 &  46 \\
\hline
\end{tabular}
\end{center}
\end{table*}  

\addtocounter{table}{-1}
\begin{table*}
\caption{Continued.}
\begin{center}
\begin{tabular}{cccccc}
  \hline
  Epoch & Comp. & $S_{\rm 43 GHz}$& $a$ & $r$ & $Theta$ \\  
\hline
06/08/2017 & C  &  497 &   -   &   -   &  - \\
           & C1 &  200 & 0.060 & 0.043 & $-$135 \\
           & N  &  153 & 0.200 & 0.368 & $-$127 \\
           & B3 &   56 & 0.300 & 1.339 & $-$141 \\
           & A1 &  178 &   -   & 0.104 &  48 \\
04/09/2017 & C  &  338 &   -   &  -    &  -  \\
           & C1 &  404 & 0.070 & 0.025 & $-$121 \\
           & N  &  153 & 0.190 & 0.389 & $-$127 \\
           & B3 &   75 & 0.420 & 1.339 & $-$142 \\
           & A1 &  206 &   -   & 0.111 &  43 \\
06/11/2017 & C  &  407 &   -   &  -    &  - \\
           & C1 &  440 & 0.080 & 0.030 & $-$128 \\
           & N  &  145 & 0.220 & 0.417 & $-$127 \\
           & B3 &   59 & 0.080 & 1.395 & $-$141 \\
           & A1 &  220 &    -  & 0.103 &  48 \\
17/02/2018 & C  &  171 &   -   &  -    &  - \\
           & C1 &  485 & 0.100 & 0.017 & 179 \\
           & N  &  117 & 0.310 & 0.452 & $-$129 \\
           & B3 &   49 & 0.280 & 1.443 & $-$142 \\
           & A1 &  124 &  -    & 0.137 &  50 \\ 
10/03/2018 & C  &  507 &  -    &  -    & - \\
           & C1 &  232 & 0.090 & 0.064 & $-$129 \\
           & N  &  109 & 0.270 & 0.497 & $-$127 \\
           & B3 &   62 & 0.370 & 1.437 & $-$141 \\ 
           & A1 &  208 &   -   & 0.123 &  38 \\
19/04/2018 & C  &  499 &   -   &   -   &  - \\
           & C1 &  394 & 0.080 & 0.073 & $-$144 \\
           & N  &  110 & 0.310 & 0.533 & $-$128 \\
           & B3 &   48 & 0.250 & 1.500 & $-$140 \\
           & A1 &  287 &   -   & 0.110 &  41 \\
11/05/2018 & C  &  390 &   -   &   -   &  - \\           
           & C1 &  451 & 0.080 & 0.070 & $-$137 \\
           & N  &  119 & 0.310 & 0.537 & $-$128 \\
           & B3 &   50 & 0.310 & 0.537 & $-$140 \\
           & A1 &  322 &   -   & 0.109 &   37 \\
\hline
\end{tabular}
\end{center}
\end{table*}  

\newpage

\section{Multi-epoch polarization images}
\label{section_polla}

\begin{figure*}
\begin{center}
\includegraphics{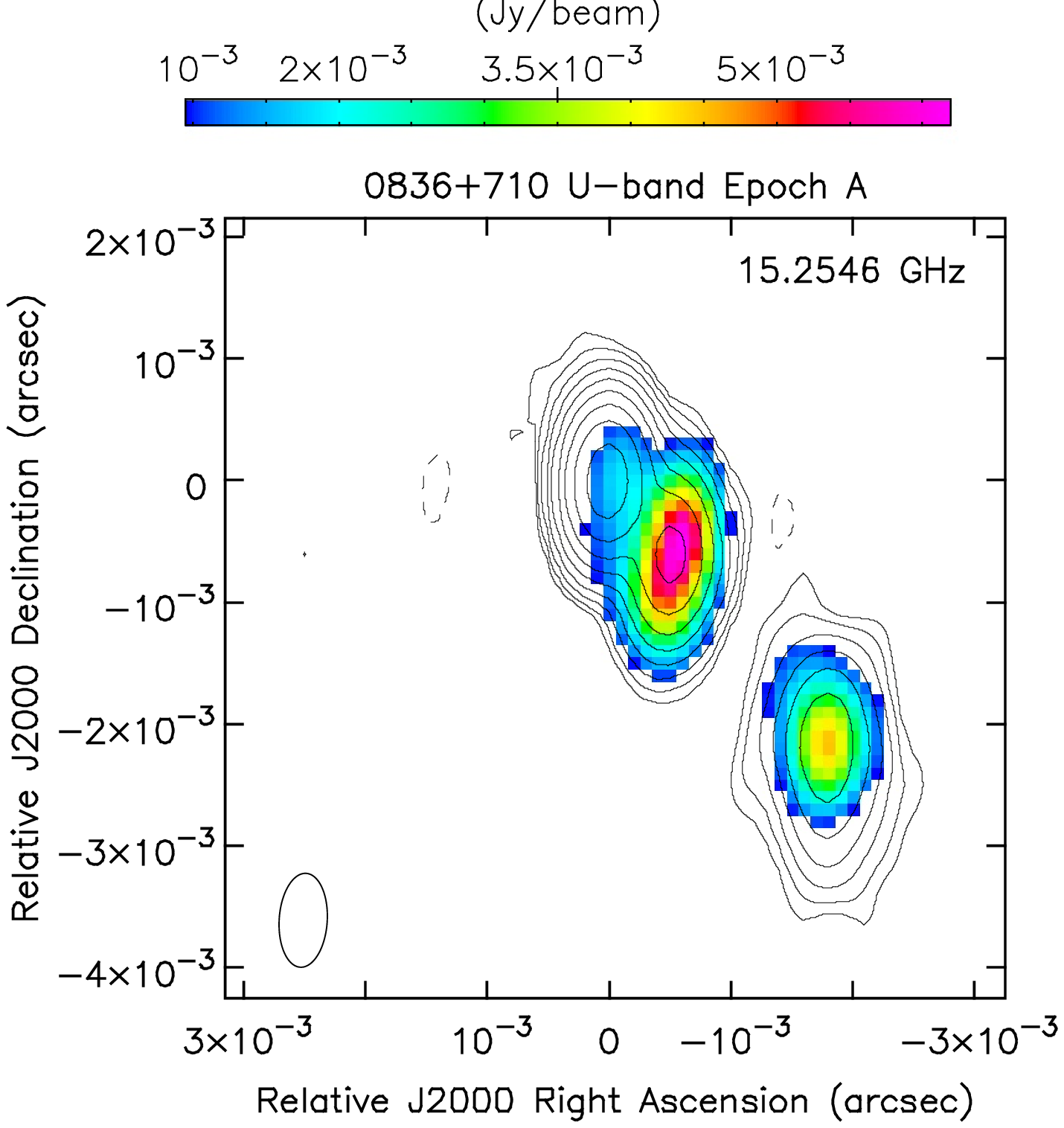}
\includegraphics{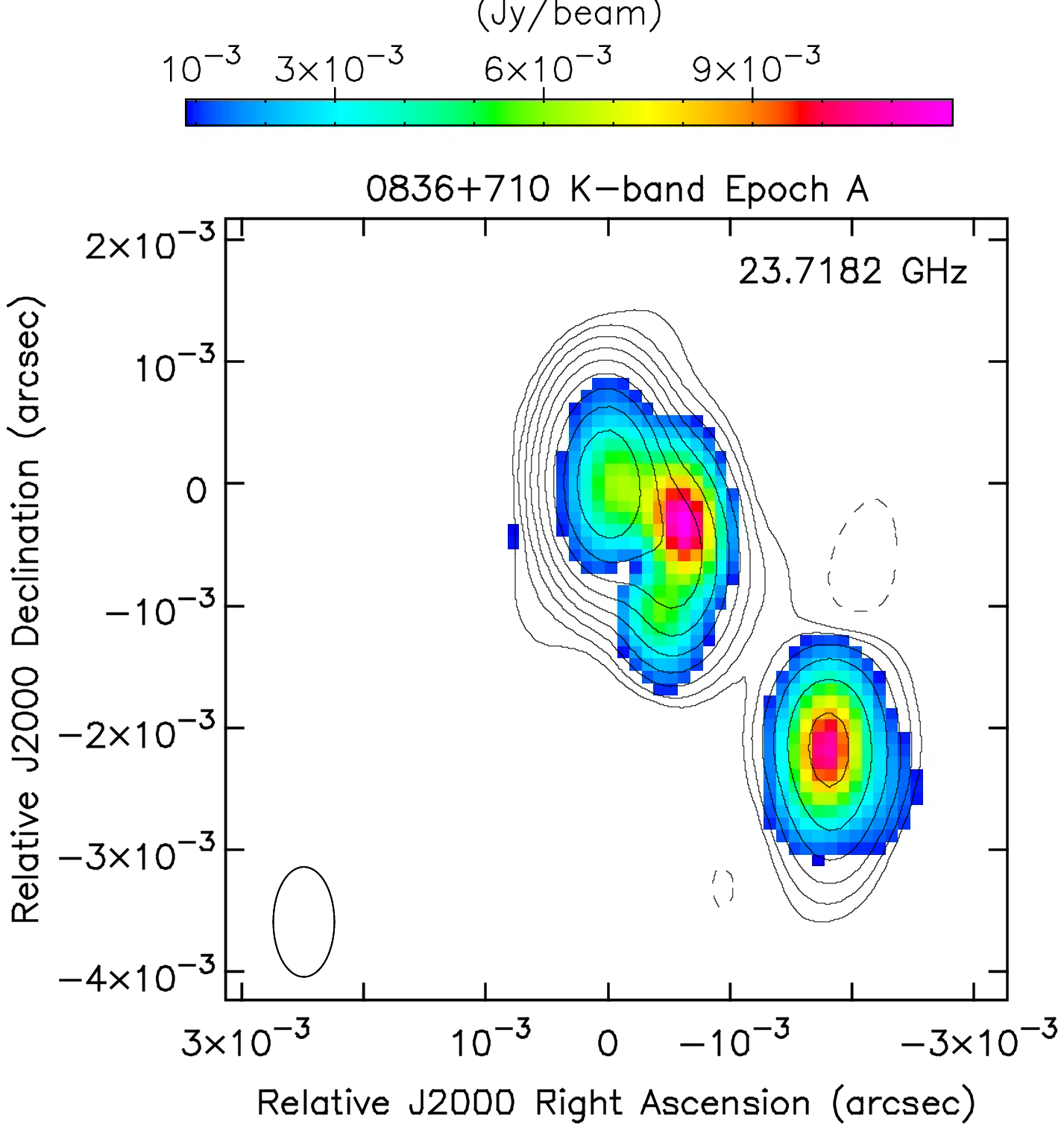}
\includegraphics{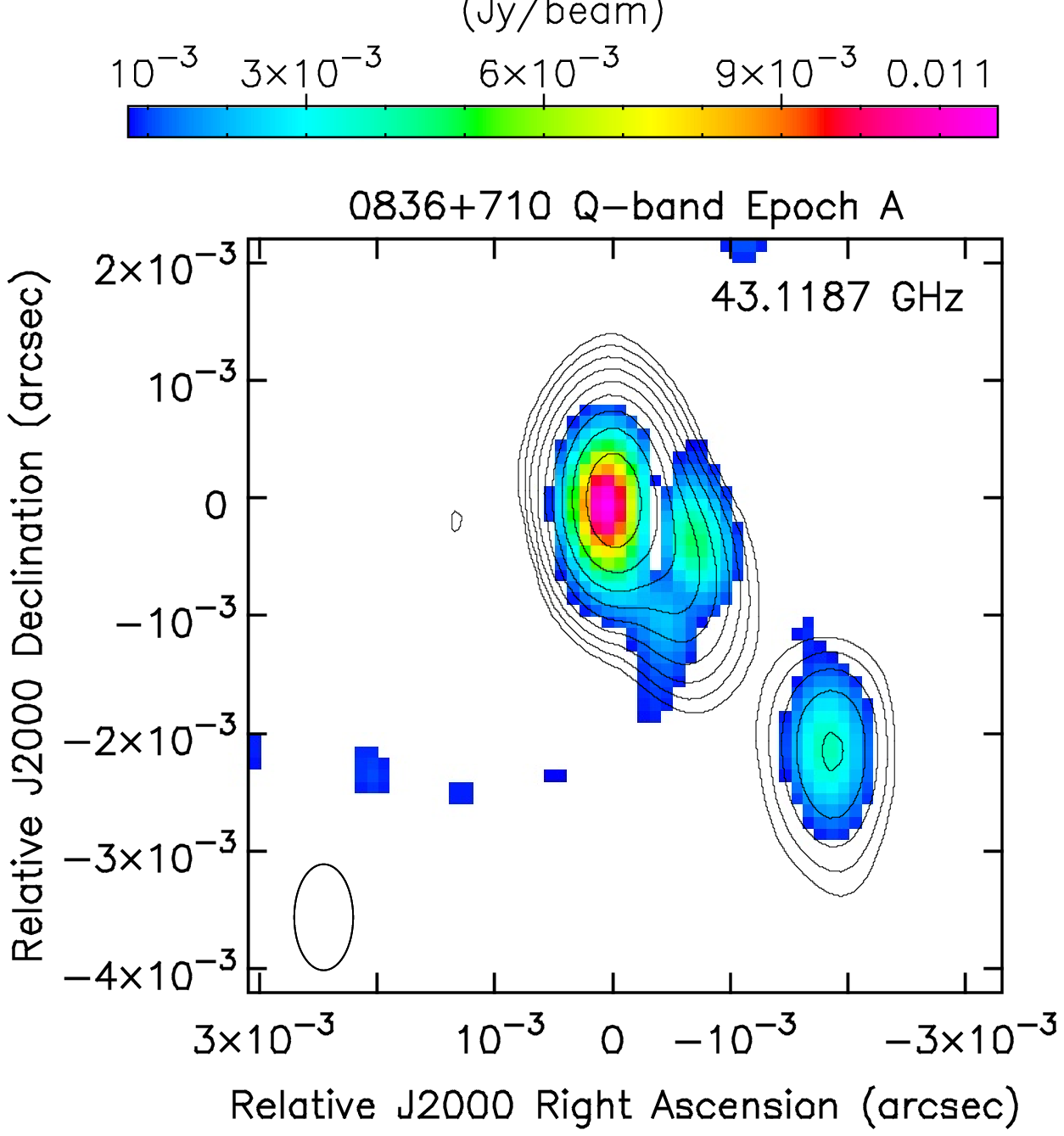}
\includegraphics{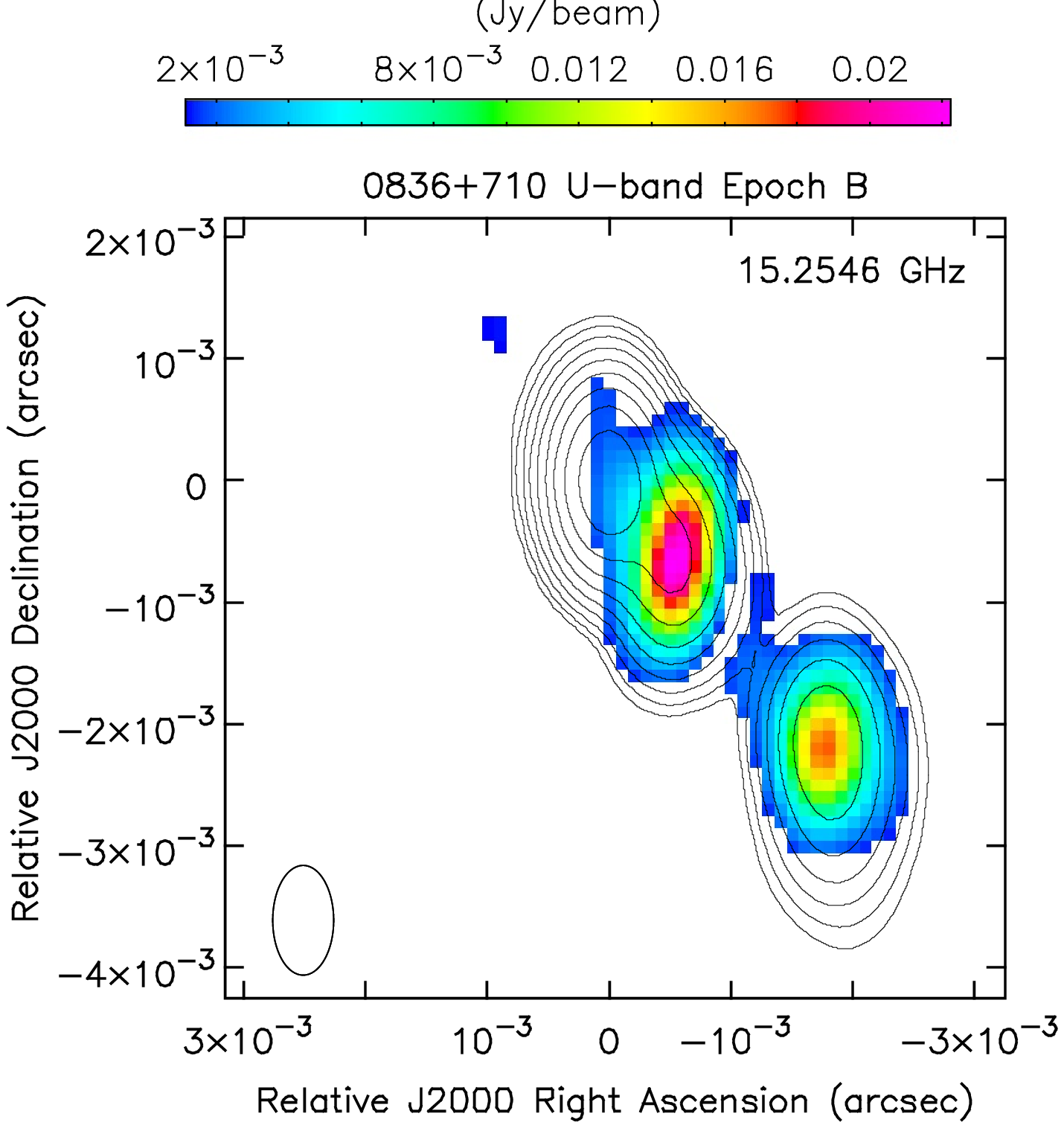}
\includegraphics{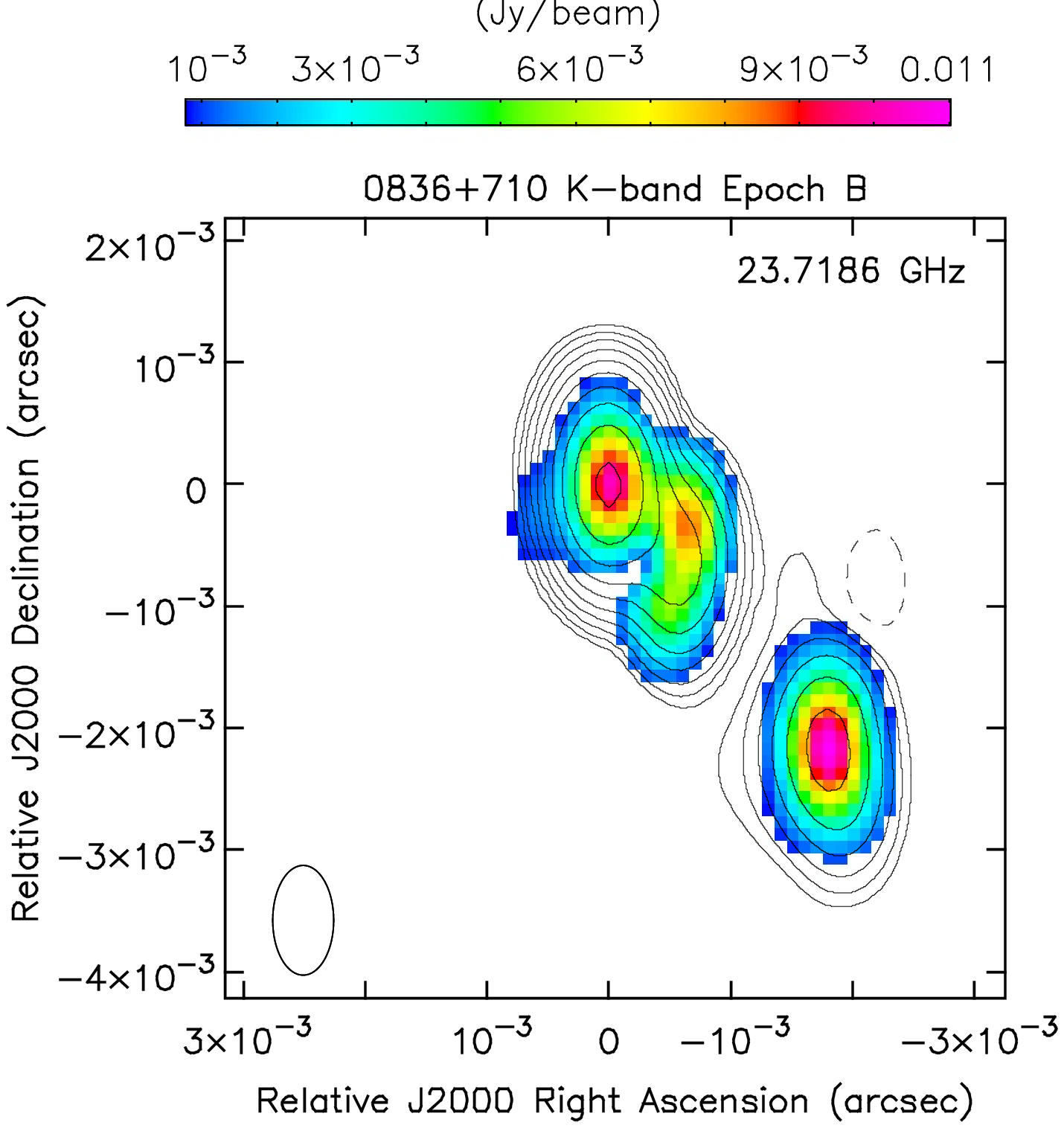}
\includegraphics{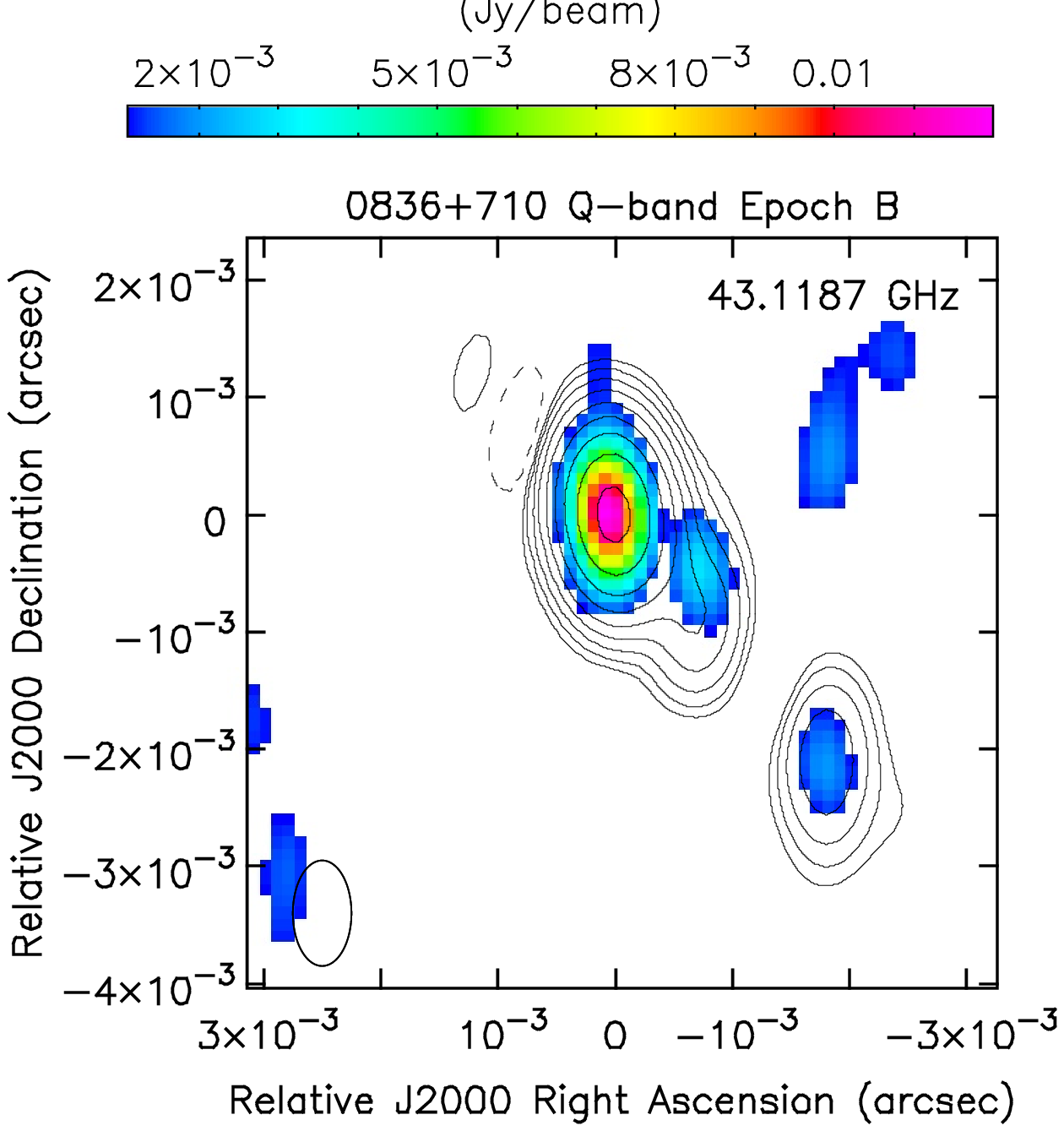}
\includegraphics{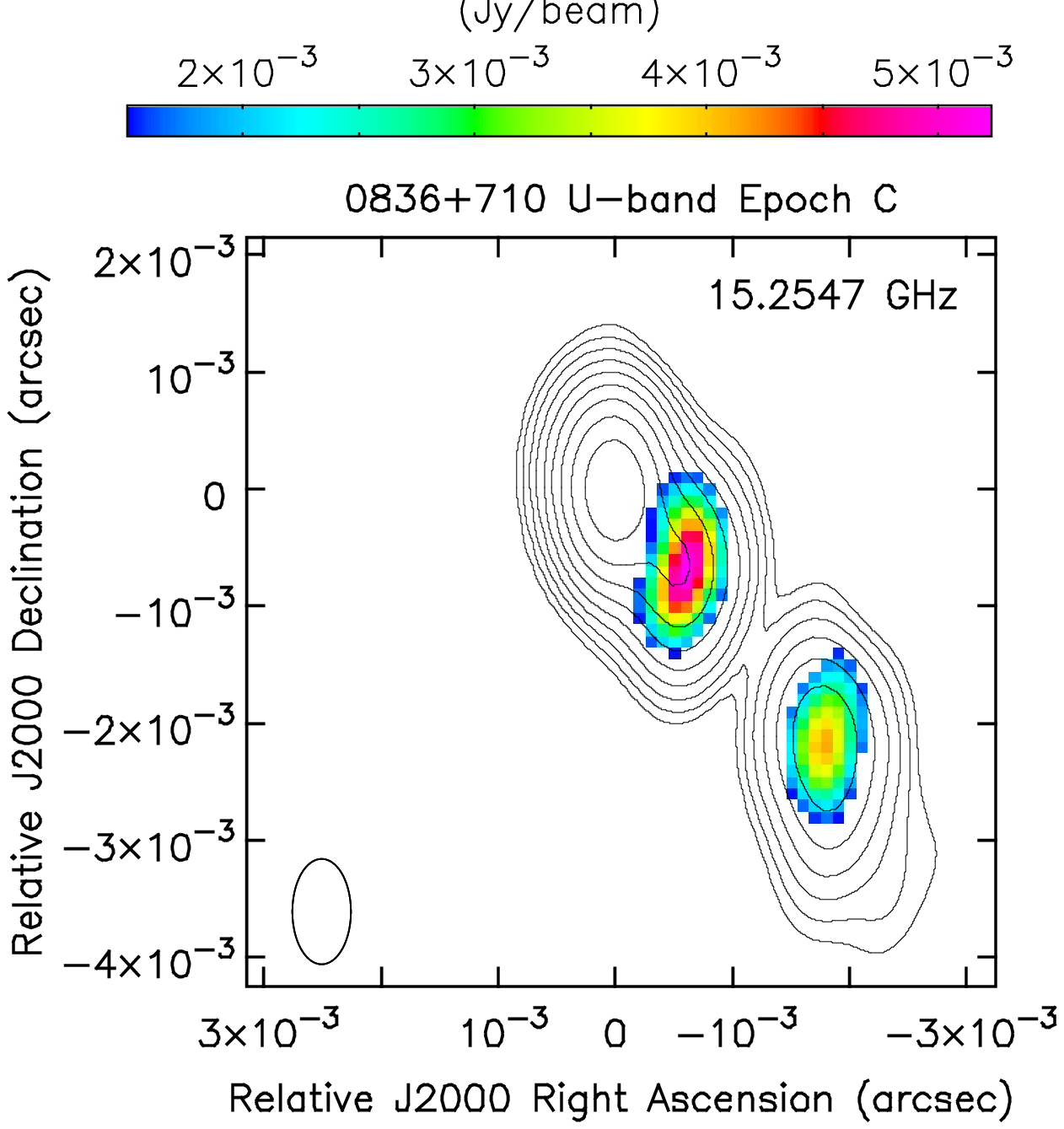}
\includegraphics{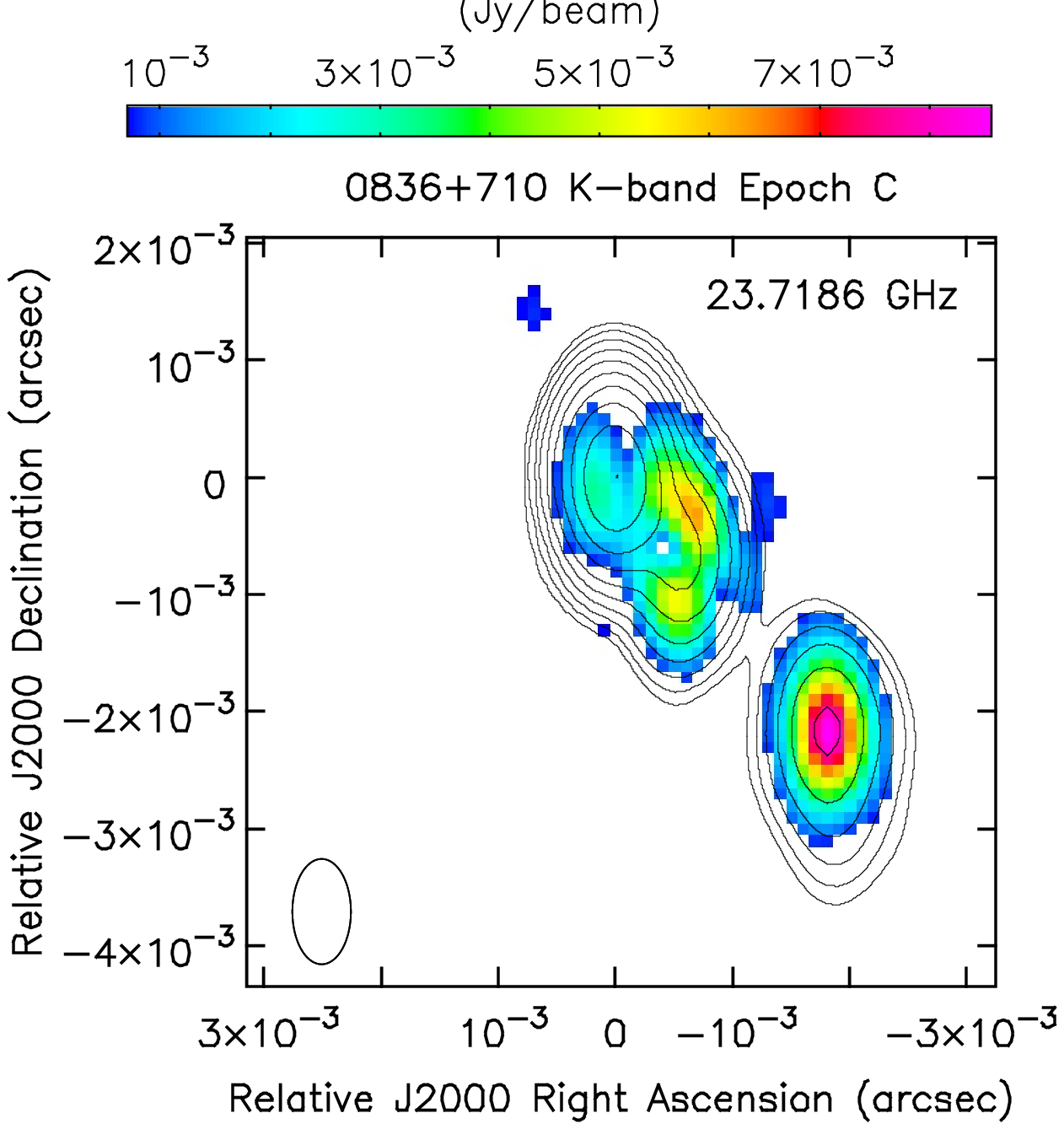}
\includegraphics{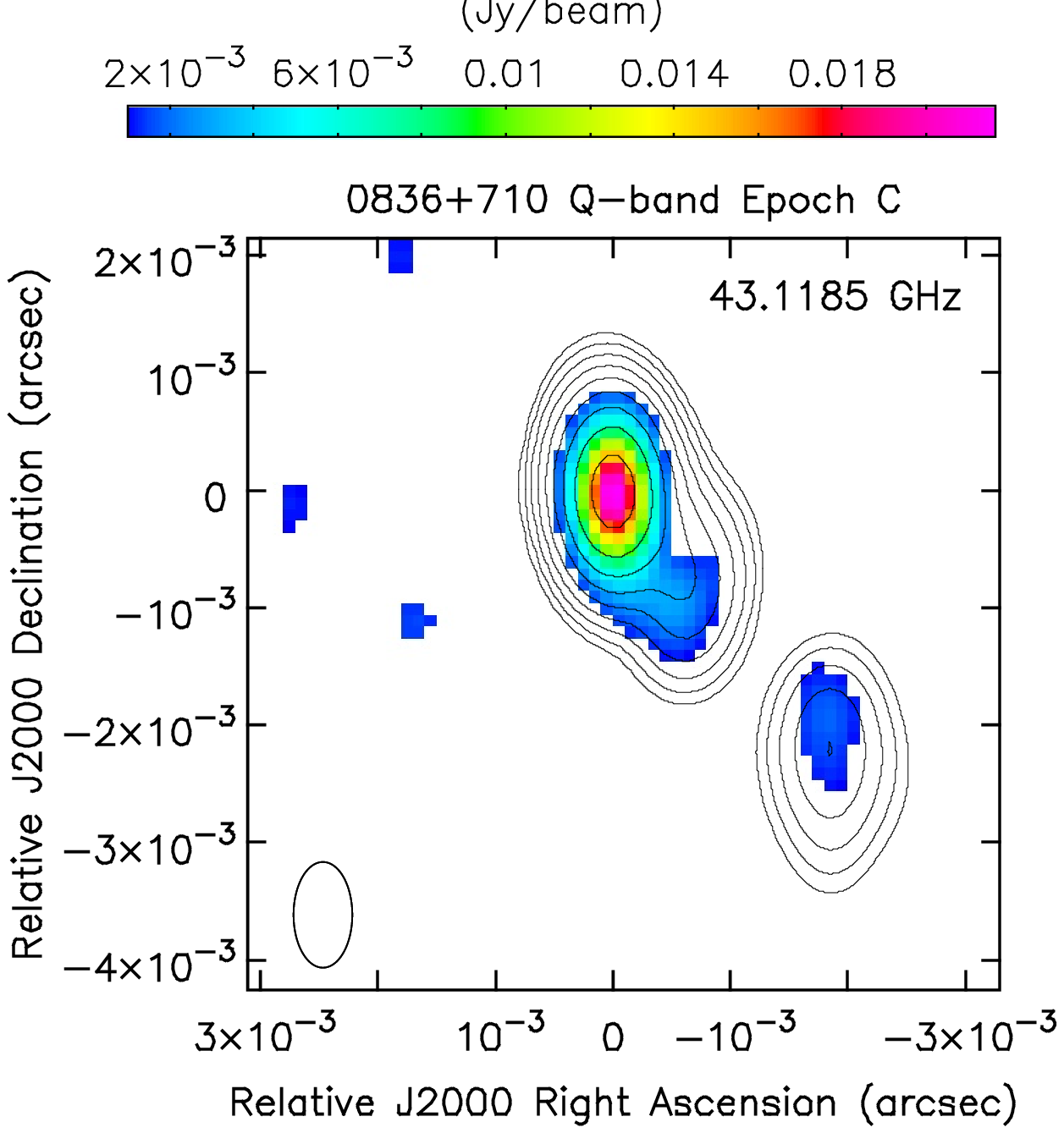}
\vspace{21cm}
\caption{VLBA images of the source S5\,0836$+$710. The colour-scale
  represents the polarization intensity, while contours are the total
  intensity. The first contour is about $\sim$3 mJy/beam, which
  corresponds to three times the off-source noise level. Contour
  levels increase by a factor of 2. All the images have been reconstructed
  with the same beam, which is plotted on the bottom left-hand
  corner. On each image we report the observing frequency (and band),
  and the observing epoch.}
\label{appendix_polla}
\end{center}
\end{figure*}

\addtocounter{figure}{-1}
\begin{figure*}
\begin{center}
\includegraphics{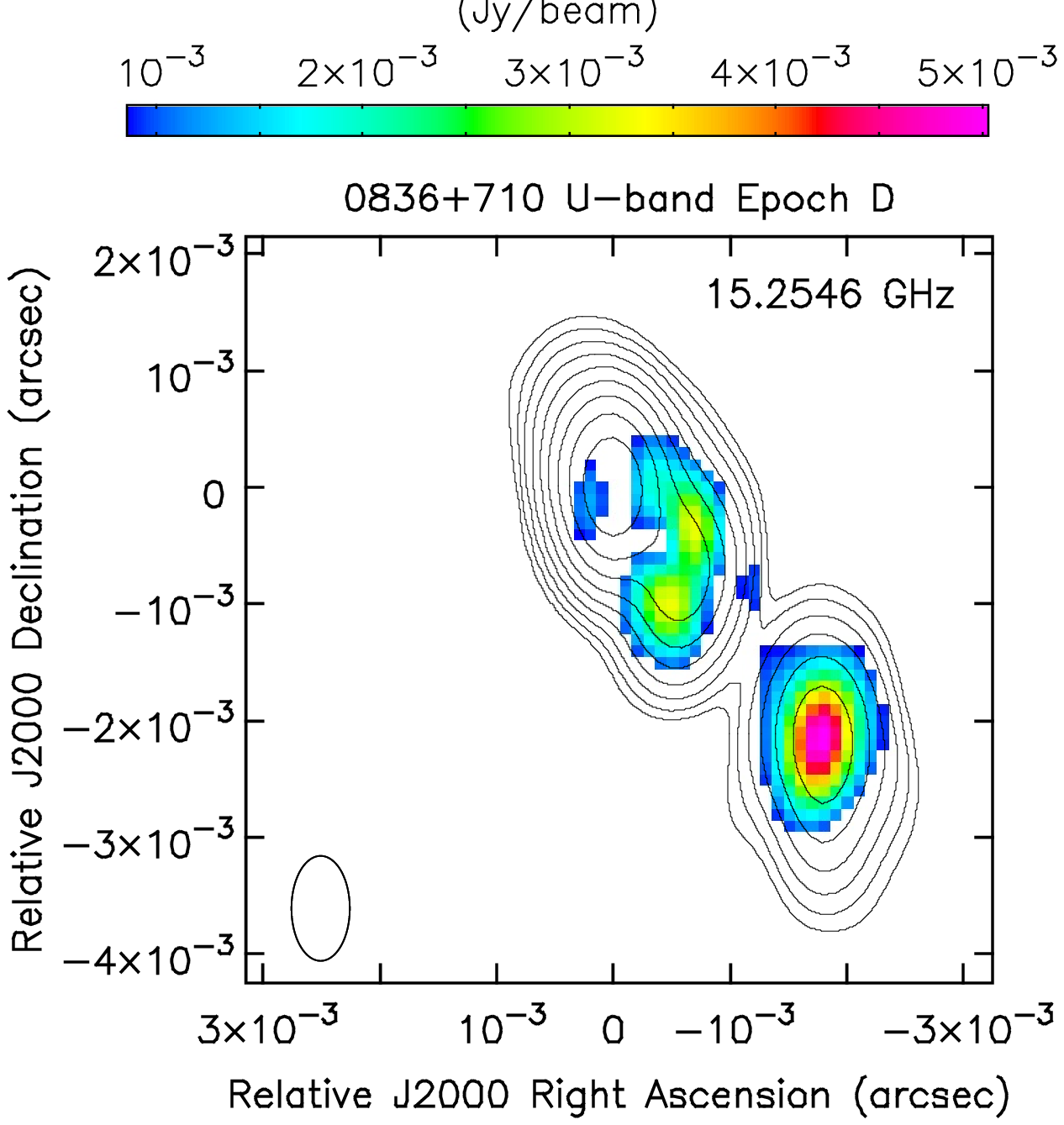}
\includegraphics{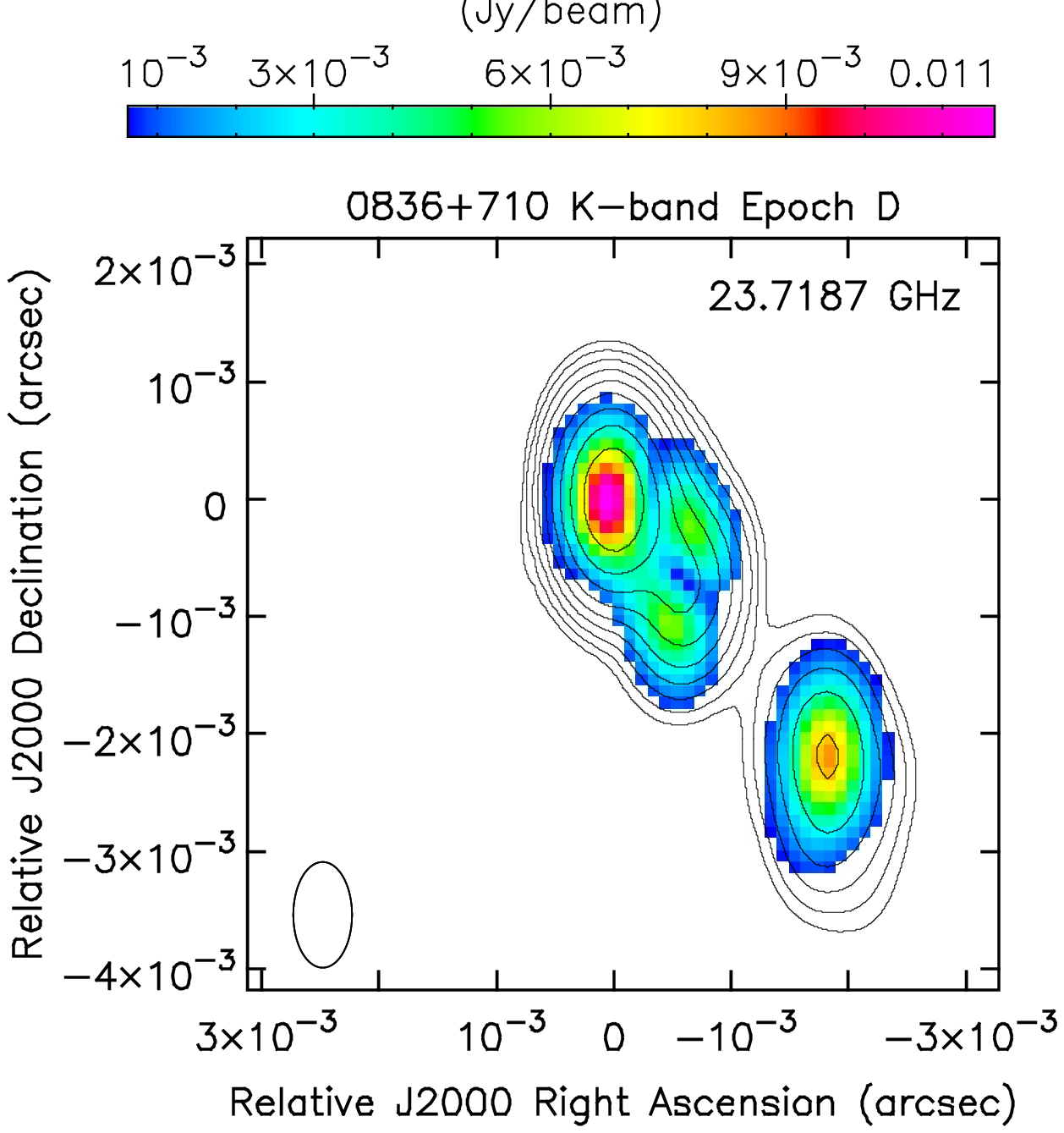}
\includegraphics{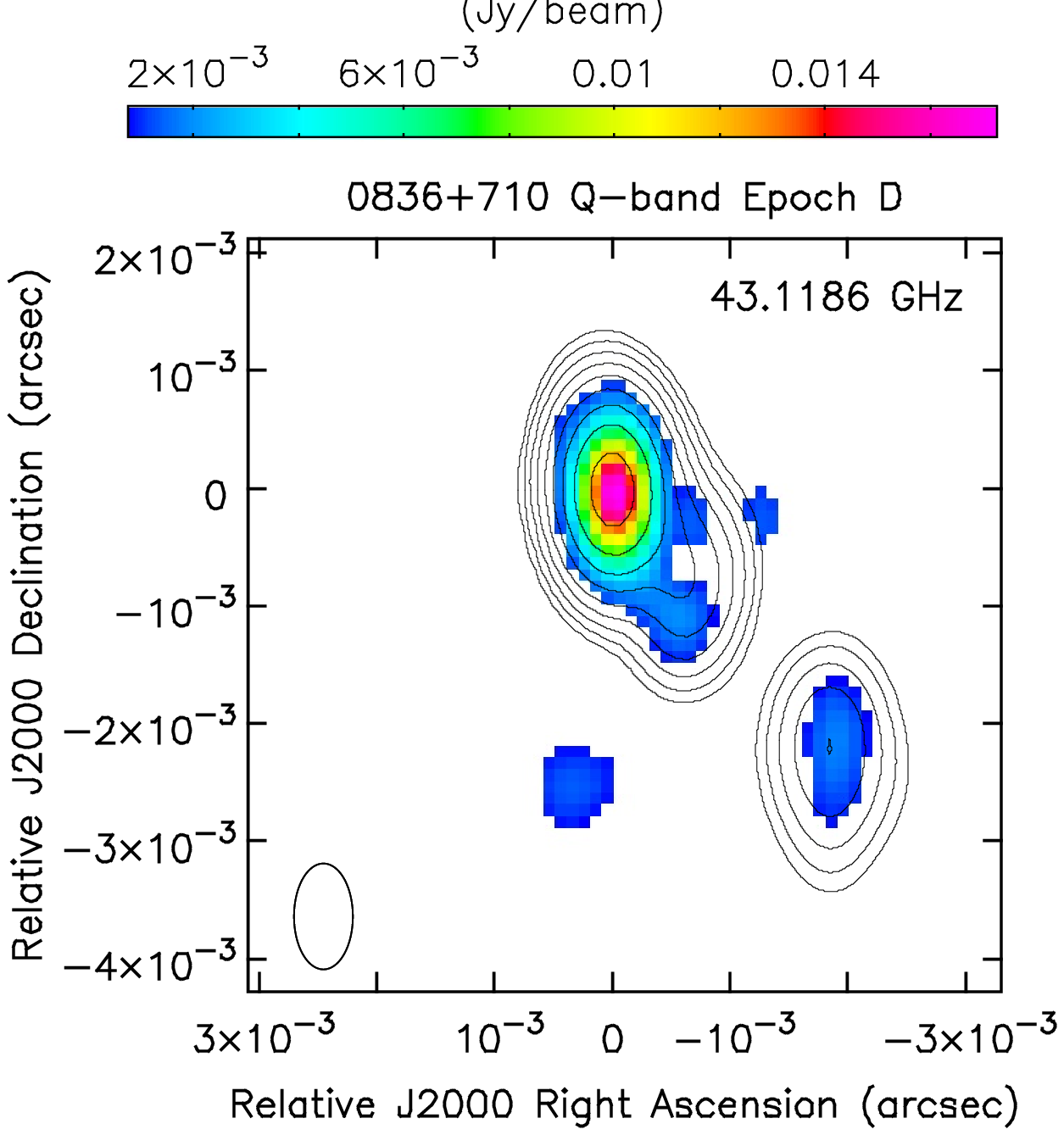}
\includegraphics{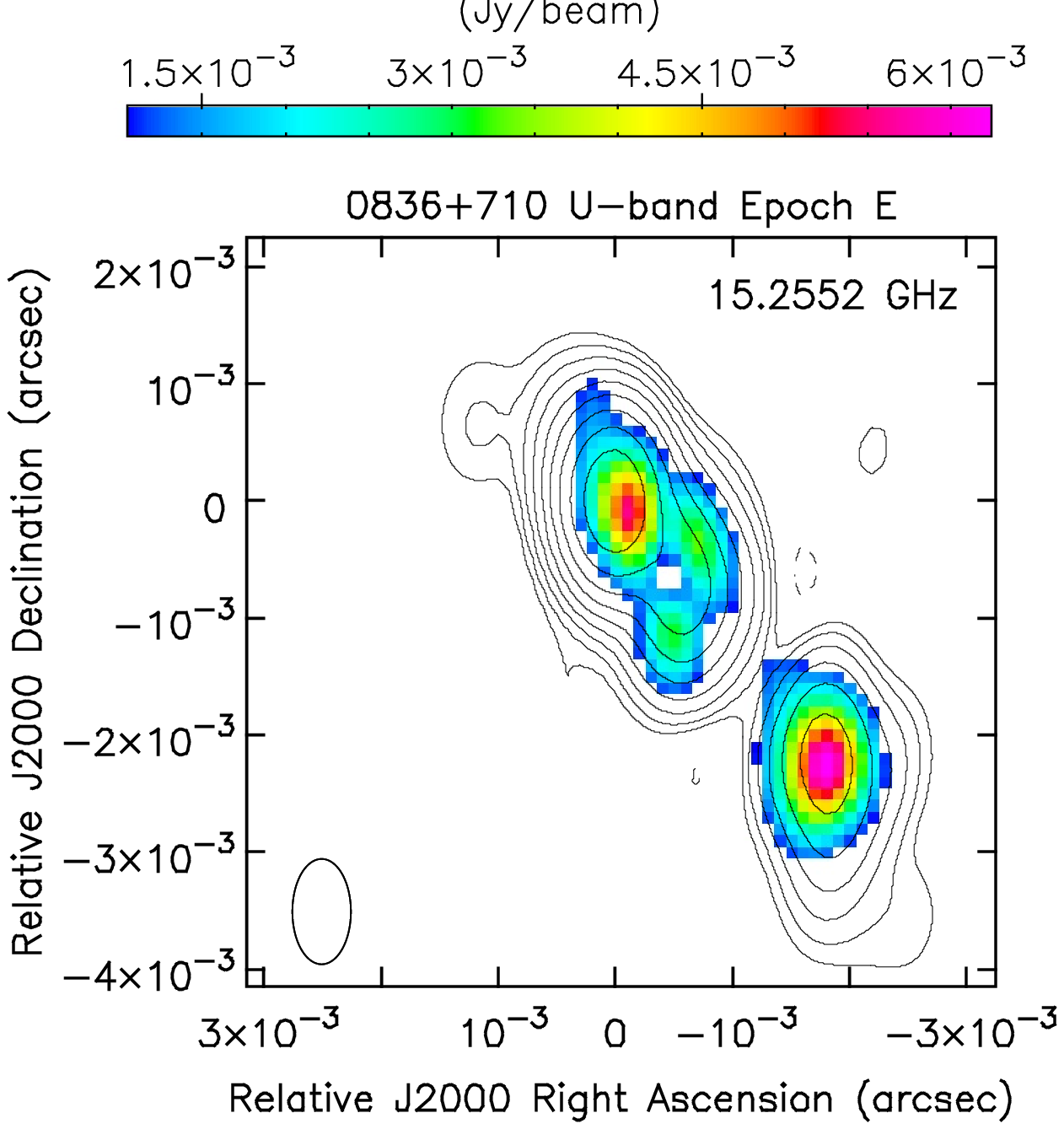}
\includegraphics{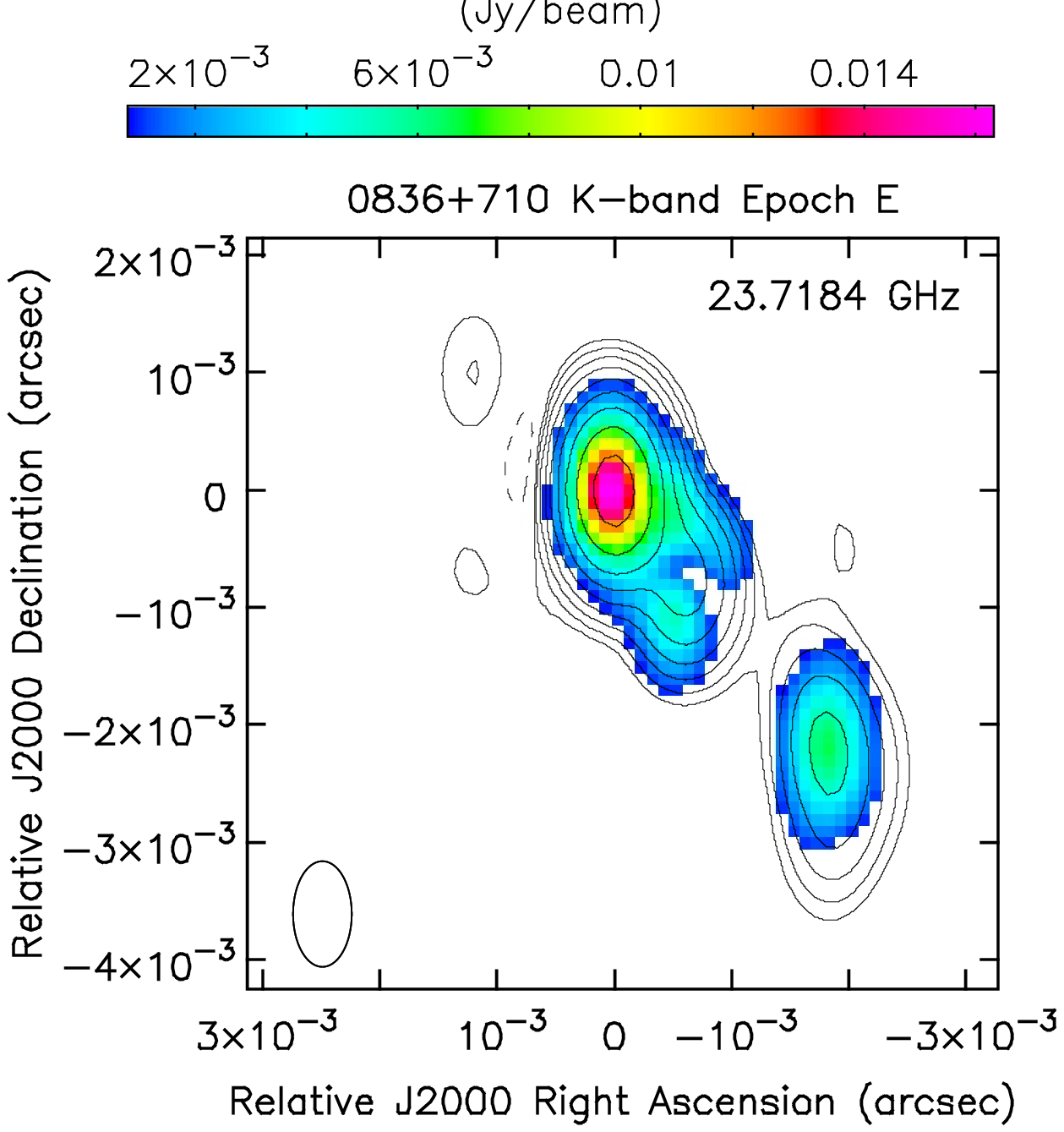}
\includegraphics{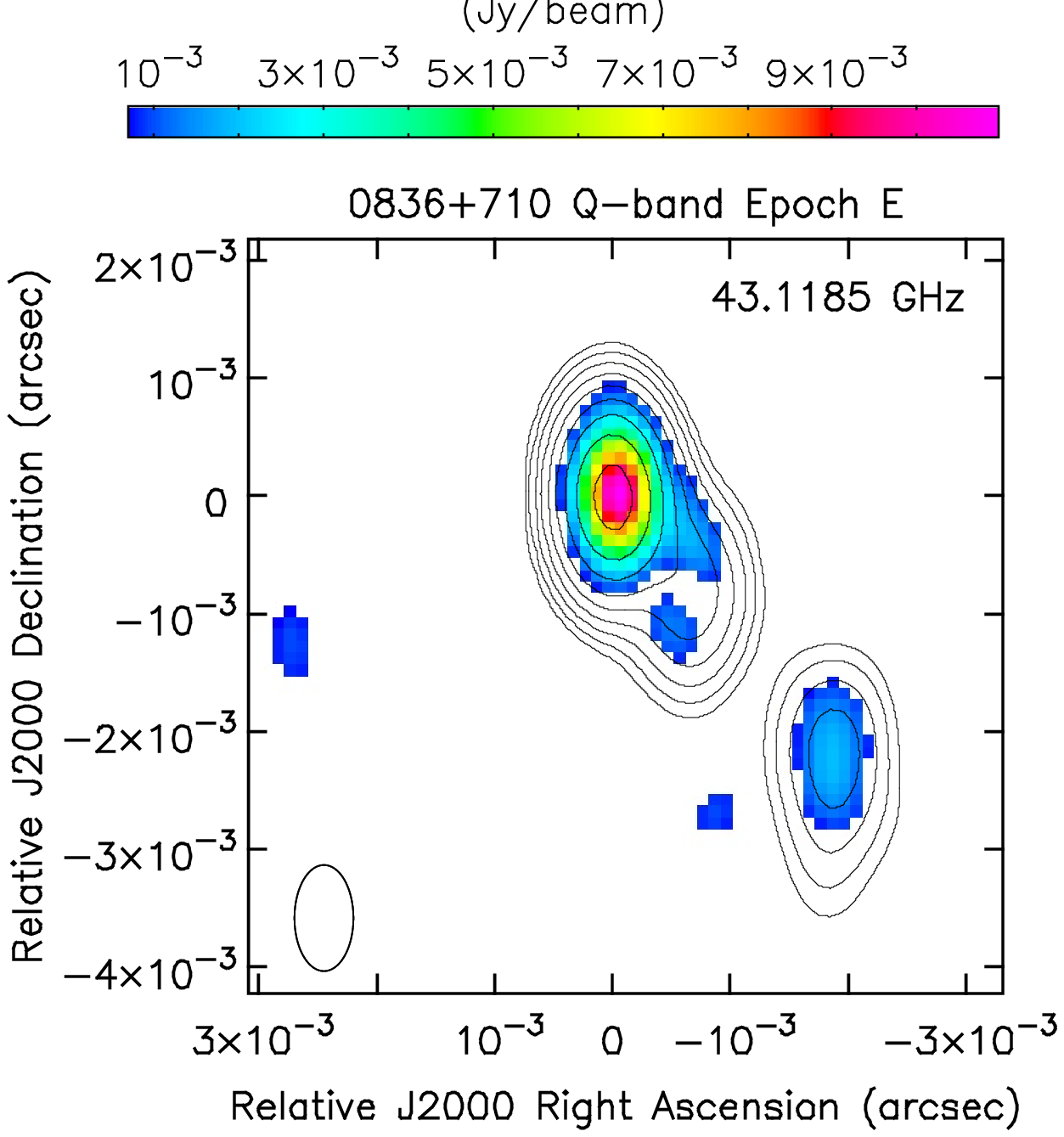}
\includegraphics{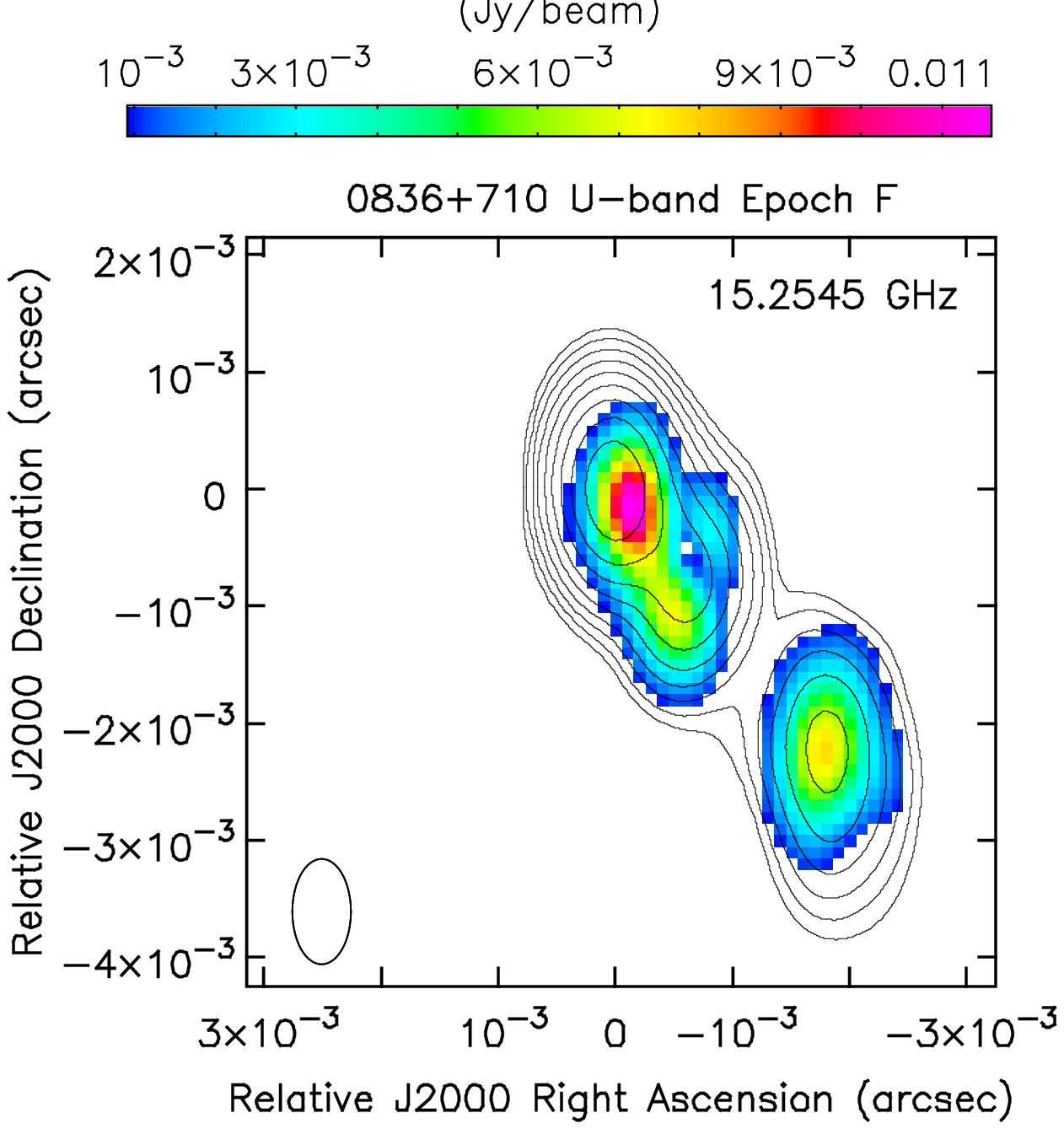}
\includegraphics{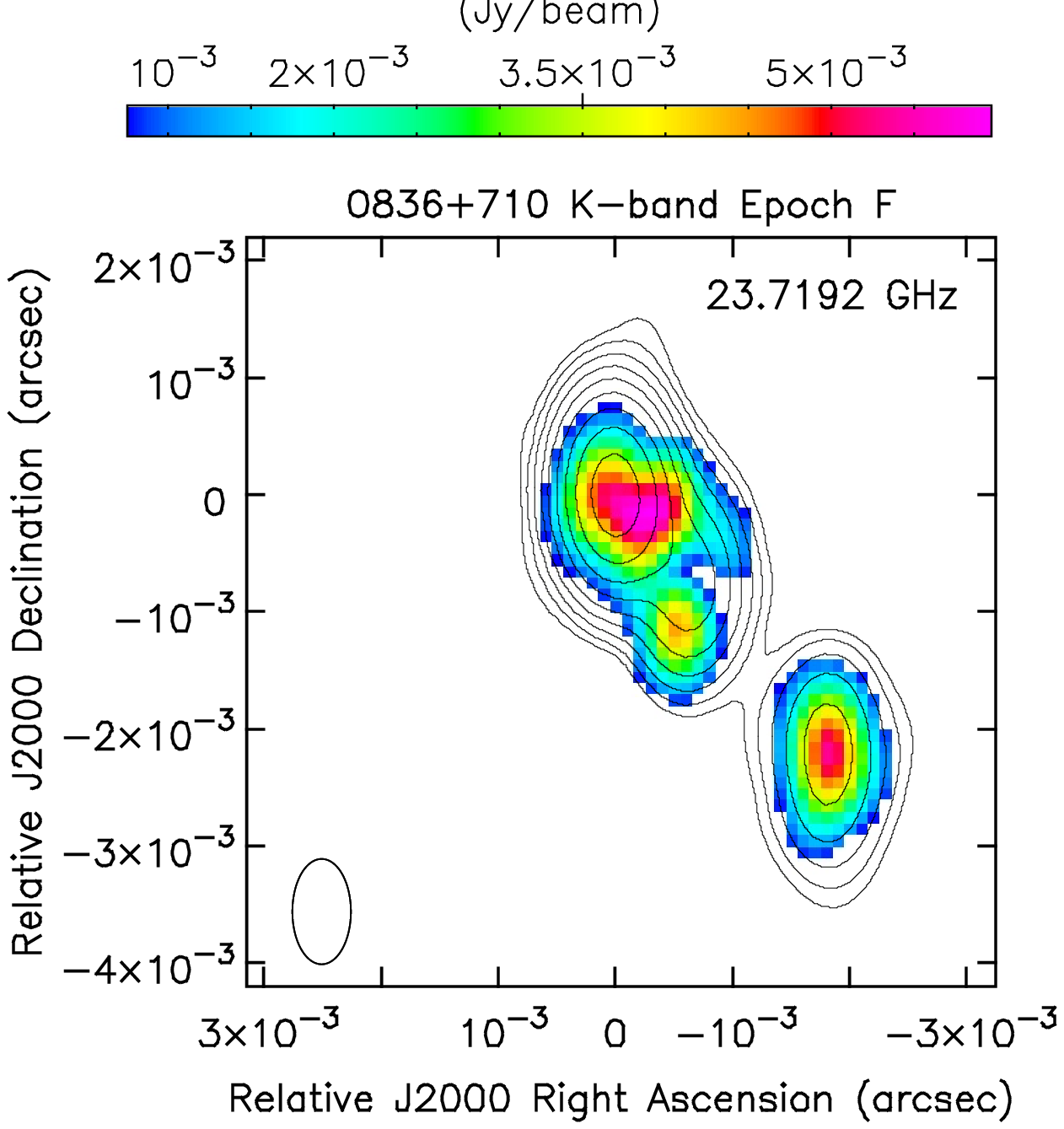}
\includegraphics{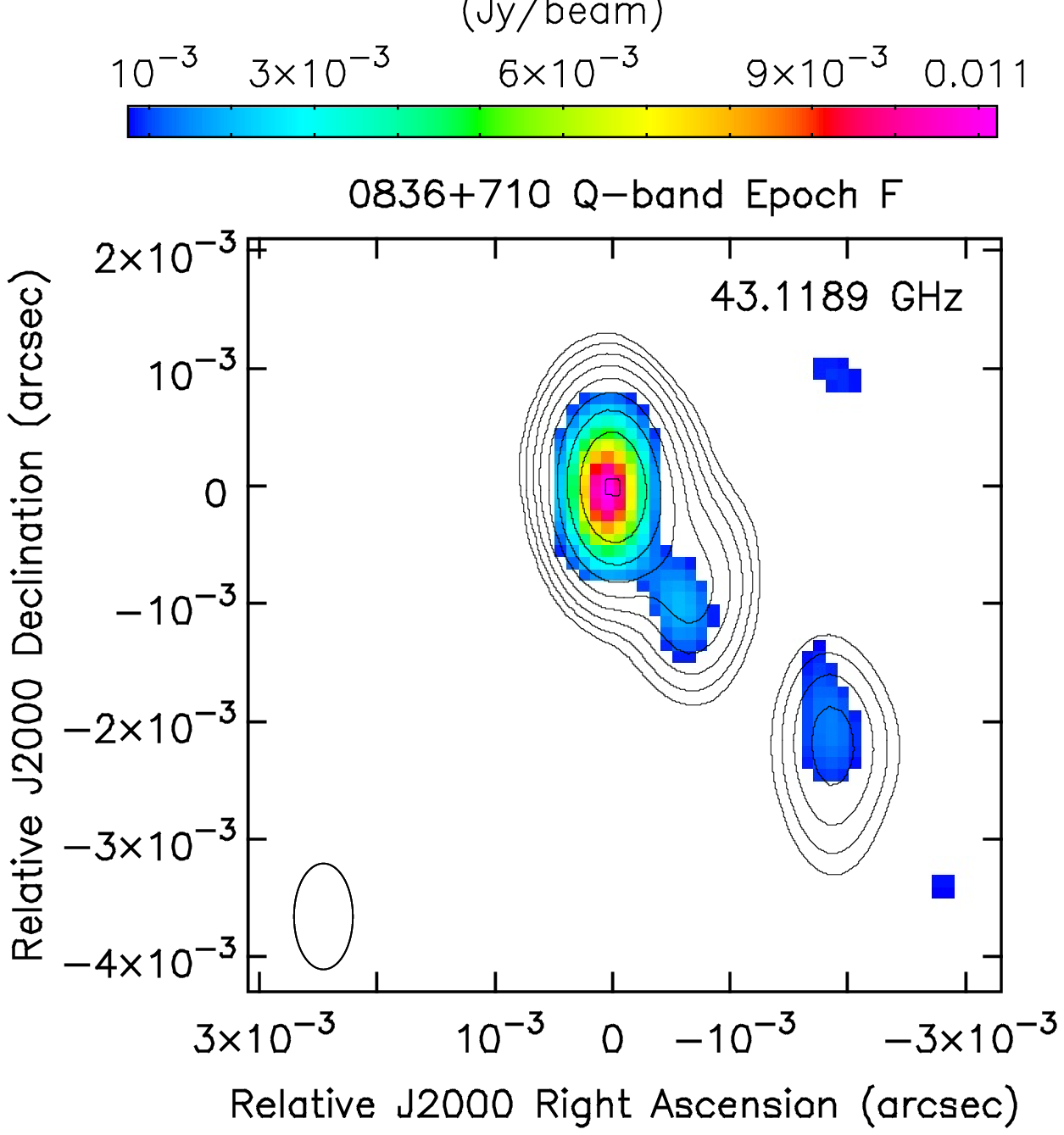}
\vspace{21cm}
\caption{continued.}
\end{center}
\end{figure*}


\begin{thebibliography}{}

\bibitem[Abdo et al.(2015)]{abdo15}
Abdo, A.A., et al. 2015, ApJ, 799, 143 

\bibitem[Abdollahi et al.(2018)]{abdollahi18}
Abdollahi, S., et al. 2019, Science, 362, 1031


\bibitem[Acero et al.(2016)]{acero16}  
Acero, F., et al. 2016, ApJS, 223, 2

\bibitem[Ackermann et al.(2015)]{ackermann15} 
Ackermann, M., et al. 2015, ApJ, 810, 14

\bibitem[Akyuz et al.(2013)]{akyuz13}
  Akyuz, A., et al. 2013, A\&A, 556,71
  
\bibitem[Asada et al.(2010)]{asada10} Asada, K., Nakamura, M., Inoue,
  M., Kameno, S., Nagai, H. 2010, ApJ, 720, 41 

\bibitem[Atwood et al.(2009)]{atwood09}
  Atwood, W. B., et al. 2009, ApJ, 697, 1071

\bibitem[Atwood et al.(2013)]{atwood13}
  Atwood, W. B., et al. 2013, 2012 Fermi Symposium proceedings - eConf C121028 (arXiv:1303.3514)

\bibitem[Ajello et al.(2017)]{ajello17}
  Ajello, M., et al. 2017, ApJS, 232, 18

\bibitem[Barthelmy et al.(2005)]{barthelmy05}
Barthelmy, S. D., et al. 2005, Space Sci. Rev., 120, 143 

\bibitem[Beckmann et al.(2009)]{beckmann09}
Beckmann, V., et al. 2009, A\&A 505, 417

\bibitem[Breeveld et al.(2010)]{breeveld10}
Breeveld, A. A., et al. 2010, MNRAS, 406, 1687

\bibitem[Broderick \& McKinney(2010)]{broderick10}
Broderick, A.E., McKinney, J.C. 2010, ApJ, 725, 750

\bibitem[Burrows et al.(2005)]{burrows05}
Burrows, D. N., et al. 2005, Space Sci. Rev., 120, 165  

\bibitem[Cardelli et al.(1989)]{cardelli89}
Cardelli, J. A., Clayton, G. C., Mathis, J. S. 1989, ApJ, 345, 245

\bibitem[Casadio et al.(2019)]{casadio19}
Casadio, C., et al. 2019, A\&A, 622, 158
  
\bibitem[Celotti \& Ghisellini(2008)]{celotti08}
Celotti, A., Ghisellini, G. 2008, MNRAS, 385, 1283

\bibitem[Ciprini(2015)]{ciprini15} 
Ciprini, S. 2015, The Astronomer's Telegram, 7870

\bibitem[Collmar et al.(2006)]{collmar06}
Collmar, W. 2006, ASPC, 350, 120	

\bibitem[Conway et al.(1992)]{conway92} 
Conway, J.E., Pearson, T.J., Readhead, A.C.S., Unwin, S.C., Xu, W., Mutel, R.L. 1992, ApJ, 396, 62
	
\bibitem[Costamante et al.(2018)]{costamante18} 
Costamante, L., Cutini, S., Tosti, G., Antolini, E., Tramacere, A. 2018, MNRAS, 477, 4749

\bibitem[D'Ammando et al.(2011)]{dammando11}
D'Ammando, F., et al. 2011, A\&A, 529A, 145

\bibitem[D'Ammando \& Orienti(2016)]{dammando16}
D'Ammando, F., \& Orienti, M. 2016, MNRAS, 455, 1881

\bibitem[Dominguez et al.(2011)]{dominguez11}
  Dominguez, A., et al. 2011, MNRAS, 410, 2556

\bibitem[Dominguez \& Ajello(2015)]{dominguez15}
  Dominguez, A., \& Ajello, M. 2015, ApJ, 813L, 34

\bibitem[Finke et al.(2010)]{finke10}
  Finke J. D., Razzaque S., Dermer C. D., 2010, ApJ, 712, 238

  \bibitem[Gabuzda et al.(2014)]{gabuzda14}
Gabuzda, D.C., Reichstein, A.R., O'Neill, E.L. 2014, MNRAS, 444, 172

\bibitem[Gabuzda et al.(2017)]{gabuzda17}
Gabuzda, D.C., Roche, N., Kirwan, A., Knuettel, S., Nagle, M.,
Houston, C. 2017, MNRAS, 472, 1792

\bibitem[Gehrels et al.(2004)]{gehrels04}
Gehrels, N., et al. 2004, ApJ, 611, 1005

\bibitem[Ghisellini et al.(2014)]{ghisellini14}
Ghisellini, G., et al. 2014, Nature, 515, 376

\bibitem[Gomez et al.(2011)]{gomez11}
G\'omez, J.L., Roca-Sogorb, M., Agudo, I., Marscher, A.P.,
Jorstad, S.G. 2011, ApJ, 733, 11

\bibitem[Hovatta et al.(2012)]{hovatta12}
Hovatta, T., Lister, M.L., Aller, M.F., Aller, H.D., Homan, D.C.,
Kovalev, Y.Y., Pushkarev, A.B., Savolainen, T. 2012, AJ, 144, 105
  
\bibitem[Hovatta et al.(2014)]{hovatta14}
Hovatta, T., et al. 2014, ApJ, 147, 143

\bibitem[Jorstad et al.(2007)]{jorstad07}
Jorstad, S.G., et al. 2007, AJ, 134, 799

\bibitem[Jorstad et al.(2013)]{jorstad13}
Jorstad, S., et al. 2013, EPJWC, 6104003
  
\bibitem[Jorstad et al.(2017)]{jorstad17}
Jorstad, S.G., et al. 2017, ApJ, 846, 98

\bibitem[Kalberla et al.(2005)]{kalberla05}
  Kalberla, P. M. W., Burton, W. B., Hartmann, D., Arnal, E. M., Bajaja, E., Morras, R., P{\"o}ppel, W. G. L 2005, A$\&$A, 440, 775

{\bibitem[Kirk et al.(1998)]{kirk98} Kirk, J. G., Rieger, F. M., Mastichiadis, A. 1998, A\&A, 333, 452}

\bibitem[Krichbaum et al.(1990)]{krichbaum90}
Krichbaum, T.P., Hummel, C.A., Quirrenbach, A., Schalinski, C.J.,
Witzel, A., Johnson, K.J., Muxlow, T.W.B., Qian, S.J. 1990, A\&A, 230, 271

\bibitem[Lister et al.(2009)]{lister09}
Lister, M.L., et al. 2009, AJ, 137, 3718

\bibitem[Lister et al.(2013)]{lister13}
Lister, M.L., et al. 2013, AJ, 146, 120

\bibitem[Lister et al.(2018)]{lister18}
Lister, M.L., Aller, M.F., Aller, H.D., Hodge, M.A., Homan, D.C.,
Kovalev, Y.Y., Pushkarev, A.B., Savolainen, T. 2018, ApJS, 234, 12
  
\bibitem[Lobanov(1998)]{lobanov98}
Lobanov, A.P 1998, A\&A, 330, 79

\bibitem[Mahmud et al.(2009)]{mahmud09}
Mahmud, M., Gabuzda, D.C., Bezrukovs, V. 2009, MNRAS, 400, 2

\bibitem[Marscher et al.(2008)]{marscher08}
Marscher, A.P., et al. 2008, Nature, 452, 966
  
\bibitem[Marscher et al.(2010)]{marscher10}
Marscher, A.P., et al. 2010, ApJL, 710, 126
  
\bibitem[Marscher(2014)]{marscher14}
Marscher, A.P. 2014, ApJ, 780, 87
  
\bibitem[Mattox et al.(1996)]{mattox96}
Mattox, J. R., et al. 1996, ApJ, 461, 396

\bibitem[Moretti et al.(2005)]{moretti05}
Moretti, A., et al. 2005, SPIE, 5898, 360

\bibitem[Oh et al.(2018)]{oh18}
Oh, K., et al. 2018, ApJS, 235, 4

\bibitem[Orienti et al.(2011)]{mo11}
Orienti, M., Venturi, T., Dallacasa, D., D'Ammando, F., Giroletti, M.,
Giovannini, G., Vercellone, S., Tavani, M. 2011, MNRAS, 417, 359

\bibitem[Orienti et al.(2013)]{orienti13}
Orienti, M., et al. 2013, MNRAS, 428, 2418
  
\bibitem[Orienti et al.(2014)]{mo14}
Orienti, M., D'Ammando, F., Giroletti, M., Finke, J., Ajello, M.,
Dallacasa, D., Venturi, T. 2014, MNRAS, 444, 3040

\bibitem[Pacholczyk(1970)]{pacho70}
Pacholczyk, A.G. 1970, Radio Astrophysics, W. H. Freeeman, San Franciso

\bibitem[Paliya(2015)]{paliya15} 
Paliya V. 2015, ApJ, 804, 74

\bibitem[Paliya et al.(2019)]{paliya19} 
Paliya V. et al. 2019, ApJ, 871, 211

\bibitem[Perucho et al.(2012)]{perucho12}
Perucho, M., Kovalev, Y.Y., Lobanov, A.P., Hardee, P.E., Agudo,
I. 2012, ApJ, 749, 55

\bibitem[Petropoulou et al.(2016)]{petropoulou16}
Petropoulou, M., Giannios, D., Sironi, L. 2016, MNRAS, 462, 3325
  
\bibitem[Poole et al.(2008)]{poole08}
Poole, T. S., et al. 2008, MNRAS, 383, 627

\bibitem[Pushkarev et al.(2009)]{pushkarev09}
Pushkarev, A. B., Kovalev, Y. Y., Lister, M., Savolainen, T. 2009, A\&A, 507, L33


\bibitem[Raiteri et al.(2014)]{raiteri14}
Raiteri, C. M., et al. 2014, MNRAS, 442, 629

\bibitem[Rohlfs(1986)]{rohlfs86}
Rohlfs, K., {\it Tools of radio astronomy}, 1986, Springer-Verlag
  
\bibitem[Roming et al.(2005)]{roming05}
Roming, P. W. A., et al. 2005, Space Sci. Rev., 120, 95 

\bibitem[Sambruna et al.(2007)]{sambruna07}
Sambruna, R., Tavecchio, F., Ghisellini, G., Donato, D., Holland, S. T.,
Markwardt, C. B., Tueller, J., Mushotzky, R. F. 2007, ApJ, 669, 884

\bibitem[Schlafly \& Finkbeiner(2011)]{schlafly11}
Schlafly, E. F. \& Finkbeiner, D. P. 2011, ApJ, 737, 103

\bibitem[Stickel \& Kuehr(1993)]{stickel93}
Stickel, M., Kuehr, H. 1993, A\&AS, 100, 395

\bibitem[Tagliaferri et al.(2015)]{tagliaferri15}
Tagliaferri, G., et al. 2015, ApJ, 807, 167

\bibitem[Tavecchio et al.(2000)]{tavecchio00} 
Tavecchio, F., et al. 2000, ApJ, 543, 535

\bibitem[Tavecchio et al.(2010)]{tavecchio10}
Tavecchio, F., et al. 2010, MNRAS, 405, L94

\bibitem[Thompson et al.(1993)]{thompson93}
  Thompson, D. J., et al. 1993, ApJ, 415, L13

\bibitem[Vercellone et al.(2010)]{vercellone10}
  Vercellone, S., et al. 2010, ApJ, 712, 405
 
\bibitem[Vercellone et al.(2019)]{vercellone19}
  Vercellone, S., et al. 2019, A\&A, 621A, 82  
 
\bibitem[Wilms et al.(2000)]{wilms00} Wilms, J., Allen, A., McCray, R. 2000, ApJ, 542, 914   

\end{thebibliography}
\end{document}